\documentclass{aa}  
\usepackage{longtable,lscape,amssymb}
\usepackage{booktabs}
\usepackage{graphicx}
\usepackage{lipsum} 
\usepackage{pdflscape}
\usepackage{comment}
\usepackage{float}
\usepackage{soul}
\usepackage{txfonts}
\makeatletter
\newcommand\footnoteref[1]{\protected@xdef\@thefnmark{\ref{#1}}\@footnotemark}
\makeatother

\def\mstar  {$M_{\star}$}
\def\macc   {$\dot{M}_{\rm acc}$}
\def\lacc   {$L_{\rm acc}$}

\def\msun {$M_{\odot}$}

\def\lstar {$L_\star$}

\def\laccnoise   {$L_{\rm acc,noise}$}

\newcommand{\Lacc}{{$L_{\rm acc}$}}
\newcommand{\Macc}{{$\dot{M}_{\rm acc}$}}
\newcommand{\teff}{$T_{\rm eff}$}
\newcommand{\logg}{$\log g$}

\usepackage{xcolor}
\usepackage{booktabs}

\begin{document} 
\defcitealias{manara13a}{MTR13}
\defcitealias{manara17b}{MFA17}
\defcitealias{HH14}{HH14}
    \title{FitteR for Accretion ProPErties of T Tauri stars (FRAPPE):\\ A new approach to use Class III spectra to derive\\ stellar and accretion properties\thanks{Based on observations collected at the European Southern Observatory under ESO programmes 085.C-0764(A), 093.C-0506(A), 106.20Z8.004, 106.20Z8.006, 106.20Z8.008, 109.23D4.001, 110.23P2.001}}

    \titlerunning{An interpolated grid of X-Shooter photospheric templates of pre-main sequence stars}
    \authorrunning{Claes et al.}
    \author{
        R.A.B. Claes \inst{\ref{instESO}}
        \and 
        J. Campbell-White \inst{\ref{instESO}}
        \and
        C.F. Manara\inst{\ref{instESO}}
        \and
        A. Frasca\inst{\ref{AFrasca}}
        \and
        A. Natta\inst{\ref{instDIAS}}
        \and
        J.M. Alcal\'{a}\inst{\ref{instNA}}
        \and \\
        A. Armeni \inst{\ref{instTubingen}}
        \and 
        M. Fang\inst{\ref{instPM},\ref{instHefei}}
        \and
        J.B. Lovell\inst{\ref{Lovell}}
        \and 
        B. Stelzer\inst{\ref{instTubingen}}
        \and
        L. Venuti\inst{\ref{instSETI}}
        \and
        M. Wyatt \inst{\ref{Wyatt}}
        \and
        A. Queitsch\inst{\ref{instTubingen}}
        }
    \institute{European Southern Observatory, Karl-Schwarzschild-Strasse 2, 85748 Garching bei M\"unchen, Germany\label{instESO}\\
              \email{Rik.Claes@eso.org}
        \and 
        INAF - Osservatorio Astrofisico di Catania, via S. Sofia, 78, 95123 Catania, Italy\label{AFrasca}
        \and
        School of Cosmic Physics, Dublin Institute for Advanced Studies, 31 Fitzwilliam Place, Dublin 2, Ireland\label{instDIAS}
        \and
        INAF -- Osservatorio Astronomico di Capodimonte, via Moiariello 16, 80131 Napoli, Italy\label{instNA}
        \and
        Purple Mountain Observatory, Chinese Academy of Sciences, 10 Yuanhua Road, Nanjing210023, China\label{instPM}
        \and
        University of Science and Technology of China, Hefei 230026, China\label{instHefei}
        \and
        Center for Astrophysics, Harvard \& Smithsonian, 60 Garden Street, Cambridge, MA 02138-1516, USA\label{Lovell}
        \and 
        Institut f\"ur Astronomie und Astrophysik, Eberhard Karls Universit\"at Tübingen, Sand 1, 72076 T\"ubingen, Germany\label{instTubingen}
        \and
        SETI Institute, 339 Bernardo Ave., Suite 200, Mountain View, CA 94043, USA\label{instSETI}
        \and
        Institute of Astronomy, University of Cambridge, Madingley Road, Cambridge CB3 0HA, UK\label{Wyatt}
        }
   \date{}

 \abstract{Studies of the stellar and accretion properties of classical T Tauri stars (CTTS) require comparison with photospheric spectral templates. The use of low-activity, slowly-rotating field dwarfs or model spectra can be advantageous for the determination of stellar parameters, but it can lead to an overestimate of mass accretion rate, since both classes of templates do not include the emission of the active chromosphere present in young stars. Observed spectra of non-accreting young stars are best suited to this purpose. Using such templates comes with the downside of a limited number of available templates and observational uncertainties on the properties of the templates.}
 {Here we aim at expanding the currently available grid of wide-wavelength coverage observed spectra of non-accreting stars with additional new spectra and an interpolation method that allows us to obtain a continuous grid of low resolution spectra ranging from spectral type G8 to M9.5, while also mitigating observational uncertainties. This interpolated grid is then implemented in the self-consistent method to derive stellar and accretion properties of CTTS. With the new templates, we aim to estimate a lower limit on the accretion luminosities that can be obtained through a study of the UV excess emission using observed templates.}
 {We analyse the molecular photospheric features present in the VLT/X-Shooter spectra of the targets to perform a spectral classification, including estimates of their extinction. 
 
 We apply a non-parametric fitting method to the full grid of observed templates to obtain an interpolated grid of templates. Both the individual templates and interpolated grid are provided to the community. We implement this grid to improve the method to self-consistently derive stellar and accretion properties of accreting stars. We use the uncertainties on our interpolated grid to estimate a lower limit on the accretion luminosity that we can measure with this method.}
 {Our new method, which uses a continuous grid of templates, provides results that are consistent with using individual templates but it significantly improves the reliability of the results in case of degeneracies associated to the peculiarities of individual observed templates. We find that the measurable accretion luminosities ranges from $\sim 2.7$ dex lower than the stellar luminosity in M5.5 stars to $\sim 1.3$ dex lower for G8 stars. For young stars with masses of $\sim 1M_{\odot}$ and ages of 3-6 Myr this limit translates into an observational limit of mass accretion rate on the order of $10^{-10} \rm M_{\odot}/yr$. This limit is higher than the lower limit on the measurable mass accretion rate when using the various emission lines present in the spectra of young stars to estimate the accretion rate. An analysis of these emission lines allows us to probe lower accretion rates, pending a revised calibration of the relationships between line and accretion luminosities at low accretion rates}
 {The implementation of an interpolated grid of observed templates allows us to better disentangle degenerate solutions, leading to a more reliable estimate of accretion rates in young accreting stars.} 
 


   \maketitle
%

\section{Introduction}

Young stars are surrounded by circumstellar disks in which planets form \citep[e.g.,][]{williams11}. These disks dissipate in a few Myr \citep[e.g.,][]{fedele10}. The processes governing their evolution and dispersal, and their impact on planet formation remain open questions \citep[e.g.,][for a review]{Manara2023}. The first proposed mechanism involves the gradual spread of the disk due to viscosity, resulting in most of the material accreting onto the central star, with a small fraction of material carrying the angular momentum outward \citep{lyndenbell74,hartmann98}. An alternative scenario sees the evolution dominated by the ejection of angular momentum from the disk in MHD winds \citep{Blandford1982,Lesur21,Lesur2021b,tabone21}.  At the same time, high-energy radiation from the star, such as far-ultraviolet (FUV), extreme ultraviolet (EUV), or X-ray radiation, removes material from the inner disk in the form of photoevaporative winds. This process may potentially create a gap in the inner disk and ultimately lead to the dissipation of the disk \citep[e.g.,][]{clarke2001,alexander14,Ercolano2017}. Understanding the impact of viscosity, MHD winds and internal photoevaporation is critical for our understanding planet formation.
Indeed, while these processes are ongoing, the dust grains present in the disks are expected to grow from small grains into planetesimals \citep[e.g.;][]{Johansen2007,Booth2016,Carrera2017,Carrera2021}. 
These planetesimals can further grow into planetary systems \citep[e.g.;][]{Alibert2005,Liu2019,Lyra2023}, or may be revealed in the form of collisionally active debris disks once the gas in the disk has dissipated \citep[e.g.;][]{Wyatt2008,Hughes2018,Lovell2021}.

Each of these disk evolution mechanisms makes specific predictions about the time evolution of accretion rates through the disk, and thus on the central star. Determining the mass accretion rate of young stars is therefore a key parameter to shed light on whether viscous or MHD-wind driven evolution dominates disk evolution and which role photevaporation plays in dissipating the disk \citep[e.g.,][]{mulders17,lodato17,somigliana20,somigliana23}.

Numerous studies attempted to constrain disk evolution by analyzing theoretically expected and observed relationships between mass accretion rate and stellar or disk properties, such as disk and stellar mass \citep[e.g.,][]{dullemond06,hartmann06,alexander06,ercolano14,mulders17,rosotti17,manara20}. Currently no definitive conclusions have been drawn in favor of either model.  

Recently, \citet{Alexander23} showed that the distribution of $\gtrsim 300$ homogeneous accretion rate measurements of young stars in the $0.5 - 1.0$ \msun\, range should suffice to distinguish between a viscous+photoevaporation and a MHD-wind driven disk evolution scenario. Currently, only $\sim 100$ accretion rate measurements are available in this mass range, and this sample lacks homogeneous upper limits for the non detections. Providing such upper limits could enable even stronger statistical approaches to further constrain model parameters. \cite{Ercolano2023} used the accretion rate distribution of a sample of potential extreme low accretors presented by \cite{Thanathibodee2022,Thanathibodee2023} to discern between different types of photoevaporation that could dissipate the disk at the end of a viscously driven evolution scenario. This sample was limited as it only contained 24 sources. A thorough characterisation of more low accretors can therefore provide valuable constraints on disk evolution processes.

Tracers of the accretion and outflow processes can be found in the spectra of young stars. 
The current paradigm for accretion in young, low-mass ($ \lesssim 1$ \msun) stars is known as magnetospheric accretion \citep[][for a review]{hartmann16}.
In this scenario, accreting material is funneled by the star's magnetosphere onto its surface, creating hotspots that emit excess continuum radiation in the ultraviolet and optical spectra \citep{Calvet1998}.
The hot gas associated with the accretion streamers leads to the appearance of various prominent emission lines (e.g., $\rm H \alpha$, $\rm H \beta$ Ca K, $\rm Pa\beta$,...) in the spectra \cite[e.g.,][]{muzerolle01,Campbell-White2021}.

The spectroscopic study of young stars can, therefore, provide vital constraints on the physics governing disk evolution. Accurately representing the emission from the central star is crucial in these investigations.
Several methods are employed to estimate accretion rates in young stars. As a first option one can fit the line profile using magnetospheric accretion flow models \citep{Muzerolle2001,Thanathibodee2023}. The downside of these models is that they assume the stellar magnetic field to be both dipolar, aligned with the stellar rotation axis and disk, which is not necessarily accurate \citep[e.g.,][]{Donati2008, Alencar2012, Singh2024}.
In a second method, the line luminosity of various emission lines is converted into an accretion luminosity using empirical relations  \citep[e.g.,][]{Herczeg2008,alcala14,alcala17}. These relations are calibrated using the third method which is the only one that directly probes the energy released at the accretion shock. This method consists of measuring the UV continuum excess emission \citep[e.g.,][]{valenti93,Calvet2004,Herczeg2008,rigliaco12,Ingleby2013,Pittman2022,Robinson2022}. In order to self-consistently determine both the stellar and accretion parameters, the UV-excess and the spectral features in the optical part of medium-resolution flux-calibrated spectra must be fit simultaneously \citep{manara13b}. For this aim, both the photospheric and chromospheric emission in the UV part of the spectrum must be correctly accounted for \citep{Ingleby2011,manara13a}.

Available synthetic spectra \citep[e.g.,][]{allard11} do not fully reproduce the observed spectra of chromospherically active low gravity objects, such as pre-main sequence low-mass stars. In particular, none of the current models contain emission originating in the chromospheres of young stars, since only the stellar photospheres are modeled, which dominate only at wavelengths longer than the Balmer Jump. At shorter wavelengths, bound free and free free emission originating in the chromosphere starts to dominate the continuum \citep{Houdebine1996,Franchini1998}. The chromosphere also emits in several spectral lines that are used to constrain accretion properties of accreting young stellar objects (YSOs,  \citealt{Stelzer2013}). The spectra of field dwarfs will also poorly represent those of young stars as the former have a significantly higher $\log{g}$. The best solution appears to use spectra of non-accreting young stars. Class III stars \citep{Greene1994}, which are defined according to their infrared classification where they display a lack of infrared excess emission $d\log(\nu F_\nu) / d\log( \nu) <-1.6$ \citep{williams11}, make ideal candidates for such templates, as this category often overlaps with Weak lined T Tauri Stars (WTTS), which present much fainter emission lines than CTTS.

A grid of Class III templates with broad wavelength coverage medium-resolution spectra obtained with the X-Shooter spectrograph on the ESO Very Large Telescope (VLT) was previously provided to the community by \citet{manara13a} (hereafter \citetalias{manara13a}) and further expanded by \citet{manara17b} (hereafter \citetalias{manara17b}). This grid includes 41 spectra and contains spectral types ranging from G4 to M8. Both \citetalias{manara13a} and \citetalias{manara17b} used these spectra to estimate the contribution of chromospheric emission to the emission lines that are commonly used for the determination of mass accretion rates. To study the contribution of the chromosphere to the UV continuum emission, \citet{ingleby11} compared the spectrum of the M0 Class III RECX-1 to the photospheric spectrum of a standard dwarf star. Here it was found that attributing this excess to accretion would result in an estimated accretion luminosity of $\log(L_{\rm acc}/L_{\star})  = -1.3$. The chromospheres of young stars can therefore significantly affect measured accretion rates in low accretors. Moreover, the Balmer continuum excess emission and the Balmer Jump are more difficult to detect in the spectra of early-type (<K3) YSOs than in the later types due to the lower contrast between photospheric emission and accretion induced continuum excess emission \citep{Herczeg2008}.  
Constraints on the influence of the chromospheric emission in the UV on measurements of (low) accretion rates is still lacking. A better understanding of the chromospheres of young stars is therefore needed to characterise the lowest accretors to better understand the late phases of disk dispersal and hence, provide constraints on disk evolution models.

Here we expand the library of X-Shooter Class III templates previously provided by \citetalias{manara13a} and \citetalias{manara17b} by an additional 18 templates and present a method for interpolating between them.
We also aim to constrain the influence of the UV continuum chromospheric emission on determination of mass accretion rates. The paper is structured as follows.  
The sample selection, observations and data reduction are described in Sect. \ref{ObsAndDat}. The analysis of the stellar properties of our sample is described in Sect. \ref{StellarProp}. In Sect. \ref{CombGrid} the final grid and our method for interpolating it are discussed. Sect. \ref{fitter} discusses an application of this interpolated grid using a self consistent method to derive the mass accretion rates and validate this method on a set spectra of accreting young stars in the Chamaeleon~I star-forming region. In Sect. \ref{sect:limit} we obtain a lower limit on the mass accretion rates that we can measure from the UV excess, taking into account uncertainties on the chromospheric emission and discuss the implications for studies of mass accretion rate (\Macc) in CTTS. Finally, we summarize our conclusions in Sect. \ref{sect:concl}.

\section{Sample, observations, and data reduction}\label{ObsAndDat}

The new targets considered in this work come mainly from the sample of \citet{Lovell2021}, who studied 30 Class III stars (age $\lesssim$10 Myr) using the Atacama Large Millimeter Array (ALMA). \citet{Lovell2021} reaffirmed that these sources are Class III YSO through a comparison of the K-band (2MASS) and either
12$\mu$m (WISE) or 24$\mu$m (Spitzer) fluxes. In addition to this \citet{Lovell2021} fitted the SEDs of these targets and found no significant NIR excess emission. These stars are likely members of of the Lupus star forming region, although their membership still needs to be confirmed \citep{Lovell2021,Michel2021}. A total of 19 objects of this sample were recently observed using X-shooter (Pr.ID. 109.23D4.001, 110.23P2.001, PI Manara), a broad-band, medium-resolution, high-sensitivity spectrograph mounted on the ESO/VLT. The two components of a binary system, THA15-36A and THA15-36B, were observed simultaneously in the slit. Three additional archival X-Shooter spectra were available for the targets MT Lup and NO Lup (Pr.ID. 093.C-0506(A), PI Caceres) and MV Lup (Pr.ID. 085.C-0764(A), PI G\"unther). Other targets in the sample of \citet{Lovell2021} were already included in the grid of \citetalias{manara13a} and \citetalias{manara17b}.  In addition to this, we use in this work two targets observed with VLT/X-shooter in the PENELLOPE Large Program (Pr.ID. 106.20Z8, \citealt{manara21}), namely RECX-6 and RXJ0438.6+1546. RECX-6 was identified as a class III YSO by \citet{Sicilia2009} based on a fit of its SED. RXJ0438.6+1546 was already included in the grid of \citetalias{manara17b}, who selected their targets based on available Spitzer data.
The latter was already observed with X-Shooter and included in the sample of \citetalias{manara17b}, but the new observations are used here. 
In total the targets considered in the analysis in this work are 24, observed in 23 observations.

The wavelength range covered with X-Shooter is divided into three arms, the UVB arm ($\lambda\lambda \sim 300-550$nm), the VIS arm ($\lambda\lambda \sim 500-1050$nm), and the NIR arm ($\lambda\lambda \sim 1000-2500$nm). Different slit widths were used for the different arms and for fainter or brighter targets. For brighter sources, we used $1\farcs0$, $0\farcs4$, $0\farcs4$ wide slits in the UVB, VIS and NIR arm respectively. These slit widths provide typical spectral resolutions of $R \sim 5400$, $18400$ and $11600$ in the three arms.  Fainter, later spectral type objects were observed using $1\farcs0$, $0\farcs9$, $0\farcs9$ in the three arms, providing resolutions of $R \sim 5400$, $8900$ and $5600$. For MV lup, RXJ0438.6+1546, MT Lup  and NO Lup, the $0\farcs5$ wide slits were used in the UVB arm, resulting in spectra with resolution $R \sim 9700$. The observations with the slits with the aforementioned widths were preceded by short exposures with a slit with the significantly larger width of $5\farcs0$ in all arms to obtain spectra not affected by slit losses to be used for absolute flux calibration. The only exception of the use of the wide slits were MV Lup, MT Lup, and NO Lup. 

Appendix \ref{obsLog} contains a log of the observations presented here. In the observing log, several targets are highlighted as spatially unresolved binaries. We exclude these spectra from our sample since they can not be used as templates to represent the stellar emission of individual targets and an analysis of these spectra is outside the scope of this paper. In particular, we exclude CD-39 10292, THA15-38, CD-35 10498, V1097 Sco and NN Lup. NN Lup was discovered to be SB2 binaries in our analysis with the ROTFIT tool (see Sect. \ref{rotfit}). CD-39 10292 is an SB2 binary first identified by \citet{Melo2001}. V1097 Sco and CD-35 10498 were identified as binaries by \citet{Zurlo21} and were not spatially resolved in the X-Shooter slit. THA15-38 was resolved as a binary in the acquisition image of ESPRESSO observations that are part of the PENELLOPE program. This was possible due to the excellent seeing conditions ($\lesssim 0\farcs6 $) during the night of the ESPRESSO observations. 

The data reduction was performed using the ESO X-Shooter pipeline v.4.2.2 \citep{xspipe} in the Reflex workflow \citep{reflex}. The pipeline executes the standard reduction steps: flat fielding, bias subtraction, extraction and combination of orders, rectification, wavelength calibration, flux calibration using a standard star observed during the same night and the final extraction of the 1D spectrum. The telluric lines were removed from the narrow slit observations using the molecfit tool \citep{molecfit1}. This was done by fitting the telluric features on the spectra themselves, rather than  by using a telluric standard star. 
As a final step, the narrow slit observations are rescaled to the continuum flux of the wide slit spectra to get flux-calibrated spectra, using the procedure developed for the PENELLOPE program \citep[see][]{manara21}. For the UVB arm this was done using a correction factor constant with wavelength. The VIS and NIR arms were rescaled using a factor with a linear dependency on wavelength. To test the flux calibration of our spectra they were compared to archival photometry. 
The overall agreement between the spectra and photometry is excellent ($\Delta {\rm mag} < 0.2$ mag) with the exception of 2MASSJ16090850-3903430 and THA15-36. This is however within the typical variation range at optical wavelength which is mostly due to starspots rotational modulation. A description of how we flux calibrated targets with $\Delta {\rm mag} > 0.2$ mag can be found in Appendix \ref{obsLog}. Appendix \ref{obsLog} also contains a description of how we obtained an accurate flux calibration for the apparent visual binary THA15-36, which was spatially resolved in the X-Shooter slit. \citet{Zurlo21} resolved this system using VLT/NACO, and argued that both components are unbound given their different {\it Gaia}  distances ($146.5 \pm 0.8$ pc for the primary and $154.3 \pm 1.8$ pc for the secondary).

\section{Stellar Parameters of the new Class III spectra}\label{StellarProp}

In this section, we derive the stellar parameters for all the 19 resolved stars whose spectra are presented here for the first time. This is done by first determining the spectral type from atomic and molecular features present in the spectra, then comparing these estimates with results from the fitting of individual absorption lines, and finally determining the stellar luminosity to place the targets on the Hertzsprung-Russel Diagram (HRD) to derive their stellar masses.

\subsection{Spectral type and extinction determination}\label{spectralTyping}

\begin{table*}[]
    \centering
        \caption{Spectral types and extinction obtained in this work. }
    \label{tab:SpT_new}
\begin{tabular}{l|lll|l|ccc}
\toprule
                    Name &   $A_V$ & SpT& Uncertainty & literature & HH14 &  TiO & Riddick et al. \\
\midrule
             CD-31\_12522 &  0.0 &      K0.5  & $\pm 1$ &        K2$^{\rm a}$ & K0.7 & K4.7 &               ... \\
          RXJ1608.9-3905 &  0.0 &      K2.0 &   $\pm 1$ &        K2$^{\rm b}$ & K1.7 & K4.5 &                ... \\
                   MV Lup &  0.0 &      K2.0 &    $\pm 1$ &       K2$^{\rm c}$ & K2.0 & K4.5 &                ... \\
RXJ0438.6+1546 &  0.0 &      K2.0 &    $\pm 1$ &       K2$^{\rm d}$  & K1.3 & K4.6 &                ... \\
  2MASSJ15552621-3338232 &  0.1 &      K6.0 &     $\pm 1$ &      K5$^{\rm e}$  & K8 & K5.6 &               ... \\
                   MT Lup &  0.1 &      K5.5 &    $\pm 1$ &       K5$^{\rm f}$& K8.7 & K4.9 &                ... \\
                   MX Lup &  0.0 &      K6.0 &     $\pm 0.5$ &      K6$^{\rm c}$ & K9.0 & K5.2 &               ... \\
          RXJ1607.2-3839 & 0.35 &      K7.5 &  $\pm 0.5$ &        K7$^{\rm c}$  & K9.0 & K7.7 &               ... \\
                   MW Lup &  0.0 &      K7.5 &  $\pm 0.5$ &        K7$^{\rm c}$ & K11.9 & K7.4 &               ... \\
                   NO Lup & 0.35 &      K7.5 &   $\pm 0.5$ &       K7$^{\rm c,g}$ & K9.9 & K7.5 &               ... \\
                THA15-43 &  0.0 &      K7.5 &  $\pm 0.5$ &        M0$^{\rm h}$ & M3.6 & K7.7 &              ... \\
               THA15-36A &  0.0 &      M0.5 &  $\pm 1$ &      ... & M1.1 & M0.8 &               ... \\
               THA15-36B &  0.5 &      M2.0 &   $\pm 1$ &     ... & M2.7 & M2.0 &               ...\\
                  RECX-6 &  0.0 &      M3.0 &   $\pm 0.5$ &       M3$^{\rm i}$  & M3.6 & M3.4 &               M3.6 \\
                    Sz67 &  0.0 &      M3.0 &    $\pm 0.5$ &      M4$^{\rm c}$ & M3.8 & M3.7 &               M3.8 \\
  2MASSJ16090850-3903430 &  0.3 &      M5.0 &  $\pm 0.5$ &        M5$^{\rm c}$ & M5.3 & M5.4 &               M5.3 \\
  2MASSJ16075888-3924347 & 0.25 &      M4.5 &   $\pm 0.5$ &       M5$^{\rm c}$ & M5.0 & M4.8 &               M5.1 \\
  2MASSJ16091713-3927096 & 0.75 &      M5.5 &   $\pm 0.5$ &       M5$^{\rm c}$ & M5.8 & M6.0 &               M5.7 \\
                V1191 Sco &  0.4 &      M5.5 &   $\pm 0.5$ &       M5$^{\rm c}$ & M6.2 & M5.8 &               M5.9 \\
\bottomrule
\end{tabular}
\tablefoot{The spectral types and extinction obtained in this work from comparing to available templates are reported in column 2 and 3, with the uncertainty on the SpT reported in column 4. The other columns report previous literature SpT values, and those obtained using the spectral indices from \citetalias{HH14}, \citet{Jeffries07}, and \citet{Riddick2007}. The uncertainty on the SpT is given in subclasses. Refrences: a: \citet{Torres2006}, b: \citet{Galli2015},c: \citet{Padgett2006}, d: \citetalias{manara17b},e: \citet{Kohler2000},f: \citet{Krautter1997},g: \citet{Hardy2015},h: \citet{merin2008},i: \citet{Rugel2018}}
\end{table*}

Determining spectral types (SpT) for (non-accreting) young stellar objects is a complex process that is best performed by comparison with other targets of well known SpT. In this case, we consider the grid of non-accreting Class III young stellar object of \citetalias{manara13a} and \citetalias{manara17b} as a starting point, and later also check and refine their grids using the new spectra presented here.

The first estimate of the SpT of our targets was obtained applying to the new spectra the same spectral indices used by \citetalias{manara13a} and \citetalias{manara17b}, namely the spectral indices presented by \citet{Riddick2007}, the TiO index of \citet{Jeffries07} and the indices presented by \citet[][hereafter HH14]{HH14}. The indices of \citet{Riddick2007} are accurate for spectral types later than M3. The indices by \citet{Jeffries07} hold for spectral types later than K6, since the used TiO feature disappears for earlier SpTs. The indices of \citetalias{HH14} are accurate for SpTs from early K to late M stars. 
The results for these different methods are listed in Table \ref{tab:SpT_new}. 
Spectral indices are less reliable estimates of the SpT in cases in which the extinction is substantial since they are based on ratios of features at different wavelength ranges. In general, extinction is low for our targets, but it can still introduce additional uncertainty to the SpT estimates obtained from the spectral indices.

\begin{figure*}
\centering
    \includegraphics[width =\textwidth]{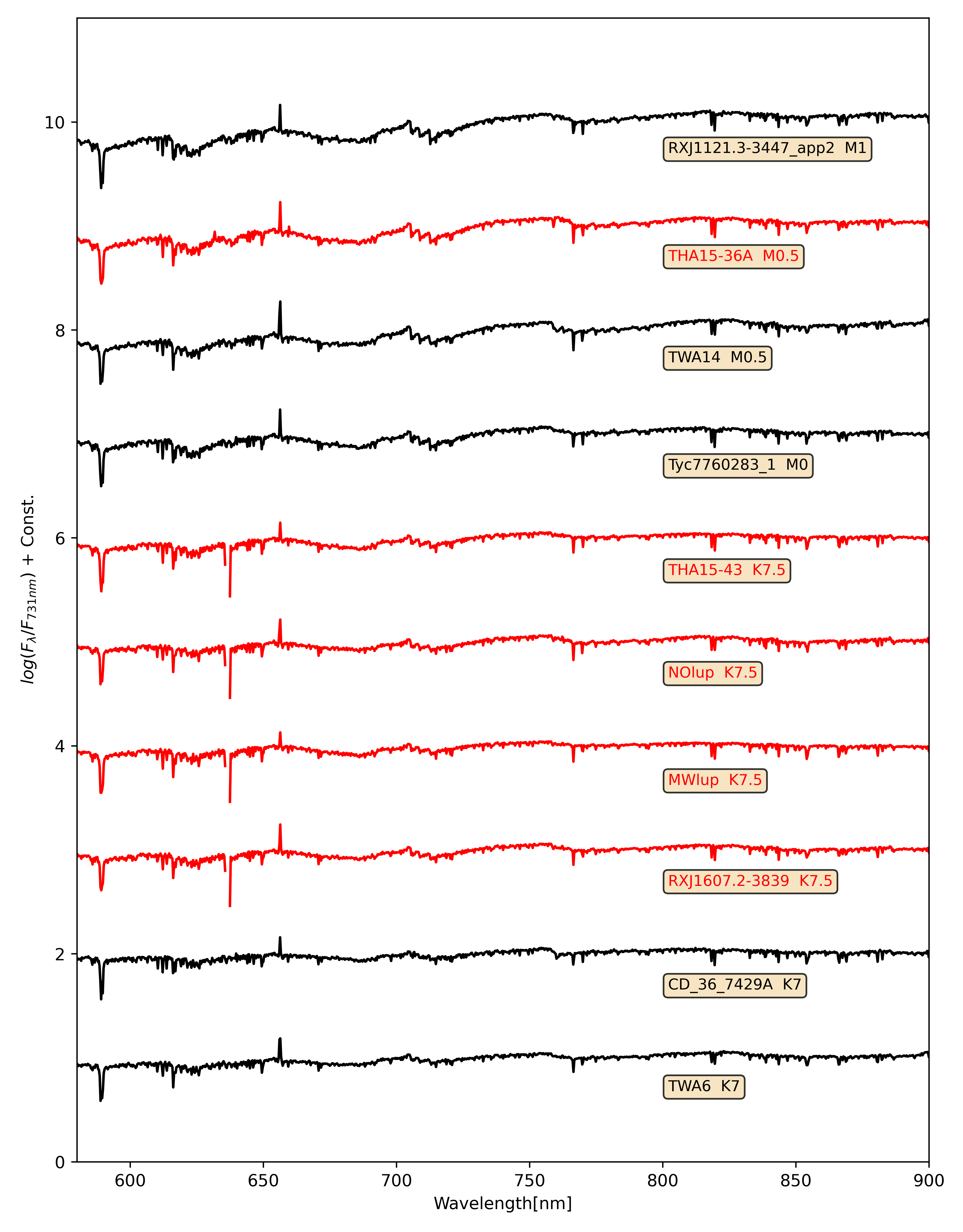}
    \caption{X-Shooter spectra of Class III YSOs with spectral type ranging from M1 to K7. All the spectra are normalized at 731 nm and offset in the vertical direction for clarity. The spectra are also smoothed to the resolution of 2500 at 750 nm to make easier the identification of the molecular features. The black colors indicate spectra presented by \citetalias{manara13a} and \citetalias{manara17b}. The red color is used for spectra presented for the first time here.}
    \label{fig:luhman750main}
\end{figure*}

Therefore, to obtain a simultaneous extinction and SpT estimate, we performed a comparison of the spectra in our sample with the spectra of \citetalias{manara13a} and \citetalias{manara17b}, which have negligible extinction. We obtained the first estimates of SpT and $A_V$ using a simple $\chi^2$-like comparison. The final values of both the spectral type and extinction are derived through a visual comparison between the spectra presented here and those presented by \citetalias{manara13a} and \citetalias{manara17b}. The main indicators used in this comparison are the depths of various molecular bands.
The K6 to M9.5 spectra include a variety of molecular features whose depth increases almost monotonically for later SpTs in the spectral region between 580 and 900 nm. This includes various absorption bands from TiO ($\lambda\lambda $ 584.7-605.8, 608-639, 655.1-685.2, 705.3-727, 765-785, 820.6-
856.9, 885.9-895 nm), CaH ($\lambda\lambda $ 675-705 nm) and VO ($\lambda\lambda $ 735-755, 785-795, 850-865 nm).
Fig. \ref{fig:luhman750main} displays spectra with SpTs ranging from M1 to K7.5. Similar figures for the other spectra with SpT later than K5 can be found in Appendix \ref{app:luhman}.  The previously mentioned molecular features can be seen to increase in depth for later spectral types.

\begin{figure*}
    \centering
    \includegraphics[width =\textwidth]{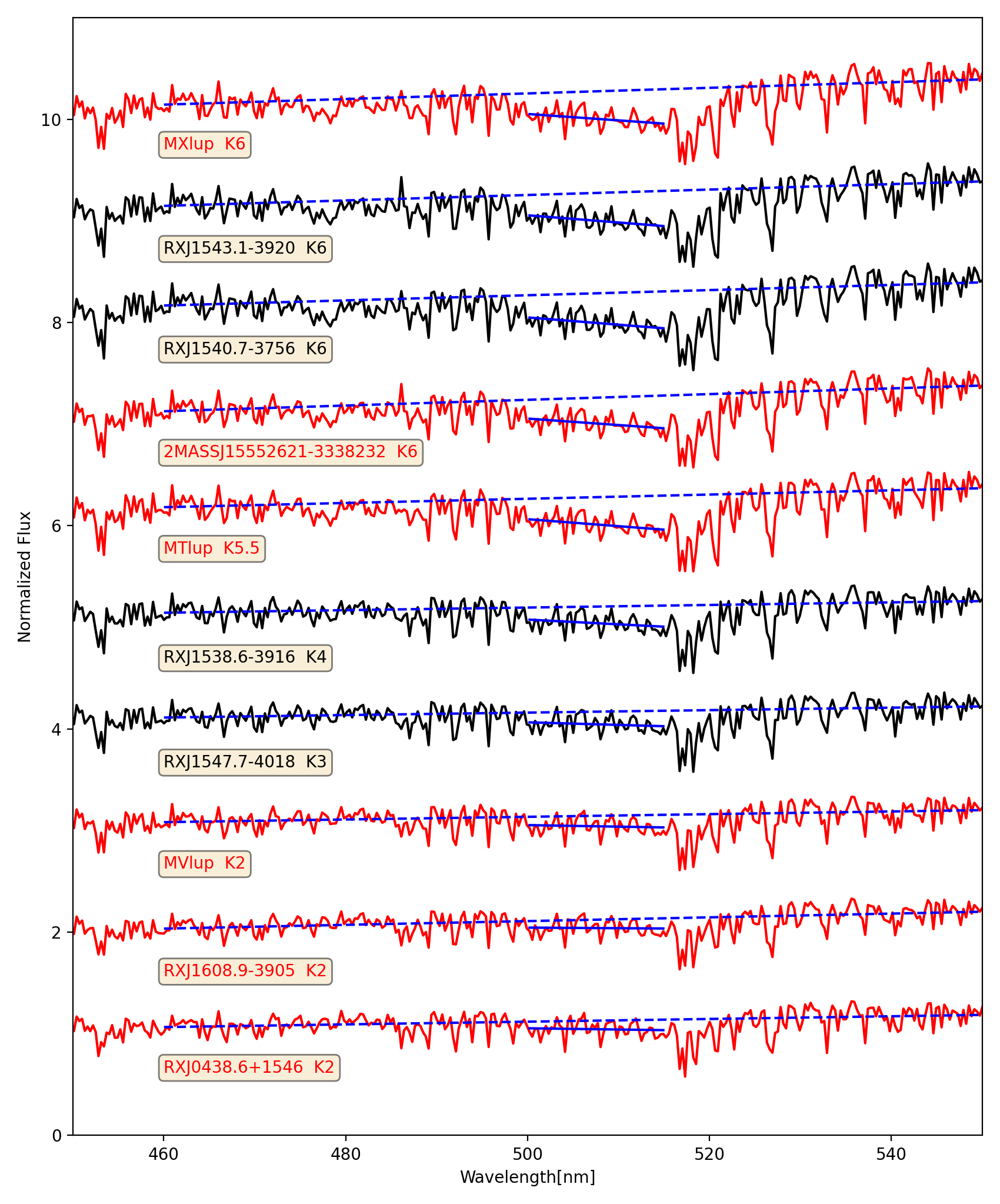}
    \caption{X-Shooter spectra of Class III YSOs with SpT ranging from K6 to K2. The spectra have been normalized at 460 nm and an offset has been added for clarity. The spectra have been smoothed to have a resolution of $\sim$2500 at 750nm.  The observations presented here are displayed in red, while the spectra of \citetalias{manara13a} and \citetalias{manara17b} are indicated in black. We highlight the R510 spectral index feature using the solid blue line. The dashed blue line indicates the slope of the surrounding continuum.}
    \label{fig:luhman510main}
\end{figure*}

For SpT earlier than K6 a MgH and Mg b absorption feature ($\lambda\lambda $ 505-515 nm) is used to assess the spectral type. The use of this feature was first discussed by \citetalias{HH14} who present it in the context of a spectral index. The depth of this absorption feature is best determined by comparing the flux at 510nm to the flux expected at the same wavelength from a linear fit between the median flux of the spectral regions of $\lambda\lambda $ 460-470 nm and $\lambda\lambda $540-550nm. Fig. \ref{fig:luhman510main} displays this feature for spectral types from K1 to K6. Here the linear fit between $\lambda\lambda $ 460-470 nm and $\lambda\lambda $540-550 nm is also indicated. Additionally, a linear fit for the regions of $\lambda\lambda $ 488-492 nm and $\lambda\lambda $514-516 nm is included in this figure to highlight the increasing depth of this feature with SpT. We perform this comparison at different values of extinction to estimate it simultaneously. The results are listed in Table \ref{tab:SpT_new}, where we have rounded the SpT to half a subclass.

Our estimates are typically consistent within at most 1.5 subclass of the literature values. The TiO spectral indices agrees within 1 subclass for targets later than K5. The spectral indices of \citetalias{HH14} agree within a subclass at all spectral types except for the K5 to M0 range. Finally the indices of \citet{Riddick2007} have a similar agreement for the targets later than M3.
We note that the use of features in the NIR region ($\lambda\lambda > 1000$\,nm) can give rise to significantly different spectral types. On top of the fact that the NIR indices are based on features covering a larger wavelength range and thus are more affected by extinction and by imperfect telluric removal and or flux calibration in that wavelength range (e.g., \citetalias{manara13a}), the likely explanation for this deviation is the presence of cold spots on the stellar surface \cite[e.g.,][]{Stauffer03,Vacca2011,Pecaut2016,GullySantiago2017,Gangi2022}.

We adopt similar uncertainties on the SpT as \citetalias{manara17b}. For objects later than K6 we estimate the uncertainties to be 0.5 subclass. For objects earlier than K6 we estimate the uncertainties to be 1 subclass. For the binary components of THA15-36 we assume larger uncertainties of 1 subclass due to the less certain flux calibration of the spectra. We estimate the uncertainty on the extinction to be 0.2 mag. A description of how we confirmed the  uncertainties on the SpT and estimated those on the extinction follows in Sect. \ref{CombGrid} and Appendix \ref{app:uncertainties}.

\subsection{Photospheric properties from ROTFIT}\label{rotfit}

A large number of absorption lines are resolved in the spectra, allowing us to obtain photospheric properties.
We analyse the absorption lines in the VIS arm of the X-Shooter spectra using the ROTFIT tool \citep[e.g.,][]{frasca15,frasca2017} to derive the effective temperature ($T_{\rm{eff}}$), surface gravity ($\log{g}$), projected rotational velocity ($v \sin{i}$), radial velocities (RV), and veiling at three wavelengths ($\lambda  \,  620  \,  \rm{nm}$, $\lambda   \, 710  \,  \rm{nm}$ and $\lambda  \,  970  \,  \rm{nm}$).  The veiling is given by $r = EW(template)/EW(observation) -1$, where EW is the equivalent width of absorption lines near the wavelength of interest.
ROTFIT searches, within a grid of templates, for the spectrum that minimizes the $\chi^2$ to the target spectrum in different spectral regions. The templates consist of a grid of BT-Settl model spectra \citep{allard11} of solar metallicity, $\log{g}$ ranging from 0.5 to 5.5 dex and an effective temperature ranging from 2000 to 6000 K. ROTFIT achieves this by first deriving the radial velocity of the star cross-correlating the target and template spectrum and uses this information to shift the observed spectrum to the rest frame. ROTFIT then convolves the templates with both a Gaussian to match the X-Shooter resolution. Then the templates are iteratively broadened by convolution with a rotational profile with increasing  $v \sin{i}$ until a minimum $\chi^2$ is attained.  ROTFIT analyses spectral intervals that contain features sensitive to $\log{g}$ and/or the $T_{\rm{eff}}$ such as the K{\small I} doublet at $\lambda \: \approx 766-770 \: \rm{nm}$ and the Na{\small I} doublet at $\lambda \: \approx 819 \: \rm{nm}$. An additional continuum emission component is added to the template in order to estimate the veiling. As for the  $v \sin{i}$, per each template, the veiling is also a free parameter in the fit. 

\begin{figure}[t]
    \centering
    \includegraphics[width =0.49\textwidth]{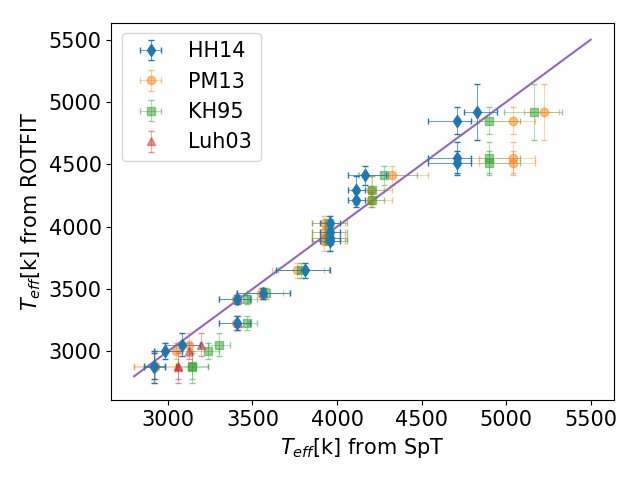}
    \caption{Comparison between the temperatures obtained with ROTFIT and from the spectral type for the targets presented here. The relation of \citet{HH14} is labeled as HH14, the one by \citet{Pecaut2013} as PM13, that of \citet{kenyon95} asn KH95 and the relationship of \citet{luhman03} as Luh03. The relations are only applied to spectral type ranges where they are valid. The solid line represents the one-to-one relation.}
    \label{fig:rotfit}
\end{figure}

We did not set this continuum component and therefore the veiling to 0 despite most of our targets being non-accretors. 
We preferred to let the veiling vary for our Class III stars as a check. As expected, we always found zero or low veiling values, which could be the result of small residuals in the scattered light subtraction or the effect of the interplay between different parameters. The largest value of r=0.4 is found at 620 nm for three K-type stars. A possible explanation could be the effect of starspots on the spectra. Another possible explanation is that of an undetected companion. Apart from these extreme values, we do not consider veiling values as small as 0.2 to be significant.
The photospheric parameters, $v \sin{i}$, veiling measurements, and RV derived with ROTFIT can be found in Table \ref{tab:rotfit}.  As stated in Sect. \ref{ObsAndDat}, during this analysis, we found two targets to be double lined spectroscopic binaries, namely NN Lup  and CD-39 10292.
We excluded these from our analysis. Fig. \ref{fig:rotfit} shows the comparison between the $T_{\rm{eff}}$ derived using ROTFIT and obtained using different SpT -$T_{\rm eff}$ conversions. Here it can be seen that the effective temperature obtained from the SpT agrees best with ROTFIT when using the SpT -$T_{\rm eff}$ relation by \citetalias{HH14}. The low \logg\, found are mostly compatible with that of young objects ($\lesssim$10 Myr).

\begin{table*}[t]
    \centering
    \caption{Photospheric parameters derived using ROTFIT.}
    \label{tab:rotfit}
    \begin{tabular}{l||ll||ll|ll|ll|lll}
\toprule
                  Name & $T_{\rm{eff}}$ & $\sigma(T_{\rm{eff}})$ & $\log{g}$ & $\sigma(\log{g})$ & $v\sin{i}$ & $\sigma(v\sin{i})$ &   $RV$ & $\sigma(RV)$ & $r_{970\rm{nm}}$ & $r_{710\rm{nm}}$ & $r_{620\rm{nm}}$ \\
                   & $[K]$ & $[K]$ & $[cgs]$ & $[cgs]$ & $[km/s]$ & $[km/s]$ & $[km/s]$ & $[km/s]$ &  &  &  \\
\midrule
           CD-31\_12522 &  4920 &    223 & 4.04 &  0.12 &  30.2 &    1.0 &  2.2 & 0.8 &   0.00 &   0.00 &   0.00 \\
        RXJ0438.6+1546 &  4851 &    111 & 4.08 &  0.23 &  21.0 &    1.0 & 18.9 & 0.7 &   0.30 &   0.00 &   0.40 \\
        RXJ1608.9-3905 &  4552 &    124 & 4.27 &  0.14 &  43.6 &    1.0 & -0.7 & 0.9 &   0.00 &   0.00 &   0.40 \\
                 MV Lup  &  4513 &     96 & 4.83 &  0.25 &   1.0 &    6.0 &  2.4 & 0.7 &   0.00 &   0.00 &   0.00 \\
                 MT Lup   &  4411 &     77 & 4.41 &  0.12 &   7.9 &    1.4 &  2.7 & 0.6 &   0.40 &   0.20 &   0.30 \\
J15552621-3338232 &  4295 &    112 & 4.88 &  0.24 &   1.0 &   11.0 &  0.8 & 0.6 &   0.10 &   0.00 &   0.20 \\
                 MX Lup  &  4216 &     61 & 4.45 &  0.21 &  10.0 &    1.0 &  2.4 & 0.6 &   0.00 &   0.00 &   0.20 \\
                 MW Lup  &  4032 &     54 & 5.03 &  0.29 &  15.2 &    3.2 &  3.4 & 0.6 &   0.00 &   0.00 &   0.00 \\
        RXJ1607.2-3839 &  3954 &     86 & 4.53 &  0.11 &  32.8 &    2.0 & -0.4 & 1.0 &   0.00 &   0.00 &   0.40 \\
                 NO Lup  &  3907 &     98 & 4.43 &  0.11 &  11.2 &    1.0 & -2.3 & 0.8 &   0.10 &   0.20 &   0.30 \\
              THA15-43 &  3885 &     23 & 5.19 &  0.18 &   9.4 &    2.0 &  1.6 & 0.7 &   0.10 &   0.00 &   0.30 \\
             THA15-36A &  3650 &     60 & 4.57 &  0.12 &  19.2 &    1.0 &  2.2 & 0.8 &   0.10 &   0.20 &   0.30 \\
             THA15-36B &  3465 &     43 & 4.92 &  0.11 &  21.1 &    1.0 &  1.8 & 0.7 &   0.10 &   0.20 &   0.30 \\
                RECX-6 &  3419 &     36 & 4.68 &  0.12 &  23.4 &    1.9 & 16.3 & 0.7 &   0.00 &   0.00 &   0.00 \\
                  Sz67 &  3226 &     57 & 4.53 &  0.20 &  70.3 &    1.0 & -0.6 & 3.6 &   0.00 &   0.00 &   0.10 \\
J16075888-3924347 &  3054 &     93 & 4.24 &  0.15 &  25.0 &   11.0 &  2.4 & 0.9 &   0.00 &   0.00 &   0.00 \\
J16090850-3903430 &  3000 &     64 & 4.25 &  0.13 &   4.1 &   10.2 & -1.6 & 1.0 &   0.20 &   0.00 &   0.10 \\
              V1191Sco &  2883 &    105 & 3.59 &  0.11 &  33.5 &    9.7 & -0.2 & 1.0 &   0.00 &   0.00 &   0.00 \\
J16091713-3927096 &  2876 &    130 & 4.73 &  0.13 &  69.0 &   13.0 &  2.8 & 4.3 &   0.00 &   0.00 &   0.00 \\
\bottomrule
\end{tabular}
\tablefoot{ $r_{970\rm{nm}}$, $r_{710\rm{nm}}$ and $r_{620\rm{nm}}$ is the veiling measured around 970nm ,710 nm and 620 nm respectively.}
\end{table*}

\subsection{Luminosity determination}\label{lumCalcSect}

The stellar luminosities are obtained using a method similar to that used by \citetalias{manara13a}. In this method, the dereddened spectra are extrapolated to wavelengths not covered by the spectra and integrated to compute the bolometric flux. 

The X-Shooter spectra are extrapolated using a BT-Settl \citep{allard11} synthetic spectrum appropriate for the target. We use the effective temperature obtained from the SpT$- T_{\rm eff}$ relation of \citetalias{HH14}. We prefer this $T_{\rm eff}$, since in Sect. \ref{sect:fitter} we will use the obtained SpT to compute the luminosity.
For the metallicity of the synthetic spectra we assumed solar values, typical for targets in the Lupus region \citep{Biazzo2017}, and for the surface gravity we chose $\log{g} = 4.0$ in all cases. This parameter has very limited impact on the global luminosity estimate, since most of the stellar emission is covered by the X-Shooter spectra and the synthetic spectra are used to measure a minor fraction of the global emission. 
The synthetic spectra are then matched to the spectra at 400.5 nm and 2300 nm and used to represent the flux at wavelengths shorter and longer than these respective values. The wavelengths chosen here differ from those used by \citetalias{manara13a}. We chose these wavelengths in order to avoid the high noise level at short wavelengths and poor flux calibration at the end of the $K$-band present in some spectra. Due to this difference, we also re-applied this method to the sample of \citetalias{manara13a} and \citetalias{manara17b}.
We also performed a linear interpolation across the strong telluric absorption features between the $J$,$H$, and $K$ band ($\lambda\lambda $\: 1330-1550 \rm{nm},  $\lambda\lambda $\: 1780-2080 nm) in order to account for the stellar flux in these regions. An example of a prolonged spectrum created in this way is shown in Appendix \ref{app:BolCor}. The bolometric flux is then computed by integrating over the prolonged spectrum.

We used the photogeometric distances of \citet{Bailer-Jones2021} to convert the fluxes to luminosities for all targets with exception of TWA 26. For this star we used the geometric distance of the same authors since a photogeometric distance is unavailable. The distances provided by \citet{Bailer-Jones2021} are based on the {\it Gaia}  DR3 \citep{gaia2016,gaia2023} parallaxes but include a photometric and geometric prior to improve the distance estimate. The inclusion of these priors significantly improves the uncertainties for the most distant targets (namely those in $\sigma$ Orionis). For more nearby targets the distance estimate and its uncertainty do not change significantly from the inverse {\it Gaia}  parallaxes.  
 
The uncertainties on these estimates are computed by propagating the uncertainties of matching the BT-Settl models to the observations, the distance, the extinction, and the photometric flux. We adopt an extinction uncertainty of $\sigma_{A_V} = 0.2 mag$. For the targets calibrated using wide slit spectra, we assumed the uncertainties on the photometric flux 5\% of the total flux, which is slightly more conservative than the 4\% of \citet{Rugel2018}. For the spectra calibrated using the available photometry we assume an uncertainty of 0.2 dex on the luminosity, the same as that of \citetalias{manara13a}.

We also computed the luminosities of the grid using the bolometric correction of \citetalias{HH14}. This correction uses the flux measured at 751 nm to estimate the total flux of the star. We notice a small but systematic discrepancy between the luminosities derived from the two methods at effective temperatures lower than $~4500 \rm K$. We therefore propose a correction to the relationship provided by \citetalias{HH14} which is further discussed in Appendix \ref{app:BolCor}. This new relation is assumed in this work.

\begin{figure}[th!]
    \centering
    \includegraphics[width =0.49\textwidth]{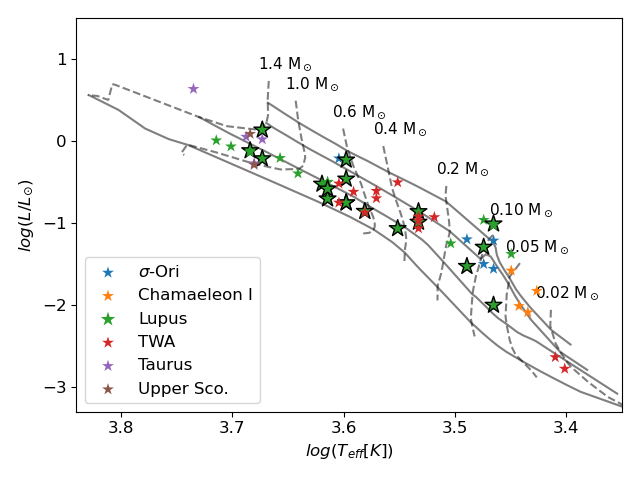}
    \caption{HR Diagram of the objects analysed here (highlighted with larger black outlined star symbols) and those analysed by \citetalias{manara13a} and \citetalias{manara17b}. The model isochrones and evolutionary tracks by \cite{B15} are also shown. The isochrones are the 1.2, 3, 10 and 30 Myr ones.}
    \label{fig:HDR}
\end{figure}

Fig. \ref{fig:HDR} shows our sample in a Hertzsprung-Russell diagram together with the samples of \citetalias{manara13a} and \citetalias{manara17b}. Here, we used the luminosities obtained by integrating the extended spectrum and the effective temperature is obtained from the spectral type using the relationship of \citetalias{HH14}. The isochrones and evolutionary tracks of \cite{B15} are also plotted. The isochronal age of the targets in our sample appears to be dependent on the stellar mass. Targets with $ M_\star \gtrsim 0.7 M_{\odot} $ tend to be $\sim 10 \rm \;to \;30 \;\rm Myrs$ old according to the isochrones, whereas those with lower mass have ages of $\sim 1 \rm\; to\; 10 \; Myrs$, despite all except one target being members of the Lupus star forming region. This trend is also present in the samples of \citetalias{manara13a} and \citetalias{manara17b} as well as other star samples such as those of \citet{Bell14}, \citet{HH15}, and \citet{Pecaut2016}. \citet{Stelzer2013} see a similar trend when comparing their sample to isochrones in the $\log{g}  - T_{\rm eff}$ diagram. It is unclear what causes this discrepancy but possible origins include the presence of starspots affecting the position in the HRD \citep[e.g.,][]{Gangi2022,Paolino2024} or affecting overall evolution \citep{Somers2015}, the effect of accretion during earlier stages of the star's evolution \citep{Baraffe17} or the effect of the stellar magnetic field on the subphotospheric convective motions in the stars \citep{Feiden2016}. The evolutionary tracks by \citet{Feiden2016} include the later effect and are similar to the models of \citet{B15} in the non-magnetic case. The SPOTS models of \citet{Somers2020} includes the influence of starspots at different filling factors on the isochrones. An analysis of the spot coverage of our sample and comparison with the SPOTS isochrones is beyond the scope of this work and deferred to a future work.

\section{New combined grid}\label{CombGrid}

The first goal of this work is to improve upon the grid of template spectra of YSOs presented by \citetalias{manara13a} and \citetalias{manara17b} by adding new templates to the grid and interpolating the spectra to generate a continuous grid. This is discussed in this section. 

\subsection{Description of the new grid}\label{CombGridDescr}

We combine the observations presented here with those of \citetalias{manara13a} and \citetalias{manara17b} to create an enhanced library of spectral templates that is made available to the community.  We add one additional spectrum presented by \citet{manara16a}, that of the K0 star HBC407. The sources selected by \citetalias{manara13a} and \citetalias{manara17b} were classified to be Class III YSOs and using Spitzer data \citep[e.g.,][]{Evans2009}. \citet{manara14} presented  additional X-Shooter observations of Class III targets, namely IC348-127, T21, CrA75. These targets were highly extincted, with $A_V \sim 6 \: \rm mag$, $ A_V = 3.2 \: \rm  mag$ and $A_V = 1.5 \: \rm mag$, respectively. This high extinction causes additional uncertainties in the dereddened spectra due to the uncertainties in the assumed extinction law. Therefore we do not include these spectra in our grid. 

There are 6 stars from our new observations that are not included in the grid. The 5 unresolved binaries previously mentioned in Sect. \ref{ObsAndDat} are excluded from our grid since they cannot be used to represent the stellar emission of individual stars. To avoid potential accretors in our sample we we use the $H\alpha$ equivalent width criterion of \citet{white04} and \citet{Barrado2003}, among others. The spectrum of 2MASSJ16075888-3924347 shows evidence of ongoing accretion, therefore we exclude it from our grid. This is further discussed in Appendix \ref{2mass888}.

From the grid of \citetalias{manara13a}, Sz121 and Sz122 are excluded since they are likely spectroscopic binaries or ultrafast rotators \citep[MTR13,][]{Stelzer2013}.  
For RXJ0438.6+1546 a spectrum was presented by \citetalias{manara17b}. This target was also observed as a part of the PENELLOPE VLT Large Program \citep{manara21}. We only use the more recent PENELLOPE spectrum since the observations were performed with better sky transparency conditions.

We confirm the spectral types of all except for one target of \citetalias{manara13a} and \citetalias{manara17b} through visual inter-comparison. The TiO absorption bands of CD 36-7429A appear to best match that of the K7 templates. We re-assign the spectral type of CD 36-7429A from K5 \citepalias{manara13a} to K7. This spectral type is further confirmed by the TiO index of \citet{Jeffries07} which yields a spectral type of K7.0 and the analysis of \citet{Fang2017,Fang2021} who also adopted a SpT of K7 for this target. \citet{Pecaut2013} also adopted a SpT of K7 for this source. Interestingly, the estimate of \teff\, from the ROTFIT analysis of the absorption lines lead to a result more consistent with a SpT of K5 \citep{Stelzer2013}. It is possible that spots covering the stellar surface lead to this discrepancy \citep[e.g.,][]{Stauffer03,Vacca2011,Pecaut2016,GullySantiago2017,Gangi2022}. We however assume K7 as the SpT for this target based on the molecular features. The spectra first presented in this work have been dereddened using the values listed in Table \ref{tab:SpT_new} to provide an extinction-less grid. This grid can be found on GitHub\footnote{\url{https://github.com/RikClaes/FRAPPE}}.

The uncertainties on both $A_V$ and SpT are estimated by fitting the spectra of targets within 1 spectral type subclass of other targets in the combined grid. In this procedure, we search for the best fitting template spectrum in the remainder of the grid at different values of extinction. The best fit template and extinction are found by searching for the minimum of a $\chi^2$-like metric that includes the spectral features discussed in Sect. \ref{spectralTyping}.

For both the difference in spectral type and extinction we find a distribution with a median value of 0 and standard deviations of $\sim 0.5 \: \rm{ subclasses}$ and $ \sim 0.25 \: \rm{ mag}$ respectively. We note that the standard deviation is larger for spectra earlier than K6, in part due to the lack of spectra of similar spectral type. We therefore estimate the uncertainties on SpT's earlier than K6 to be $\sim 1 \: \rm{subclasses}$, those at later SpT to be $\sim 0.5 \: \rm{subclasses}$. For the extinction we adopt an uncertainty of $0.2 \: \rm{mag}$. THA15-36B appears as a strong outlier with $\Delta \rm SpT =-1.5$ and $\Delta \rm A_V =0.8$. This supports our decision to provide higher uncertainties on both THA15-36A and THA15-36B. Because of this we also exclude THA15-36A and THA15-36B when constructing our interpolated grid in Sect. \ref{interpol}.

\begin{figure}
    \centering
    \includegraphics[width=0.5\textwidth]{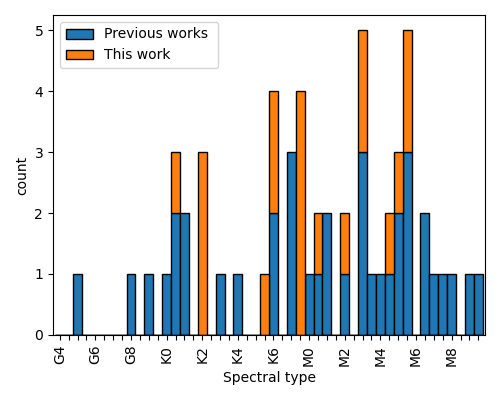}
    \caption{Histogram of SpT for the grid of Class III templates.}
    \label{fig:histSptGrid}
\end{figure}

The final grid of templates is reported in Table 3. Fig. \ref{fig:histSptGrid} shows the spectral type distribution of all the spectra in our final grid. Our final grid of templates includes 57 targets and spans a range of spectral types from G5 to M9.5. The spectral types later than K6 appear to be well represented. The earlier spectral types appear less well sampled, with the most prominent gaps between G5 to G8 and K4 to K5.5.

\begin{table*}[]
    \centering
        \caption{The final grid of spectral templates and their stellar properties.}
        
\begin{tabular}{lllrrrr}
\toprule
                    Name &   Region &  SpT &  $\rm T_{\rm eff} [K]$  &  distance [pc] &  $\log(L/L_\sun)$ & reference \\
\midrule
          RXJ0445.8+1556 &      Taurus &   G5 &  5430 &   $167.3_{-3.8}^{+3.2}$ &          $0.64\pm 0.07$  & \citetalias{manara17b}\\
          RXJ1508.6-4423 &     Lupus&   G8 &  5180 &          $145.7\pm 0.4$ &          $0.01\pm 0.07$ & \citetalias{manara17b}\\
          RXJ1526.0-4501 &     Lupus&   G9 &  5025 &          $150.1\pm 0.3$ &         $-0.07\pm 0.06$ & \citetalias{manara17b}\\
                  HBC407 &      Taurus &   K0 &  4870 & $249.1_{-29.7}^{+43.5}$ &  $0.05_{-0.12}^{+0.17}$ & \cite{manara14}\\
    PZ99J160843.4-260216 &     Upper Scorpius & K0.5 &  4830 &   $137.4_{-1.0}^{+1.1}$ &          $0.09\pm 0.06$ & \citetalias{manara17b}\\
             CD-31\_12522 & Lupus & K0.5 &  4830 &          $127.4\pm 0.5$ &         $-0.12\pm 0.06$ & This work\\
          RXJ1515.8-3331 &     Lupus& K0.5 &  4830 &          $115.4\pm 0.2$ &         $-0.14\pm 0.06$ & \citetalias{manara17b}\\
    PZ99J160550.5-253313 &     Upper Scorpius &   K1 &  4790 &          $105.1\pm 0.2$ &         $-0.29\pm 0.06$& \citetalias{manara17b} \\
          RXJ0457.5+2014 &      Taurus &   K1 &  4790 &          $122.2\pm 0.3$ &         $-0.27\pm 0.06$ & \citetalias{manara17b} \\
RXJ0438.6+1546 &      Taurus &   K2 &  4710 &          $139.1\pm 0.3$ &          $0.02\pm 0.06$ & This work \\
          RXJ1608.9-3905 & Lupus &   K2 &  4710 &          $156.0\pm 0.4$ &          $0.14\pm 0.06$ & This work \\
                   MV Lup &   Lupus &   K2 &  4710 &          $137.1\pm 0.3$ &         $-0.21\pm 0.2$ & This work\\
          RXJ1547.7-4018 &     Lupus&   K3 &  4540 &          $130.2\pm 0.3$ &         $-0.21\pm 0.06$ & \citetalias{manara17b} \\
          RXJ1538.6-3916 &     Lupus&   K4 &  4375 &          $122.5\pm 0.4$ &         $-0.40\pm 0.06$ & \citetalias{manara17b} \\
                   MT Lup &   Lupus & K5.5 &  4163 &          $133.1\pm 0.3$ &         $-0.52\pm 0.2$ & This work \\
  2MASSJ15552621-3338232 &   Lupus &   K6 &  4115 &          $119.0\pm 0.2$ &         $-0.70\pm 0.05$ & This work \\
          RXJ1540.7-3756 &     Lupus&   K6 &  4115 &          $134.3\pm 0.3$ &         $-0.50\pm 0.05$ & \citetalias{manara17b} \\
          RXJ1543.1-3920 &     Lupus&   K6 &  4115 &   $133.6_{-6.2}^{+5.4}$ & $-0.50_{-0.07}^{+0.06}$ & \citetalias{manara17b} \\
                   MX Lup & Lupus &   K6 &  4115 &   $128.8_{-0.2}^{+0.3}$ &         $-0.58\pm 0.05$ & This work \\
                   SO879 &  $\sigma$ Ori. &   K7 &  4020 &   $395.6_{-2.2}^{+3.4}$ &         $-0.21\pm 0.2$ & \citetalias{manara13a} \\
                    TWA6 &    TW Hya &   K7 &  4020 &           $65.5\pm 0.1$ &         $-0.75\pm 0.2$ & \citetalias{manara13a} \\
             CD -36 7429A &    TW Hya &   K7 &  4020 &           $76.3\pm 0.1$ &         $-0.52\pm 0.2$ & \citetalias{manara13a} \\
          RXJ1607.2-3839 & Lupus & K7.5 &  3960 &          $167.9\pm 2.7$ &         $-0.23\pm 0.05$ & This work\\
                   MWlup &   Lupus & K7.5 &  3960 &   $129.6_{-0.3}^{+0.2}$ &         $-0.75\pm 0.05$ & This work\\
                   NO Lup &   Lupus & K7.5 &  3960 &   $132.8_{-0.2}^{+0.3}$ &         $-0.46\pm 0.2$ & This work\\
                THA15-43 & Lupus & K7.5 &  3960 &          $125.1\pm 0.3$ &         $-0.75\pm 0.05$& This work\\
            Tyc7760283\_1 &    TW Hya &   M0 &  3900 &           $53.5\pm 0.1$ &         $-0.62\pm 0.2$ & \citetalias{manara13a}\\
                   TWA14 &    TW Hya & M0.5 &  3810 &    $92.0_{-0.1}^{+0.2}$ &         $-0.87\pm 0.2$ & \citetalias{manara13a}\\
               THA15-36A & Lupus & M0.5 &  3810 &   $145.6_{-0.6}^{+0.9}$ &         $-0.86\pm 0.2$ & This work \\
     RXJ1121.3-3447\_app2 &    TW Hya &   M1 &  3720 &           $59.7\pm 0.1$ &         $-0.70\pm 0.2$ & \citetalias{manara13a} \\
     RXJ1121.3-3447\_app1 &    TW Hya &   M1 &  3720 &           $59.7\pm 0.1$ &         $-0.61\pm 0.2$ & \citetalias{manara13a}\\
               THA15-36B & Lupus &   M2 &  3560 &   $150.9_{-1.5}^{+1.6}$ &         $-1.07\pm 0.2$ & This work\\
             CD -29 8887A &    TW Hya &   M2 &  3560 &    $45.9_{-0.5}^{+0.4}$ &         $-0.50\pm 0.2$ & \citetalias{manara13a} \\
              CD -36 7429B &    TW Hya &   M3 &  3410 &           $76.1\pm 0.2$ &         $-1.06\pm 0.2$ & \citetalias{manara13a} \\
              TWA15\_app2 &    TW Hya &   M3 &  3410 &          $114.7\pm 0.2$ &         $-0.92\pm 0.2$ & \citetalias{manara13a} \\
                    TWA7 &    TW Hya &   M3 &  3410 &           $34.05\pm 0.03$ &         $-0.97\pm 0.2$ & \citetalias{manara13a}\\
                    Sz67 &   LupusI &   M3 &  3410 &   $103.2_{-0.9}^{+1.0}$ &         $-0.86\pm 0.04$& This work \\
                 RECX-6 &   LupusI &   M3 &  3410 &           $97.9\pm 0.1$ &         $-0.99\pm 0.04$ & This work\\
              TWA15\_app1 &    TW Hya & M3.5 &  3300 &          $114.2\pm 0.2$ &         $-0.93\pm 0.2$ & \citetalias{manara13a}\\
                    Sz94 &     Lupus&   M4 &  3190 &          $114.6\pm 0.3$ &         $-1.25\pm 0.2$ & \citetalias{manara13a}\\
                   SO797 &  $\sigma$ Ori. & M4.5 &  3085 & $385.9_{-10.6}^{+11.5}$ &         $-1.20\pm 0.2$ & \citetalias{manara13a}\\
                   SO641 &  $\sigma$ Ori. &   M5 &  2980 & $374.6_{-14.8}^{+10.6}$ & $-1.50\pm 0.2$ & \citetalias{manara13a}\\
              Par\_Lup3\_2 &     Lupus&   M5 &  2980 &          $157.2\pm 1.0$ &         $-0.96\pm 0.2$ & \citetalias{manara13a}\\
   2MASSJ16090850-3903430 & LupusIII &   M5 &  2980 &   $156.5_{-0.9}^{+1.3}$ &         $-1.29\pm 0.$ & This work\\
                   SO925 &  $\sigma$ Ori. & M5.5 &  2920 & $371.5_{-19.4}^{+25.8}$ & $-1.56 \pm 0.2$ & \citetalias{manara13a}\\
                   SO999 &  $\sigma$ Ori. & M5.5 &  2920 & $387.7_{-15.7}^{+16.5}$ &         $-1.21\pm 0.2$ & \citetalias{manara13a} \\
                V1191Sco & LupusIII & M5.5 &  2920 &   $168.4_{-2.4}^{+2.2}$ &         $-1.01\pm 0.04$ & This work\\
  2MASSJ16091713-3927096 & LupusIII & M5.5 &  2920 &   $133.7_{-2.8}^{+2.6}$ &         $-2.00\pm 0.04$ & This work\\
                   Sz107 &      Lupus & M5.5 &  2920 &   $152.4_{-1.8}^{+1.9}$ &         $-1.03\pm 0.2$ & \citetalias{manara13a}\\
              Par\_Lup3\_1 &      Lupus & M6.5 &  2815 &   $159.7_{-4.1}^{+4.2}$ &         $-1.38\pm 0.2$ & \citetalias{manara13a} \\
                   LM717 &     ChaI & M6.5 &  2815 &   $190.6_{-2.5}^{+2.6}$ &         $-1.58\pm 0.04$ & \citetalias{manara17b} \\
       J11195652-7504529 &     ChaI &   M7 &  2770 &   $189.2_{-4.2}^{+4.7}$ &         $-2.01\pm 0.04$ & \citetalias{manara17b}\\
                  LM601 &     ChaI & M7.5 &  2720 &   $186.1_{-4.6}^{+4.8}$ &         $-2.09\pm 0.04$  & \citetalias{manara17b}  \\
               CHSM17173 &     ChaI &   M8 &  2670 &   $191.5_{-4.2}^{+4.6}$ &         $-1.83\pm 0.04$ & \citetalias{manara17b} \\
                   TWA26 &    TW Hya &   M9 &  2570 &           $46.6\pm 0.5$ &         $-2.63\pm 0.2$ & \citetalias{manara13a}\\
               DENIS1245 &    TW Hya & M9.5 &  2520 &    $83.1_{-2.8}^{+2.3}$ & $-2.78\pm 0.2$& \citetalias{manara13a} \\
\bottomrule
\end{tabular}
\label{tab:fullGrid}
\end{table*}

\subsection{Interpolation}\label{interpol}
The grid presented in Sect. \ref{CombGridDescr} is limited in two key ways. First, the grid is sparse and possesses gaps at a number of spectral types. Secondly, individual spectra have intrinsic uncertainties. Directly applying one of these spectra to represent the photosphere of a Class II star or to analyse other Class III targets may therefore bias the results. In order to mitigate these downsides we interpolate the grid of template spectra for the characterization of young stellar photospheres.
\begin{figure*}[th!]
    \centering
    \includegraphics[width = \textwidth]{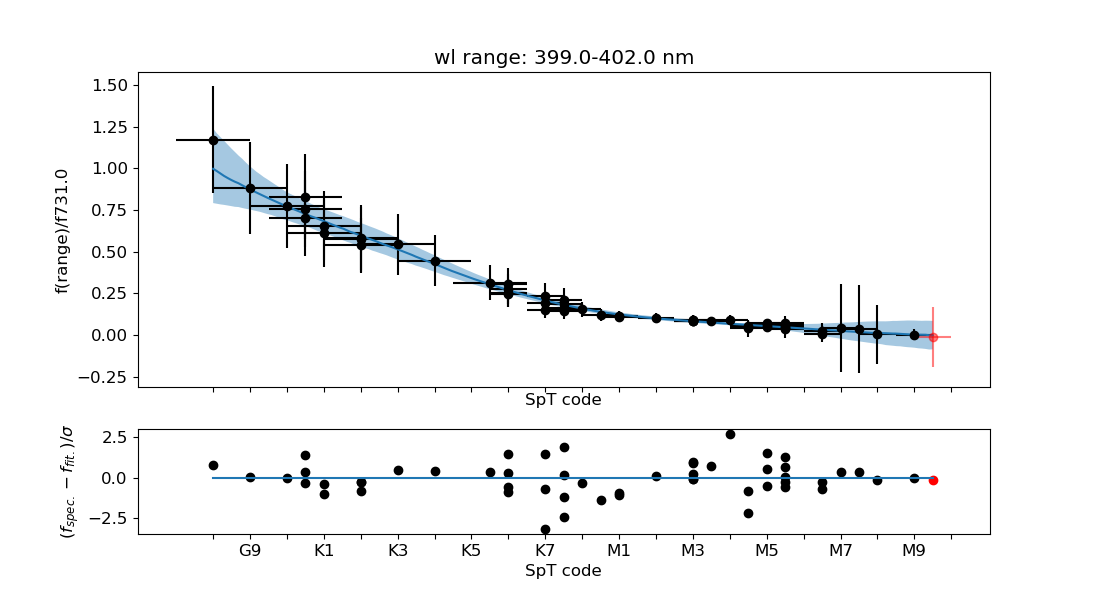}
    \caption{Example of a local polynomial fit to the normalized fluxes (black points) extracted in the wavelength range of 399 to 402 nm. The red point have been excluded from the fit due to the low SNR of this spectrum. The blue line indicates the median fit resulting from the Monte Carlo simulation. The transparent blue region indicates the 1-$\sigma \,$ uncertainty interval. The residuals are computed using only the uncertainties on the non-parametric fit.}
    \label{fig:exInterp}
\end{figure*}

We first start by normalizing the dereddened spectra to the flux measured at 731 nm in order to remove the dependency of individual target's distance and luminosity. We choose 731 nm as our reference region, since it is free of strong telluric features, avoids features dependent on \logg\; and the photospheric emission dominates at this wavelength.
Median normalized fluxes are then computed in a wavelength interval of choice for each of the spectra. To compute the uncertainty on this flux, we propagate the uncertainties on the flux in the normalization range ($\sim 731$ nm) and the flux in the wavelength range. We also propagate the uncertainties on the extinction. The spectra of some of the later spectral type stars in our sample have a signal to noise of $\sim 0$ at the shortest wavelengths in the UVB arm. In this case, we exclude these normalised fluxes from the procedure. This procedure results in a set containing a normalized flux $f_{\lambda}/f_{731 \rm nm}$ and associated uncertainty for each template in our library. 

We use a non-parametric local polynomial fit \citep{cleveland1979robust} to interpolate between these fluxes as a function of SpT. In this type of interpolation, the value at a given position (spectral type in our case) is obtained by fitting a polynomial to a weighted version of the data points, this is expressed in Eq. \ref{eq:LocPol}.. The polynomial at a given $x$, representing SpT subclass, is defined by the parameters ${\hat{a}_1(x),...,\hat{a}_n(x)}$ as:

\begin{equation}
     {\hat{a}_1(x),...,\hat{a}_n(x)} = argmin_{a_0,...,a_n} \sum_i K\left(\frac{x-x_i}{h}\right) \left(y_i -\sum_{k=0}^n a_n\cdot x_i^k\right)^2
    \label{eq:LocPol}
\end{equation}

Here $x_i$ represents the SpTs of our set of median fluxes in a given wavelength range and $y_i$ is the values of the normalized fluxes themselves. $K$ is a chosen kernel used for weighting the values that are fitted and $h$ is the selected bandwidth. We used a polynomial of degree 2, a bandwidth of 2.5 SpT subclasses, and a Gaussian kernel. 

To obtain the value of the local polynomial fit ($y$) at spectral type position (x) the expression:
\begin{equation}
    y = \sum_{k=0}^n \hat{a}_n\cdot x
\end{equation}
is evaluated.
This procedure is repeated at several equally spaced $x$ values representing spectral types from G8 to M9.5, allowing us to retrieve the interpolated value at any SpT within this range. We limited our interpolation to range from G8 to M9.5 because of the large gap between our only G5 spectrum and the rest of our grid.  We make use of the implementation provided by the localreg PYTHON package \footnote{https://github.com/sigvaldm/localreg/tree/master}.

To account for the heteroscedastic uncertainties we used a Monte-Carlo simulation of 1000 iterations. In each iteration, we resample the normalised flux and SpT of each data point by adding a value sampled from a Gaussian distribution with a standard deviation equal to the respective error and a mean of zero. The non-parametric fit is computed for each iteration. 
At each of the spectral type points we adopt the median values of these fits as the final interpolated model spectrum and for the error we adopt the 1-$\sigma \,$ interval around this median. Fig. \ref{fig:exInterp} shows an example of this procedure for one wavelength range.

\begin{figure*}
    \centering
    \includegraphics[width=0.85\textwidth]{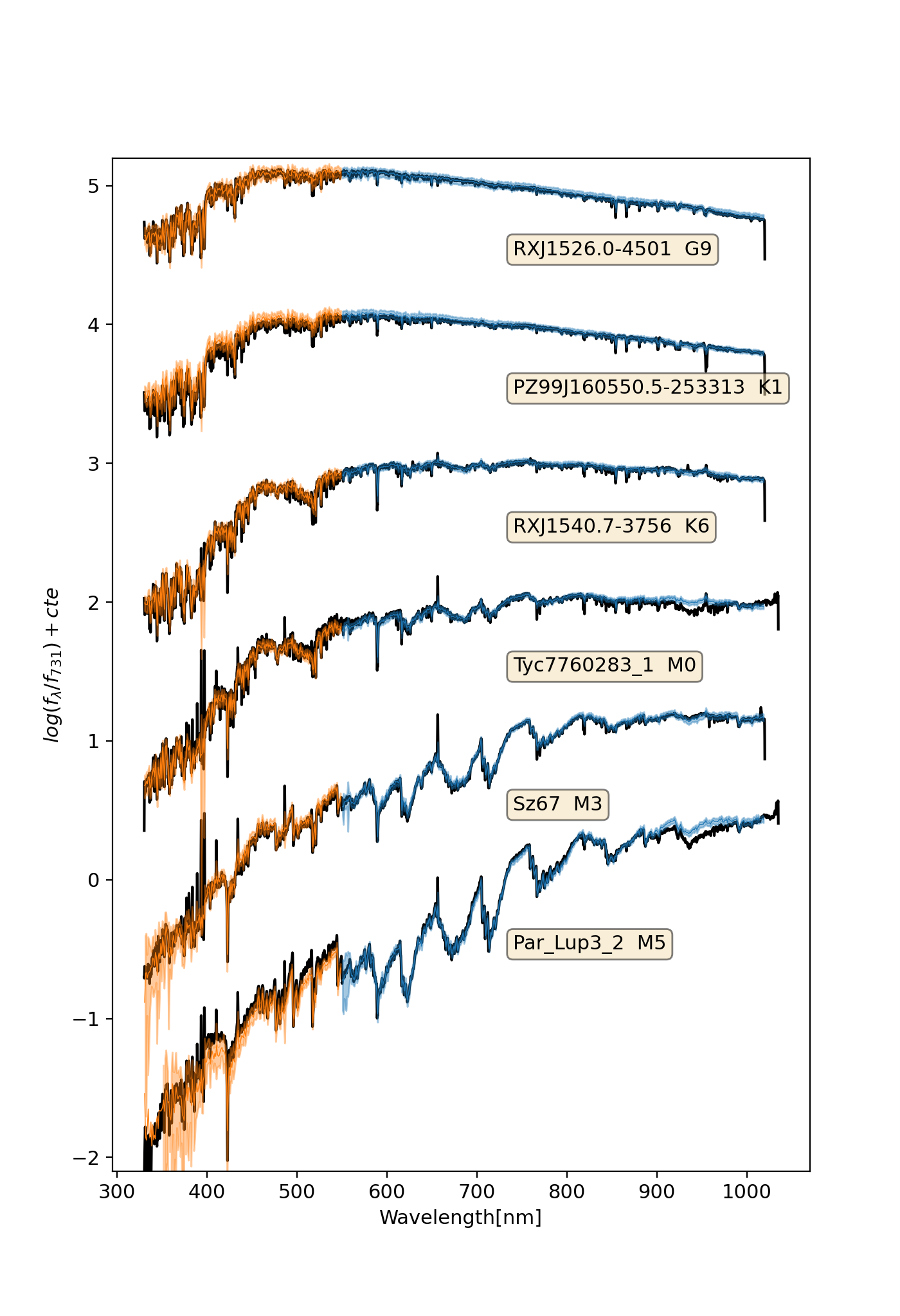}
    \caption{Comparison between individual Class III templates and the interpolated spectra of the same spectral type. The templates are plotted in black and the interpolated spectra in blue/orange. The transparent blue/orange regions represent the 1-$\sigma$ uncertainties in the interpolated spectra. The blue and orange colours correspond to the wavelength ranges of the X-Shooter VIS and UVB arms respectively. The templates have been convolved with a Gaussian kernel for clarity. }
    \label{fig:compInterpCl3}
\end{figure*}

We repeat this procedure for multiple wavelength ranges. An interpolated spectrum is then obtained by evaluating the values of these multiple non-parametric fits at a given spectral type.
We applied it to wavelength ranges of 1 nm width over the entire UVB and VIS arms. A comparison between our interpolated spectra and Class III templates of the same spectral type is shown in Fig. \ref{fig:compInterpCl3}. Here it can be seen that individual templates can have a flux that is lower (PZ99J160843.4-260216) or higher (Par Lup 2) than that of the interpolated templates at short wavelengths. This effect could be a consequence of the uncertainties on the SpT and/or $A_V$ of the spectra and/or differing levels of chromospheric activity. We find that roughly half of the Class III spectra have a Balmer continuum flux higher than that of the interpolated templates while the other half have a lower flux. Fig. \ref{fig:compInterpCl3} also shows that the interpolation can deviate significantly from the observed spectra in the region around 950 nm. This is a consequence of a poor telluric correction in several template spectra. This extinction-less interpolated grid can be found on GitHub \footnote{\url{https://github.com/RikClaes/FRAPPE}} along with the Python script used to generate it and obtain the interpolated templates.

\section{Interpolated spectra for spectral fitting}\label{fitter}

In this section we will apply the new Class III interpolated grid to the classic problem of deriving in a self-consistent way the stellar  (SpT, $A_V$, \lstar) and accretion properties (\Lacc) of an actively accreting TTS. In particular, we will compare our results to those obtained using a method in which the individual template spectra (those listed in Table \ref{tab:fullGrid}) are used to account for the stellar contribution. 

Determining accretion and stellar properties in YSOs is not trivial. Excess continuum emission due to accretion and extinction alters the observed spectrum in opposite ways, although with a different wavelength dependence.
A self consistent method for deriving stellar properties, accretion luminosity, and extinction simultaneously was previously presented by \citet{manara13b}. The method presented here is a further development of this prescription, but with a different approach to the way the stellar photosphere is accounted for when modeling an observed spectrum. This also implies changes in how the stellar luminosity is calculated. We refer to this method as FRAPPE: FitteR for Accretion ProPErties. We apply here this method to the 14 accreting T Tauri stars in the Chamaeleon~I star-forming region observed with X-Shooter during the VLT/PENELLOPE Large Program \citep{manara21}. 

\subsection{Fitting the UV excess}\label{sect:fitter}

The fitting procedure used here combines three components to fit an observed spectrum: 1) A continuum slab model is used to represent the emission from the accretion shock; 2) a reddening law to account for extinction; 3) our interpolated templates are used to represent the stellar contribution. Our fitting procedure consists of a Python code that searches in a grid of parameters the combination that best reproduces the observed spectrum of the input accreting YSO.

For the continuum excess emission we use the grid of isothermal hydrogen slab models developed by \citet{rigliaco12} and \cite{manara13b}, the details of which details are described in \citet{ManaraPhD}. Prior to these works, such an approach was used by \citet{valenti93} and \citet{Herczeg2008}. Here we describe the key parts of the model. The models assume local thermodynamic equilibrium (LTE) and include emission from both $\rm H$ and $\rm H^-$. A given models are described by an electron density ($n_{ e}$), electron temperature ($ T_{\rm e^-}$), and the optical depth at 300 nm ($\tau$). The grid contains slab models for $n_{ e}$ ranging from $10^{-11} \rm cm^{-3}$ to $10^{-16} \rm cm^{-3}$,  $ T_{\rm e^-}$ ranging from 5000 to 11000 K and $\tau$ from 0.01 to 5. The slab models are matched to the observed, dereddened spectrum of an accreting YSO using a scaling factor ($H_{\rm slab}$). The models extend below the minimum wavelength of X-Shooter ($\lesssim 330 \rm nm$), allowing us to compute $L_{\rm acc}$ from the total flux of the slab model when scaled to match the observed spectrum during the fitting procedure. 

We use the Cardelli extinction law \citep{cardelli98} to account for extinction in our procedure. We fix the total-to-selective extinction value $R_V$ to the average interstellar value of $R_V = 3.1$ for all of the fits performed in this work. To find the best fitting $A_V$ value we let the extinction free to vary from 0.0 to 2.0 mag in steps of 0.1 mag. When an extinction value is found at the upper edge of this range, we re-run the procedure for larger $A_V$.

The main difference with respect to the method presented by \citet{manara13b} is the use of an interpolated grid of templates created with the method described in Sect. \ref{interpol}., while \citet{manara13b} used the individual spectra in the grid of Class III templates from \citetalias{manara13a} (and later \citetalias{manara17b}) to represent the photospheric and chromospheric contribution of the central star. During the fitting procedure, this grid is sampled at specific SpT values. We run our procedure for SpTs at steps of 0.5 subclasses. The interpolated spectra are scaled to the observed, dereddened spectrum of the accreting YSO using a scaling factor ($K_{\rm cl3}$) which sets the stellar luminosity.

The best fit is found by minimizing the likelihood function given in equation \ref{likelihood}. 
\begin{equation}
    \chi^{2}_{like} = \sum_{W.L. ranges} \frac{(f_{mod.}-f_{obs,dered})^2}{\sigma_{obs,dered}^2+\sigma_{mod.}^2}
    \label{likelihood}
\end{equation}

Here $f_{mod.} $ is the flux density of the model spectrum, which comprises the sum of the scaled slab model and scaled interpolated Class III template, within one of the selected wavelength range. $f_{obs,dered}$ is the flux of the observed spectrum in the corresponding range after dereddening, $\sigma_{obs}$ is the noise of the dereddened observation in the same range. 
This expression is different than that from \citet{manara13b} in that it includes $\sigma_{mod.}$, the uncertainties on the interpolated grid, taking into account the scaling applied to match the observations (i.e. $\sigma_{mod.} = K_{\rm cl3}\cdot  \sigma_{interp.}$ ). At the edges of the interpolations SpT range its uncertainties become larger in all wavelength ranges. This will bias the results towards the edges of our spectral type range. To prevent this, we set the errors at SpT's earlier than K0 equal to those obtained at K0. 

We select a limited number of wavelength ranges to create an interpolated grid of templates using the method presented in Sect. \ref{interpol}.
These wavelength ranges are selected starting from those used by \citet{manara13b}, but with some additions, as they carry information about the SpT and about the excess emission due to accretion. Wavelength ranges not considered by \citet{manara13b} are highlighted in Table \ref{tab:usedFeat}. We note that the way \citet{manara13b} used the information in these ranges differs from ours. In the best fit determination of \citet{manara13b} the slopes of the observed Balmer and Paschen continuum and the flux ratio at both sides of the Balmer jump are matched to the model. We, however, choose to directly match the flux within these wavelength ranges. We also modified some of the regions adopted from \citet{manara13b} to prevent an overlap in wavelength. We included a number of TiO bandheads around $\sim 710 \: \rm nm$ to constrain the spectral type of targets ranging from K6 to M9.5. To constrain targets with SpT from G5 to K6 we include wavelength ranges around $\sim 465$,  $\sim 510$, and $\sim 540 \: \rm nm$. These are based on the ones used in the R515 index presented by \citetalias{HH14} with slight adjustments being made to avoid the sharp edges of features which are present in the spectra of later type stars.

To constrain the accretion spectrum we include two wavelength ranges that carry information about the size of the Balmer jump to be constrained ($\sim 361 \:\rm nm$  and $\sim 400.5 \:\rm nm$), three additional ranges to further constrain the Balmer continuum ($\sim 335\: \rm nm$, $\sim 340 \:\rm nm$ and $\sim 355 \:\rm nm$ ) and three more to constrain the Paschen continuum ($\sim 414.5 \:\rm nm$, $\sim 450\: \rm nm$, and $\sim 475\: \rm nm$). At longer wavelengths, the inner disk of Class II YSO can contribute to the observed flux \citep[e.g.;][]{Pittman2022}. 
Since our models do not include such a component, we omit the use of wavelength ranges in the NIR arm.
Suboptimal telluric correction of both the templates and spectra being analysed can affect the analysis when included. Wavelength regions that include telluric lines are therefore also omitted. The exact wavelength ranges can be found in Table \ref{tab:usedFeat}. 

The accretion emission is known to veil photospheric absorption features in the observed YSO spectrum. Therefore we compare the Ca I line at $\sim 420 \rm nm$ in the observed spectrum with that of a veiled Class III template to verify the goodness of our fits. We use a random Class III template with the SpT nearest to the best fit solution and the best fitting slab model for this test.

\begin{figure}
    \centering
    \includegraphics[width =0.45\textwidth]{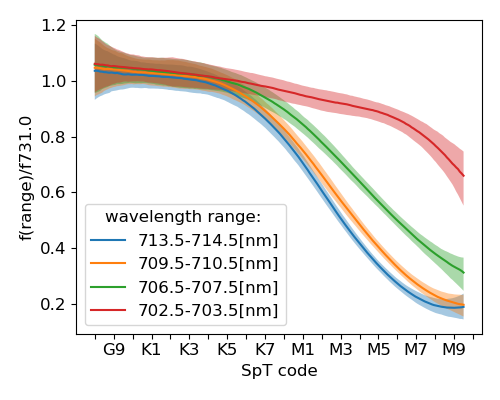}
    \caption{Example of several interpolated normalized fluxes and their associated uncertainties. The values of the fluxes in the TiO 710 nm absorption band can be seen to diverge at late spectral types, indicating its usefulness to constrain the spectral type of these later than K5. The shaded areas indicate the respective 1 sigma uncertainties.}
    \label{fig:allFeatInterp}
\end{figure}

Fig. \ref{fig:allFeatInterp} shows how the normalized TiO 710 nm features vary as a function of SpT in our interpolated grid. Here it can be seen that the different wavelength rages disperse at late spectral types, showing the usefulness of this feature to classify M stars. 

\begin{table}[]
    \centering
    \caption{Wavelength ranges included in the interpolated grid used in our fitting procedure.}
    \begin{tabular}{cc}
    \hline\hline
        Name & wavelength ranges\\
    \hline
        Balmer jump & 359-362$^b$\\
        &399-402 $^b$\\
        \hline
        Balmer continuum & 337.5-342.5 \\
        &352-358 $^b$\\
        \hline
        Paschen continuum &  414-415 $^a$\\
        &448-452 \\
        &474-476 \\
        \hline
        TiO 710 nm & 702.5-703.5 \\
        &706.5-707.5  \\
        &709.5-710.5  \\
        &713.5-714.5  \\
        \hline
        R515 &  464-466 $^b$\\
        &510-515 $^a$ \\
        &539-544 $^a$  \\
        \hline
    \end{tabular}

    \tablefoot{Wavelength ranges that are newly introduced with respect to \cite{manara13b} are indicated with $a$. Wavelength ranges that have been modified with respect to \citet{manara13b} are reported with $b$.}
    
    \label{tab:usedFeat}
\end{table}

The procedure works as follows. For each spectral type we dereddened the observed spectrum for each value of $A_V$. Then for each of these combinations, we consider all of the slab models. The scaling factor for each combination is obtained by having the combined model match the dereddened observation at $\sim 355 \: \rm nm$ and $\sim 731 \: \rm nm$. We note that the latter wavelength range is the same one as used to normalise our grid of interpolated spectra. $K_{clIII}$ therefore corresponds to the stellar flux at 731 nm. In other words, we solve the following system of equations numerically for $K_{\rm cl3}$ and $H_{\rm slab} $.
\begin{equation}
\begin{cases}
    f_{355} &= K_{\rm cl3} \cdot f_{clIII,355} + H_{\rm slab} \cdot f_{slab,355} \vspace{{0.3cm}} \\
    
    f_{731} &= K_{\rm cl3} \cdot f_{clIII,731} + H_{\rm slab} \cdot f_{slab,731} \\
     &= K_{\rm cl3} \cdot 1 + H_{\rm slab} \cdot f_{slab,731}
\end{cases}
\end{equation}
We then use the parameters to calculate $\chi^{2}_{like}$. This is done for every point in the grid. When this is done for the entire grid of parameters we select the model with the lowest $\chi^{2}_{like}$ as the best fit. This optimization procedure is the same as that used by \citet{manara13b}.

From this procedure, we directly obtain SpT and $A_V$. The accretion luminosity is calculated by first integrating the scaled slab model over its entire wavelength range ($\lambda\lambda =$ 50
nm to 2500 nm). This results in the total accretion flux $F_{acc}$ which can then be converted to $L_{acc}$ using $L_{acc} = 4\pi d^2 F_{acc}$, with $d$ being the distance to the target. 
$ T_{\rm eff}$ is obtained from the SpT using the relationship by \citetalias{HH14}.

Due to our use of interpolated spectra, our stellar luminosity calculation differs from that of \citet{manara13b}.
To obtain the stellar luminosity we first compute the stellar flux at 751 nm, the reference wavelength of our bolometric correction. We obtain $f_{751}$ by first computing the flux at this wavelength in the dereddened observed spectrum and then subtracting the scaled slab model flux at the same wavelength. For sources with $T_{\rm eff} < 3500 \rm K$ we apply the same interpolation over the VO feature as discussed in Appendix \ref{app:BolCor}. The bolometric flux $F_{bol.}$ is then computed using our adjusted relationship of \citetalias{HH14} and converted to $L_{\star}$ using $L_{\star} = 4\pi d^2 F_{\rm bol}$. 

The stellar radius is computed from $T_{\rm eff}$ and $L_{\star}$ using $R_{\star}  = L_{\star} / (4\pi \sigma_{\rm Boltz}T_{\rm eff}^4)$. The stellar mass ($M_{\star}$) and age are derived by placing the target on the HRD and interpolating evolutionary tracks at its position. We preferred to use to more recent models of \citet{Baraffe2015}. However, for 5 targets the $T_{\rm eff}$ and/or \lstar range of the \citet{Baraffe2015} models did not extend to the found values, in these cases we used the tracks of \citet{siess00} instead.
Finally, we can obtain the mass accretion rate \Macc\: using 
\begin{equation}
    \dot{M}_{acc} = \frac{L_{acc}R_{\star}}{GM_{\star}}\Bigg(1-\frac{R_{\star}}{R_{in}}\Bigg)^{-1},
\end{equation}
where we use the typical assumption for the inner disk radius of $R_{in} = 5R_{\star}$ \citep{Gullbring98}.

\begin{figure}
    \centering
    \includegraphics[width = 0.5\textwidth]{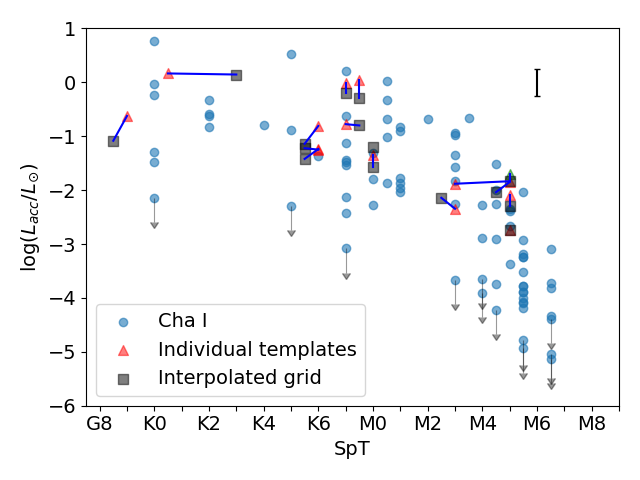}
    \caption{Accretion luminosities vs. spectral type as obtained using FRAPPE (gray squares) and obtained using the method of \cite{manara13a} (red triangles). Results obtained for the same target with the two methods are connected with a blue line. The green triangle indicates the M5 solution of WZ Cha. The light blue circles are the nearly complete Chamaeleon I sample presented by \cite{manara13b}. The black errorbars indicate the uncertainties on the accretion luminosity. The gray arrows indicate sources listed as upper limits by \cite{Manara2023}.}
    \label{fig:ChaILaccSpt}
\end{figure}

\subsection{Accretion rates in accreting young stars in the Chamaeleon I region}\label{sect:chaI}
\begin{figure}
    \centering
    \includegraphics[width = 0.5\textwidth]{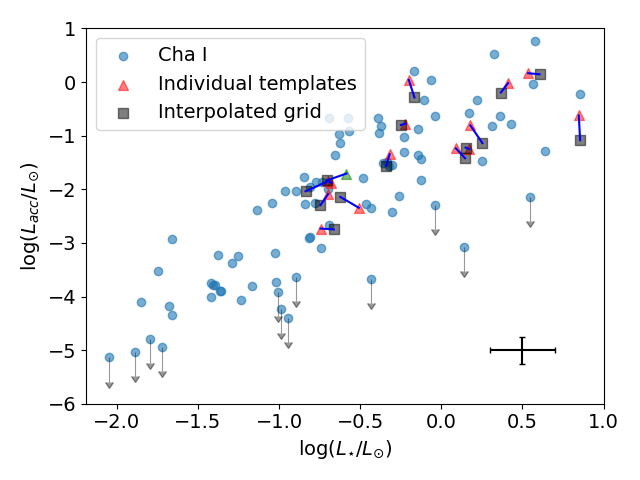}
    \caption{Accretion luminosities vs. stellar luminosity as obtained using FRAPPE (red triangles) and obtained using the method of \cite{manara13a} (gray squares). Results obtained for the same target with the two methods are connected with a blue line. The green triangle indicates the M5 solution of WZ Cha. The light blue dots is the nearly complete Chamaeleon I sample presented by \cite{manara13b}. The black errorbars indicate the uncertainties on the stellar luminosity and accretion luminosity. The gray arrows indicate sources listed as upper limits by \cite{Manara2023}.}
    \label{fig:ChaILaccLstar}
\end{figure}

To test our method we apply it to the sample of 14 Chamaeleon I targets observed with the VLT/X-Shooter spectrograph as part of the VLT/PENELLOPE Large Programme \citep[see][for details on the survey]{manara21}. The results are listed in Table \ref{tab:chamIcontFit}, and are presented here for the first time. In Appendix \ref{FitChaIBalmerPaschen} comparisons between the dereddened observations and best fit models are shown. For the purpose of comparison, we applied the procedure of \citet{manara13b} to the same sample and list the results in Table \ref{tab:chamIdiscrFit}.  In Fig. \ref{fig:ChaILaccSpt} the results of both methods are plotted on a $L_{\rm acc}$-SpT diagram. Fig. \ref{fig:ChaILaccLstar} displays both the obtained stellar luminosity and accretion luminosity. Results obtained for the same target are connected with a blue line to make the comparison easier. The values for the nearly complete sample of accreting young stars in the Chamaeleon I star-forming region presented by \citet{manara17b}, using the values from \citet{Manara2023}, is also shown with blue circles. 

It can be seen that the results obtained here are generally in agreement within the typical uncertainties for the fitting procedure of \citet{manara13b} ($\sigma_{A_V} \approx 0.2 {\,\rm mag}$, $\sigma_{L_{\rm acc}} \approx 0.25 {\,\rm dex}$ and $\sigma_{\dot{M}_{\rm acc}} \approx 0.35 {\,\rm dex}$, \citealt{Manara2023}). For the SpT we expect typical uncertainties of half a subclass for SpTs later than K6, for earlier SpTs we expect uncertainties of about 1 subclass. For 3 targets we find differences larger than the typical uncertainties. These targets are WZ Cha, CV Cha and VZ Cha. We describe the results for these targets in Appendix \ref{FitChaI}.

The derived accretion luminosities of all targets agree within the aforementioned uncertainties. The obtained stellar luminosities are also in good agreement. 

\begin{figure}
    \centering
    \includegraphics[width = 0.5\textwidth]{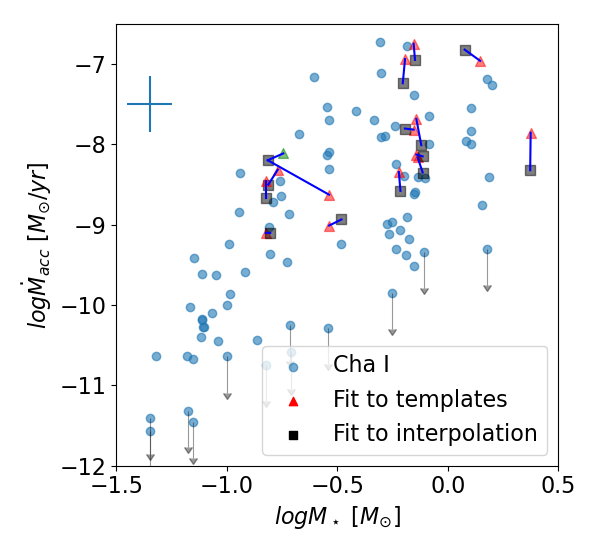}
    \caption{Mass accretion rate vs. stellar mass as obtained using FRAPPE (red triangles) and obtained using the method of \cite{manara13a} (gray squares). Results obtained for the same target with the two methods are connected with a blue line. The green triangle indicates the M5 solution of WZ Cha. The light blue dots are the nearly complete Chamaeleon I sample presented by \cite{manara13b}. The blue errorbars indicate the uncertainties on the stellar mass and mass accretion rate. The gray arrows indicate sources listed as upper limits by \cite{Manara2023}.}
    \label{fig:ChaImaccVsmstar}
\end{figure}

Fig. \ref{fig:ChaImaccVsmstar} displays the resulting \macc - \mstar\, diagram. For WZ Cha we obtained two degenerate solutions using the method of \citet{manara13b} one at SpT M3 (indicated in red) and one at SpT M5 (indicated in green). We prefer the M5 solution as the Class III template to which the M3 solution was fitted appears to be an outlier (see Appendix \ref{FitChaI}). Here it can be seen that, with exception of the M3 solution of WZ Cha, our results are within the typical uncertainties.

The result obtained for WZ Cha and VZ Cha illustrates the benefits of our method (see Appendix \ref{FitChaI}). For WZ Cha, FRAPPE breaks a degeneracy between spectral types and in the case of VZ Cha the Paschen continuum appears better reproduced. The analysis for CV Cha highlights the need for improved constraints when fitting late G to early K stars. Such constraints can not only be obtained from the inclusion of Class III spectra around these spectral types in our grid but also from additional information on the emission and absorption lines present in the spectra of the Class II YSOs.

\begin{table*}[]
    \centering
    \caption{Accretion properties of the Chameleon I PENELLOPE targets obtained with FRAPPE. }
\begin{tabular}{lllllllllll}
\toprule
      Name &  Dist  &  SpT & $T_{\rm eff}$ &   $A_V$&  $L_\star $  & $\log(L_{acc}/L_\odot)$ & $M_\star$\_B15 &  $M_{acc}$\_B15  & $M_\star$\_S00 & $M_{acc}$\_S00 \\
       &   [pc] &   & [K] &    [mag]&  $ [L_\odot]$  &  & [$M\odot$]&   [$M_\odot$/yr] &  [$M\odot$] & [$M_\odot$/yr]\\
\midrule
    CHX18N & 192.0 & K5.5 & 4164 & 0.70 & 1.40 &  -1.42 &    0.77 &  $4.48\cdot 10^{-9}$ &       0.85 &    $4.06\cdot 10^{-9}$ \\
CHX18N\_ep2 & 192.0 & K5.5 & 4164 & 0.70 & 1.41 &  -1.22 &    0.77 &  $7.11\cdot 10^{-9}$ &       0.86 &    $6.40\cdot 10^{-9}$ \\
     CSCha & 190.0 & K5.5 & 4164 & 0.40 & 1.79 &  -1.13 &    0.76 &  $9.81\cdot 10^{-9}$ &       0.85 &    $8.75\cdot 10^{-9}$ \\
     CVCha & 192.0 & K3.0 & 4540 & 1.10 & 4.03 &   0.14 &    1.19 &  $1.52\cdot 10^{-7}$ &       1.47 &    $1.23\cdot 10^{-7}$ \\
       Hn5 & 195.0 & M5.0 & 2980 & 0.80 & $1.79\cdot 10^{-1}$ &  -2.30 &      ... &       ... &       0.15 &    $2.14\cdot 10^{-9}$ \\
     INCha & 193.0 & M5.0 & 2980 & 0.00 & $2.18\cdot 10^{-1}$ &  -2.75 &      ... &       ... &       0.16 &    $7.91\cdot 10^{10}$ \\
     IPTau & 129.4 & M0.0 & 3900 & 1.10 & $3.96\cdot 10^{-1}$ &  -1.21 &    0.63 &  $5.43\cdot 10^{-9}$ &       0.64 &    $5.31\cdot 10^{-9}$ \\
     SYCha & 181.0 & K7.5 & 3960 & 0.70 & $5.63\cdot 10^{-1}$ &  -0.80 &    0.63 &  $1.58\cdot 10^{-8}$ &       0.67 &    $1.52\cdot 10^{-8}$ \\
      Sz10 & 184.0 & M4.5 & 3084 & 0.70 & $1.46\cdot 10^{-1}$ &  -2.04 &    0.15 &  $3.17\cdot 10^{-9}$ &       0.17 &    $2.87\cdot 10^{-9}$ \\
      Sz19 & 189.0 & G8.5 & 5104 & 1.50 & 7.17 &  -1.083 &      ... &       ... &       2.36 &   $ 4.79\cdot 10^{-9}$ \\
      Sz45 & 192.0 & M0.0 & 3900 & 0.50 & $4.57\cdot 10^{-1}$ &  -1.56 &    0.61 &  $2.64\cdot 10^{-9}$ &       0.61 &  $  2.62\cdot 10^{-9}$ \\
     VWCha & 190.0 & K7.0 & 4020 & 2.10 & 2.32 &  -0.194 &      ... &       ... &       0.71 &    $1.13\cdot 10^{-7}$ \\
     VZCha & 191.0 & K7.5 & 3960 & 1.20 & $6.83\cdot 10^{-1}$ &  -0.29 &    0.63 &  $5.79\cdot 10^{-8}$ &       0.67 &    $5.40\cdot 10^{-8}$ \\
     WZCha & 193.0 & M5.0 & 2980 & 0.60 & $1.97\cdot 10^{-1}$ &  -1.83 &      ... &       ... &       0.15 &   $ 6.37\cdot 10^{-9} $\\
     XXCha & 192.0 & M2.5 & 3484 & 0.50 & $2.38\cdot 10^{-1 }$&  -2.14 &    0.33 &  $1.16\cdot 10^{-9}$ &       0.35 &   $ 1.11\cdot 10^{-9}$ \\
\bottomrule
\end{tabular}
\tablefoot{The stellar mass and mass accretion rates are computed with both the isochrones of \cite{Baraffe2015} (marked as B15) and \citet{siess00} (marked as S00). CV Cha and CS Cha are binaries \citep[][ respectively]{Reipurth2002,Guenther2007}. We do not report values that are outside of the ranges of the \citet{Baraffe2015} isochrones.}
    \label{tab:chamIcontFit}
\end{table*}

\begin{table*}[]
\caption{Accretion properties of the Chameleon I PENELLOPE targets obtained with the method presented by \citet{manara13b}. }
    \centering
        \begin{tabular}{llllllllll}
\toprule
      Name & Dist. [pc] &  SpT & T$_{\rm eff}$ [K] &  $A_V$ [mag] & \lstar\, [$L_\odot$] & $\log(L_{\rm acc}/L_\odot)$ & $M_\star$ &  \macc\, [$M_\odot$/yr] & Model  \\

\midrule
    CHX18N &  192 &   K6 & 4115 & 0.7 &  1.23 &  -1.236 &  0.73 & $6.90\cdot 10^{-9}$ &   B15 \\
CHX18N\_ep2 &  192 &   K6 & 4115 & 0.8 &  1.48 &  -1.248 &  0.72 & $7.49\cdot 10^{-9}$ &   B15 \\
     CSCha &  190 &   K6 & 4115 & 0.4 &  1.51 &  -0.808 &  0.72 & $2.08\cdot 10^{-8}$ &   B15 \\
     CVCha &  192 & K0.5 & 4830 & 1.1 &  3.43 &   0.166 &  1.40 & $1.10\cdot 10^{-7}$ &   B15 \\
       Hn5 &  195 &   M5 & 2980 & 1.1 &  0.20 &  -2.093 &  0.15 & $3.55\cdot 10^{-9} $&   S00 \\
     INCha &  193 &   M5 & 2980 & 0.2 &  0.18 &  -2.732 &  0.15 & $7.93\cdot 10^{-10}$ &   S00 \\
     SYCha &  181 &   K7 & 4020 & 0.8 &  0.60 &  -0.775 &  0.70 & $1.53\cdot 10^{-8}$ &   B15 \\
      Sz10 &  184 &   M5 & 3125 & 1.0 &  0.20 &  -1.849 &  0.17 & $4.87\cdot 10^{-9}$ &   B15 \\
      Sz19 &  189 &   G9 & 5025 & 1.5 &  7.04 &  -0.623 &  2.37 & $1.40\cdot 10^{-8}$ &   S00 \\
      Sz45 &  189 &   M0 & 3900 & 0.7 &  0.48 &  -1.344 &  0.60 & $4.54\cdot 10^{-9}$ &   B15 \\
     VWCha &  190 &   K7 & 4020 & 2.3 &  2.58 &  -0.019 &  0.70 & $1.80\cdot 10^{-7}$ &   S00 \\
     VZCha &  191 & K7.5 & 3960 & 1.7 &  0.63 &   0.040 &  0.64 & $1.16\cdot 10^{-7}$ &   B15 \\
     WZCha &  193 &   M3 & 3410 & 1.0 &  0.21 &  -1.881 &  0.29 & $2.37\cdot 10^{-9}$ &   B15 \\
     *WZCha &  193 &   M5 & 3060 & 0.8 &  0.26 &  -1.706 &  0.18 & $7.73\cdot 10^{-9}$ &   S00 \\
     XXCha &  192 &   M3 & 3410 & 0.3 &  0.31 &  -2.344 &  0.29 & $9.78\cdot 10^{-10}$ &   B15 \\
\bottomrule
\end{tabular} 

    \tablefoot{The degenerate solution of WZ Cha is marked as *WZ Cha.}
    \label{tab:chamIdiscrFit}
\end{table*}

\begin{figure*}
    \includegraphics[width = 0.49\textwidth]{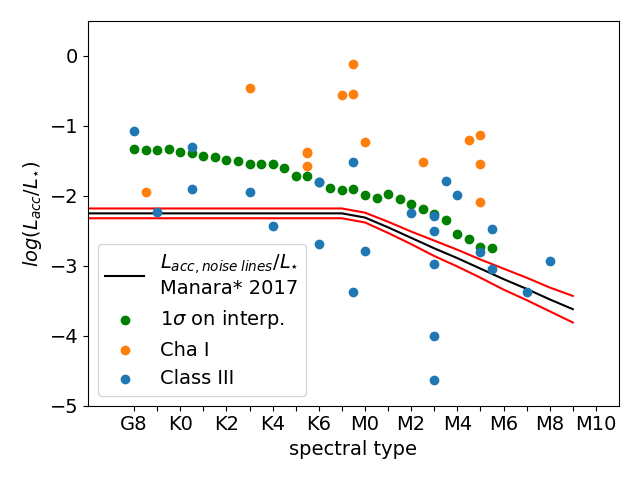}
    \quad
    \includegraphics[width = 0.49\textwidth]{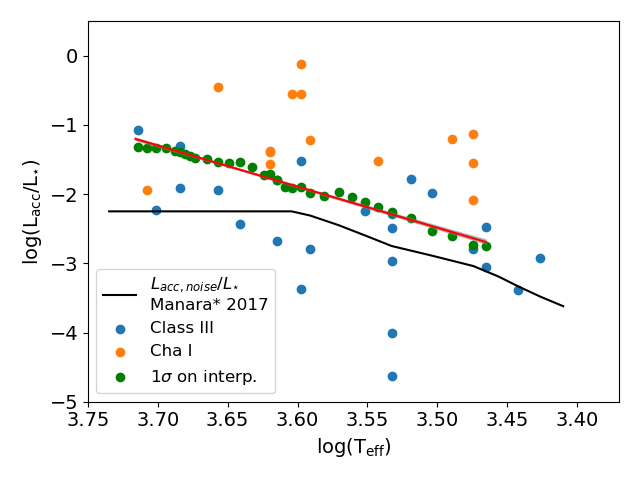}
    \caption{SpT and $T_{\rm eff}$ depedency of the accretion noise $\log(L_{acc,noise}/L_{\star})$ values obtained from interpreting the uncertainties in the Balmer and Paschen continuum ($<500$ nm) on our model Class III spectra as accretion (Green dots). The blue dots illustrate the $\log(L_{acc,noise}/L_{\star})$ values obtained when fitting the Class III spectra with an accretion slab model. The results obtained for the PENELLOPE Chamaeleon I sample is indicated with orange dots. The black line shows the \laccnoise\, relationship measured from the emission lines by \citetalias{manara17b} and the red lines on the left show the uncertainty thereon. On the right, the red line illustrates the best fit to the $\log(L_{acc,noise}/L_{\star})$ values.}
    \label{fig:money}
\end{figure*}

\section{Limits on accretion luminosity measurements from the UV excess} \label{sect:limit}

The study of the Class III spectra is important not just because they are used as templates of the photospheres of accreting stars, but also because they provide essential information on the limitations introduced by the stellar chromospheric activity to the measurements of low accretion luminosities. \citetalias{manara13a} and \citetalias{manara17b} discussed how chromospheric activity in the form of line emission defines the lower boundaries to \lacc\, derived from the well-established correlation between line emission and accretion luminosity, and named this \laccnoise. Here we will examine how the noise in the Balmer continuum emission in Class III objects sets an upper boundary to the values of \lacc\, derived from the Balmer excess.

A variety of factors such as the uncertainties on the spectral type, extinction, and levels of chromospheric activity may contribute to the uncertainties on our interpolated model spectra. While such uncertainties may only cause minor contributions to the accretion luminosity measured in strong accretors, in weakly accreting objects such noise may have an important impact. Therefore, to characterise the limitations of our method we derive the accretion luminosities that correspond to the 1 $\sigma$ uncertainties on our interpolated spectra, and we refer to this value as $L_{\rm acc,noise}$. To do this, we create a set of artificial spectra that include a UV excess equivalent to the 1 $\sigma$ uncertainties and fit these with FRAPPE.

Templates later than SpT M5.5 have a S/N < 1 in the Balmer and Paschen continuum. Therefore, including these templates in our analysis would strongly affect the uncertainty of the continuum UV flux. We therefore limit the analysis presented here to spectral types ranging from G8 to M5.5. We generate another interpolated grid based on our templates but excluding templates later than M5.5. We use this grid to create a set of blue enhanced artificial spectra. For wavelength ranges in the Balmer and Paschen continuum (< 500nm) we set the value of the artificial spectrum equal to the interpolation spectra plus  $1\sigma$ of the uncertainty thereon. The $1\sigma$ uncertainty is obtained from the Monte Carlo simulation discussed in Sect \ref{interpol}. For longer wavelengths, we simply adopt the values of the interpolated spectrum. The artificial spectra are unitless since our interpolation is normalized to have the flux at 731 nm equal to 1. We therefore multiply the artificial spectra with a random value on the order of $0.1 \times 10^{-14}$ to $100 \times 10^{-14}$ erg\,nm$^{-1}$s$^{-1}$cm$^{-2}$ so that they have a "realistic" flux. We create such artificial spectra for each half subclass ranging from G8 to M5.5. 

During this fitting procedure, we keep the SpT fixed to the value used for constructing the spectrum. The extinction is fixed to $A_V = 0$ mag. In doing so we obtain the accretion luminosities corresponding to spectra whose Balmer and Paschen continuum are 1$\sigma$ interval above the interpolation. Here we assumed that the uncertainties in the Balmer and Paschen continuum are fully correlated. 

The $L_{\rm acc,noise}$ and $L_{\star}$ obtained with this method are arbitrary as they scale with the multiplicative factor used to create the artificial spectra. Therefore we do not report the obtained values of $L_{\rm acc,noise}$ and $L_{\star}$. Instead, we report the ratio $L_{\rm acc,noise}/L_{\star}$ which is independent of this scaling and representative of the contrast down to which we can accurately measure a UV excess. The results are listed in Table \ref{tab:fitUncertRes}. The dependence of $L_{acc,noise}/L_{\star}$ as a function of SpT is shown in  Fig.~\ref{fig:money}. Assuming the SpT-$T_{\rm eff}$ relation of \citetalias{HH14}, we also plot $L_{acc,noise}/L_{\star}$ as a function of $T_{\rm eff}$ in Fig. \ref{fig:money}. Here it can be seen that the trend as a function of $ T_{\rm eff}$ can be well approximated as a simple linear relation. 
This relationship can be expressed as:

\begin{equation}
        \log(L_{\rm acc,noise}/L_{\star}) = (5.9\pm 0.2) \cdot \log(\rm T_{\rm eff}[K]) - (23.3 \pm 0.7) 
    \label{eq:Criterion}
\end{equation}

The relationships in Eq. \ref{eq:Criterion} and Table \ref{tab:fitUncertRes} can be seen as a criterion for the lowest measurable accretion luminosities. If the $\log(L_{\rm acc,noise}/L_{\star})$ measured for a CTTS is of a similar order of magnitude or falls below it, then we can not conclude if a target is accreting based on the UV excess measurement. Several relations between the effective temperature and spectral type have been presented in the literature \citep[e.g.; HH14,][]{Pecaut2013,kenyon95,luhman03}. Due to this lack of uniformity, we recommend to directly use the values of \laccnoise\, at each SpT reported in Table \ref{tab:fitUncertRes}. 

Figure \ref{fig:money} also includes results obtained by fitting the Class III templates themselves with FRAPPE. Here we once again kept the SpT fixed to the SpT of the respective Class III and kept $A_V = 0$ mag. In this way, we obtain a similar \laccnoise\, measurements. Only half of the Class III templates provide a \laccnoise\, measurement since half have a Balmer continuum below the interpolation. The \laccnoise\, values obtained from the class III templates display a large scatter, this is a natural consequence of the scatter in the residuals between the templates and the interpolation in the Balmer and Paschen continuum. These results demonstrate how different the Balmer continua, and therefore the inferred UV-excess, of the various templates are. Using individual templates one may derive accretion luminosity that can change by fractions of the stellar luminosity of up to 10\% relative to that found by using the interpolated spectra. The most extreme case is for late G SpT targets. This uncertainty is significantly weaker at later SpTs and in most cases it is only on the order of 1\% of the stellar luminosity or lower.

We also plot the results for the Class II targets presented in Sect. \ref{sect:chaI}.  One target, Sz19, clearly falls below the relationship in Eq. \ref{eq:Criterion}. Its measured mass accretion rate can therefore be highly affected by uncertainties on the chromospheric emission. Finally, for comparison, we also plot a similar relationship derived by \citetalias{manara17b} for the lowest accretion luminosities that one can measure by studying the emission lines. \citetalias{manara17b} measured the luminosities of several chromospheric emission lines in Class III stars and converted them to accretion luminosities using the relations of \cite{alcala14}. We converted their results obtained as a function of spectral type to effective temperature using the SpT-$T_{\rm eff}$ of \citetalias{HH14} for consistency.  
It can be seen that the $1\sigma$ threshold that we derive for the UV continuum excess is higher in the entire SpT/$T_{\rm eff}$ range than the threshold obtained from the line luminosities. 
This implies that an analysis of accretion tracing emission lines allows us to measure lower accretion luminosities than the UV excess. However, the empirical relationships used to convert line luminosities to accretion luminosities \citep{alcala14,alcala17} have been calibrated using measurements of the UV excess. It is therefore uncertain whether these relationships can be extrapolated to accretion luminosities below the threshold on the measurable UV excess. A more detailed modeling of the $\rm H\alpha$ emission lines, such as performed by \cite{Thanathibodee2023} may yet allow for the characterisation of accretion properties of stars with low accretion rates, even well below the threshold of \citetalias{manara17b}. Here we need to add the caveat that new work is currently being done on the influence of chromospheric emission lines on accretion rate measurements (Stelzer et al. in prep.), which will update the limit of \citetalias{manara17b} using the relationship between line and accretion luminosity of \citet{alcala17}.  

We note that our criterion for measurable accretion luminosities and that of \citetalias{manara17b} are different in nature. The criterion of \citetalias{manara17b} represents typical chromospheric line luminosities, whereas ours represents the 1 sigma uncertainty on the stellar and chromospheric UV continuum emission in our interpolated Class III model spectrum. The interpretation of both therefore differs. If the line luminosity measured in a CTTS is of a similar order of magnitude to the criterion of \citetalias{manara17b}, then the chomospheric emission likely provides a dominant contribution to the measured \Lacc. On the other hand, if a \Lacc\, obtained from the UV excess is smaller than the limit presented here, then the uncertainties on this \Lacc\, is dominated by the uncertainties on the stellar emission.

\begin{table}[]
    \centering
        \caption{Accretion noise values as a function of SpT and $T_{\rm eff}$.}
    \label{tab:fitUncertRes}
    \begin{tabular}{ccc}
\toprule
            SpT & $\rm T_{\rm eff}[K]$ &$\log(L_{\rm acc,noise}/L_{\star})$ \\
\midrule
G8.0  &  5180   & -1.3   \\
G8.5  &  5102   & -1.3   \\
G9.0  &  5025   & -1.3   \\
G9.5  &  4947   & -1.3   \\
K0.0  &  4870   & -1.4   \\
K0.5  &  4830   & -1.4   \\
K1.0  &  4790   & -1.4   \\
K1.5  &  4750   & -1.5   \\
K2.0  &  4710   & -1.5   \\
K2.5  &  4625   & -1.5   \\
K3.0  &  4540   & -1.5   \\
K3.5  &  4457   & -1.5   \\
K4.0  &  4375   & -1.5   \\
K4.5  &  4292   & -1.6   \\
K5.0  &  4210   & -1.7   \\
K5.5  &  4162   & -1.7   \\
K6.0  &  4115   & -1.8   \\
K6.5  &  4067   & -1.9   \\
K7.0  &  4020   & -1.9   \\
K7.5  &  3960   & -1.9   \\
M0.0  &  3900   & -2.0   \\
M0.5  &  3810   & -2.0   \\
M1.0  &  3720   & -2.0   \\
M1.5  &  3640   & -2.0   \\
M2.0  &  3560   & -2.1   \\
M2.5  &  3485   & -2.2   \\
M3.0  &  3410   & -2.3   \\
M3.5  &  3300   & -2.3   \\
M4.0  &  3190   & -2.5   \\
M4.5  &  3085   & -2.6   \\
M5.0  &  2980   & -2.7   \\
M5.5  &  2920   & -2.7   \\
\bottomrule
\end{tabular}
\tablefoot{The accretion noise values are obtained from interpreting the 1$\sigma$ level of uncertainty on the Balmer and Paschen continua ($<500$ nm) on our interpolated Class III spectra at a given spectral type as accretion luminosity. $T_{\rm eff}$ obtained from the relationship from \citetalias{HH14} are also listed for convenience.}
\end{table}

\begin{table}[]
    \centering
    \caption{The stellar and accretion luminosities obtained for the Class III templates when fitted with FRAPPE using $A_V = 0$ and the spectral type of the respective Class III template.}
    \begin{tabular}{lccc}
\toprule
                  Name &  $\log{L_{\star}}$ &   $\log{L_{acc}}$ &   $\log(L_{\rm acc}/L_{\star\rm})$ \\
\midrule
        RXJ1508.6-4423 &  -0.06 & -1.09 &  -1.07 \\
        RXJ1526.0-4501 &  -0.03 & -2.293 &  -2.23 \\
  PZ99J160843.4-260216 &  0.09& -1.817 &  -1.91 \\
           CD-31\_12522 &  -0.14 & -1.449 &  -1.31 \\
        RXJ1547.7-4018 &  -0.18 & -2.128 &  -1.94\\
        RXJ1538.6-3916 &  -0.36 & -2.795 &  -2.43 \\
2MASSJ15552621-3338232 &  -0.69 & -2.485 &  -1.80 \\
        RXJ1543.1-3920 &  -0.48& -3.165 &  -2.68 \\
        RXJ1607.2-3839 &  -0.21 & -3.586 &  -3.38 \\
                 NOlup &  -0.45 & -1.968 &  -1.52 \\
          Tyc7760283\_1 &  -0.61 & -3.401 &  -2.79 \\
             THA15-36B &  -1.11 & -3.352 &  -2.24 \\
                  Sz67 &  -1.16 & -4.837 &  -2.97 \\
                RECX-6 &  -1.05& -3.237 &  -2.50\\
           CD\_36\_7429B &  -0.99 & -4.134 &  -4.63 \\
                  TWA7 &  -0.84 & -5.616 &  -4.00 \\
            TWA15\_app2 &  -0.95 & -3.553 &  -2.29 \\
            TWA15\_app1 &  -1.01 & -2.786 &  -1.78 \\
                  Sz94 &  -1.23& -3.225 &  -1.99 \\
            Par\_Lup3\_2 &  -0.96& -3.753 &  -2.79 \\
2MASSJ16091713-3927096 &  -1.83& -4.887 &  -3.05 \\
                 Sz107 &  -1.06& -3.538 &  -2.48 \\
     J11195652-7504529 &  -2.07 & -5.451 &  -3.38 \\
             CHSM17173 &  -1.86 & -4.790 &  -2.93 \\
\bottomrule
\end{tabular}
    
    \label{tab:fitCL3res}
\end{table}

\subsection{Implications for mass accretion rate estimates.}
\begin{figure}
    \centering
    \includegraphics[width = 0.49\textwidth]{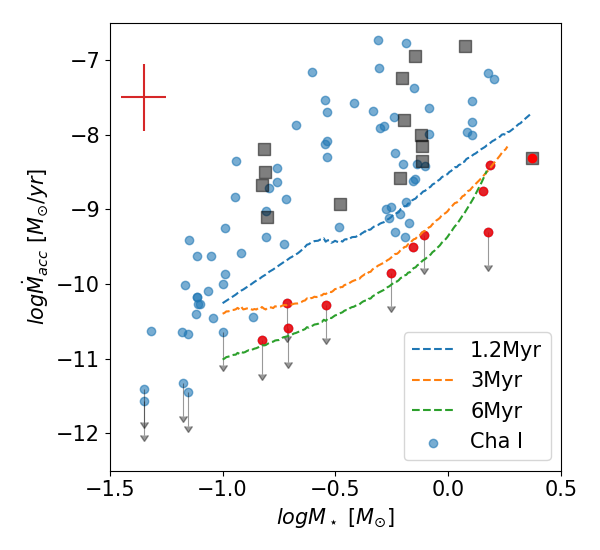}
    \caption{Mass accretion rate vs. stellar mass for the Chamaeleon I sample presented by \citet{Manara2023} (blue dots) and the Chamaeleon I sample analysed in this work. Objects that have an accretion luminosity lower than the 
    criterion given in equation \ref{eq:Criterion} are marked with a red dot. The limits we derived based on the uncertainties on the chromospheric emission are shown with the blue, orange, and green lines for 1.2, 3, and 6 Myr old objects respectively. The red errorbars indicate the uncertainties on the stellar mass and mass accretion rate. The gray arrows indicate sources listed as upper limits by \cite{Manara2023}.}
    \label{fig:EffectMacc}
\end{figure}

The limitations on measuring \Lacc\, through the UV excess in turn reflects on the $\dot{M}_{\rm acc}$ values that we can reliably measure. It is of particular interest to investigate how these limitations affect the $\dot{M}_{\rm acc} - M_\star$ relation. We use the values of \laccnoise\, from Eq. \ref{eq:Criterion} to compute the lower limit of \Macc\: as a function of stellar mass using  the non-magnetic isochrones of \citet{Feiden2016} at 1, 2, 3, and 6 Myr. The results are shown in Fig.~\ref{fig:EffectMacc}. We use these isochrones since they extend to higher masses than those of \citet{Baraffe2015} and therefore better cover the spectral types in Table \ref{tab:fitUncertRes}. The isochrones of \cite{Feiden2016} are mostly consistent with those of \cite{Baraffe2015}.  

We also applied the relationship of Eq.~\ref{eq:Criterion} directly to the measured \lacc\, and \lstar\, of the Chamaeleon I sample from \citet{Manara2023} and the sample presented in Sect.~\ref{sect:chaI}. The results are also plotted in Fig.~\ref{fig:EffectMacc}. The majority of Class II targets fall well above the UV excess measurement limit. However, a considerable fraction appears to fall below the limit. The measured accretion rates of these targets are therefore dominated by the uncertainties on the stellar UV-continuum emission. More detailed analyses are needed to disentangle the possible low accretion rate in these targets from the stronger chromospheric emission. 

\citet{manara17a} used the Chamaeleon I sample shown here in combination with a sample of targets in Lupus \citep[from][]{alcala17} to discuss whether the correlation between \macc\, and \mstar\, is driven by the detection limits and if the observed scatter in \macc\, fills the observable range. Only 3 sources from the sample of \cite{Manara2023} that fall below our detection limit were previously not indicated as upper limits. These targets all fall in a locus were upper limits were already present. Our limit on the measurable mass accretion rate therefore does not change the conclusion reached by \citet{manara17a}. Namely, that this correlation is real and not a consequence of detection limits.

\section{Summary and conclusions}\label{sect:concl}

We presented the analysis of 24 new VLT/X-shooter spectra of Class III young stars for their use as photospheric templates to complement the previous samples of \citetalias{manara13a} and \citetalias{manara17b}. All spectra are available in reduced and dereddened form on GitHub \footnote{\label{note1}\url{https://github.com/RikClaes/FRAPPE_ClassIII}}. We obtained the spectral type through the use of spectral indices and by comparing the relative strength of photospheric features to the samples of \citetalias{manara13a} and \citetalias{manara17b}.

We employed ROTFIT to derive the photospheric properties ($\log{g}$, $\rm T_{eff}$, $v\sin{i}$, RV) of the sample. The results of both methods are generally in agreement with each other and the photospheric properties confirm the young nature of these targets. As previously noted in the literature \citep[e.g.,][]{Bell14,HH15,manara17b, Pecaut2016}, we see a mass dependent trend of the derived isochronal ages for targets in the same star forming region. We compared the stellar luminosities obtained through extending and integrating the X-Shooter spectra in our sample with luminosities obtained by applying the bolometric correction of \citetalias{HH14}. We found a small but systematic difference between the results of both methods for effective temperatures lower than about 4500 K, leading us to propose an adjustment to the bolometric correction of \citetalias{HH14}. 

To remove the non-uniform SpT sampling of the templates and mitigate the uncertainties associated with their use, we proposed a method of interpolating between the templates. We applied this type of interpolation to the entire wavelength range covered by the spectra in 1nm wide bins and made the results publicly available on GitHub\footref{note1}. We also applied the method to specific regions in the spectra that are useful to obtain the stellar and accretion properties of Class II targets. This empirical method has been implemented in a code that self consistently measures the extinction, stellar, and accretion properties. We verified this framework by analysing the Chamaeleon I sample observed as part of the PENELLOPE program and compared the results to those obtained using the method presented by \citet{manara13b} on the same spectra. We found good agreement in the results obtained with the two methods. Several future improvements, such as additional Class III templates at early SpTs or the inclusion of veiling measurements in the best fit determination, can still be made to FRAPPE. Such improvements are particularly useful to constrain the stellar properties stars with a spectral type earlier than K6 and will make FRAPPE more user friendly.

We measured the typical accretion to stellar luminosity ratios that would be obtained if the 1 $\sigma$ uncertainties in our model spectra were interpreted as excess emission due to accretion. This ratio is representative of the typical lower limit on the measurable mass accretion rates in young stars when using continuum UV excess. The value has a strong dependency on the spectral type. This trend is likely a consequence of the worse constraints on earlier SpT stellar emission within our sample of Class III sources, as well as of a worse contrast with the UV excess due to earlier stars having the peak of the photospheric emission at bluer wavelengths, hampering the detection of an excess for contrast reasons. This limit is higher than the lower limit presented for accretion rates derived from emission lines presented by \citetalias{manara17b}. We, therefore, conclude that an in-depth analysis of
emission lines is needed to obtain accurate measurements of mass accretion rates in low accreting objects. We applied our lower limit to the Chamaeleon I sample of \citet{Manara2023}. The majority of objects in this region were found to have accretion rates well above our lower limit.

This shows that the uncertainties associated with the use of Class III templates do not significantly affect the observed correlations between mass accretion rate and disk mass or stellar mass (see \citealt{Manara2023} for a review). Our improved methodology for self consistently deriving accretion properties will facilitate further studies into these relations.

\begin{acknowledgements}
We thank the anonymous referee for the useful comments to our submitted manuscript. 
R. Claes acknowledges the PhD fellowship of the International Max-Planck-Research School (IMPRS) funded by ESO
This work was partly funded by the Deutsche Forschungsgemeinschaft (DFG, German Research Foundation) in the framework of the YTTHACA Project 469334657 under the project code MA 8447/1-1.
Funded by the European Union (ERC, WANDA, 101039452). Views and opinions expressed are however those of the author(s) only and do not necessarily reflect those of the European Union or the European Research Council Executive Agency. Neither the European Union nor the granting authority can be held responsible for them.

J. B. Lovell acknowledges the Smithsonian Institute for funding via a Submillimeter Array (SMA) Fellowship.

J.M. Alcal\'{a} acknowledges financial support from PRIN-MUR 2022 20228JPA3A “The path to star and planet formation in the JWST era (PATH)” funded by NextGeneration EU and by INAF-GoG 2022 “NIR-dark Accretion Outbursts in Massive Young stellar objects (NAOMY)”. 

J.M. Alcal\'{a} and A. Frasca acknowledge financial support from Large Grant INAF 2022 “YSOs Outflows, Disks and Accretion: towards a global framework for the evolution of planet forming systems (YODA)”.

This work benefited from discussions with the ODYSSEUS team (HST AR-16129), \url{https://sites.bu.edu/odysseus/}.
This work has made use of data from the European Space Agency (ESA) mission
{\it Gaia} (\url{https://www.cosmos.esa.int/gaia}), processed by the {\it Gaia}
Data Processing and Analysis Consortium (DPAC,
\url{https://www.cosmos.esa.int/web/gaia/dpac/consortium}). Funding for the DPAC
has been provided by national institutions, in particular the institutions
participating in the {\it Gaia} Multilateral Agreement.
This research has made use of the SIMBAD database, operated at CDS, Strasbourg, France.

\end{acknowledgements}

\bibliography{main.bib}

\appendix
\section{Observations and data reduction}\label{obsLog}
\begin{table*}[t]
\caption{Night log of the observations}
\label{tab:nightlog}
\begin{tabular}{l|c|ccc|cc|c}
\toprule
                           &  &\multicolumn{3}{c|}{Slit width [$" \times 11 "$]} & \multicolumn{2}{c|}{Exp. Time[$N_{\rm exp} \times {\rm s}$]} &  Exp. time [$N_{\rm exp} \times NDIT \times {\rm s}$]        \\
                    Name & Date of observation [UT] &UVB & VIS & NIR &          UVBexpT &          VISexpT &                      NIRexpT \\
\midrule
             CD-31\_12522 & 2022-06-03T00:41:42.9278 &     1.0 &     0.4 &     0.4 & 2 $\times$ 250 & 2 $\times$ 160 &  2 $\times$ 2 $\times$ 150 \\
          RXJ1608.9-3905 & 2022-05-03T08:35:32.8475 &     1.0 &     0.4 &     0.4 & 2 $\times$ 180 &  2 $\times$ 90 &  2 $\times$ 2 $\times$ 100 \\
                   MV Lup & 2010-05-05T08:30:21.1291 &     0.5 &     0.4 &     0.4 & 4 $\times$ 150 &  8 $\times$ 60 & 12 $\times$ 1 $\times$ 100 \\
RXJ0438.6+1546 & 2021-12-08T02:10:32.3906 &     0.5 &     0.4 &     0.4 & 2 $\times$ 110 &  2 $\times$ 20 &   2 $\times$ 4 $\times$ 30 \\
  J15552621-3338232  & 2022-05-03T06:04:01.0328 &     1.0 &     0.4 &     0.4 & 4 $\times$ 180 &  4 $\times$ 90 &  4 $\times$ 2 $\times$ 100 \\
                   MT Lup & 2014-04-28T05:33:01.1666 &     0.5 &     0.4 &     0.4 & 4 $\times$ 179 &  4 $\times$ 80 &  4 $\times$ 20 $\times$ 10 \\
                 MX Lup & 2022-06-03T02:34:14.6760 &     1.0 &     0.4 &     0.4 & 4 $\times$ 180 &  4 $\times$ 90 &  4 $\times$ 2 $\times$ 100 \\
         RXJ1607.2-3839 & 2022-06-24T04:15:05.3699 &     1.0 &     0.4 &     0.4 & 4 $\times$ 180 &  4 $\times$ 90 &  4 $\times$ 2 $\times$ 100 \\
                   MW Lup & 2022-06-03T01:52:44.9932 &     1.0 &     0.4 &     0.4 & 4 $\times$ 200 & 4 $\times$ 110 &  4 $\times$ 2 $\times$ 100 \\
                   NO Lup & 2014-04-28T08:07:12.0366 &     0.5 &     0.4 &     0.4 & 4 $\times$ 219 & 4 $\times$ 120 &  4 $\times$ 20 $\times$ 12 \\
                THA15-43  & 2022-07-19T03:13:14.3819 &     1.0 &     0.4 &     0.4 & 4 $\times$ 200 & 4 $\times$ 110  &  4 $\times$ 2 $\times$ 100 \\
    THA15-36A & 2022-06-24T05:03:13.4256 &     1.0 &     0.4 &     0.4 & 4 $\times$ 210 & 4 $\times$ 120 &   4 $\times$ 3 $\times$ 80 \\      
    THA15-36B & 2022-06-24T05:03:13.4256 &     1.0 &     0.4 &     0.4 & 4 $\times$ 210 & 4 $\times$ 120 &   4 $\times$ 3 $\times$ 80 \\
                  RECX-6 & 2022-03-02T03:22:05.3962 &     1.0 &     0.4 &     0.4 & 2 $\times$ 440 & 2 $\times$ 350 &  2 $\times$ 5 $\times$ 100 \\
                    Sz67  & 2022-06-16T03:55:17.2951 &     1.0 &     0.4 &     0.4 & 2 $\times$ 280 & 2 $\times$ 190 &  2 $\times$ 3 $\times$ 100 \\      
  J16090850-3903430 & 2022-06-24T05:34:42.3279 &     1.0 &     0.9 &     0.9 & 4 $\times$ 600 & 4 $\times$ 510 &  4 $\times$ 3 $\times$ 200 \\
  16075888-3924347  & 2022-07-13T23:46:15.8273 &     1.0 &     0.9 &     0.9 & 4 $\times$ 600 & 4 $\times$ 510 &  4 $\times$ 3 $\times$ 200 \\
  J16091713-3927096 & 2022-06-03T04:01:34.0592 &     1.0 &     0.9 &     0.9 & 4 $\times$ 700 & 4 $\times$ 610 &  4 $\times$ 3 $\times$ 250 \\
                V1191Sco & 2022-06-02T07:50:32.4620 &     1.0 &     0.9 &     0.9 & 4 $\times$ 550 & 4 $\times$ 460 &  4 $\times$ 3 $\times$ 200 \\
\midrule
              \bf{Binaries} &     &      &      &  &  &  \\   
\midrule
             CD-39\_10292 & 2022-07-25T01:02:17.1203 &     1.0 &     0.4 &     0.4 & 2 $\times$ 250 & 2 $\times$ 160 &  2 $\times$ 3 $\times$ 100 \\
                THA15-38 & 2022-07-22T02:23:20.5225 &     1.0 &     0.4 &     0.4 & 4 $\times$ 210 & 4 $\times$ 120 &   4 $\times$ 3 $\times$ 80 \\
             CD-35\_10498 & 2023-03-06T07:52:55.5080 &     1.0 &     0.4 &     0.4 & 2 $\times$ 200 & 2 $\times$ 110 &  2 $\times$ 2 $\times$ 100 \\
                V1097Sco & 2022-06-05T00:47:31.7504 &     1.0 &     0.4 &     0.4 & 4 $\times$ 210 & 4 $\times$ 120 &   4 $\times$ 3 $\times$ 80 \\
                   NN Lup  & 2022-06-16T04:16:09.4443 &     1.0 &     0.4 &     0.4 & 4 $\times$ 200 & 4 $\times$ 110 &  4 $\times$ 2 $\times$ 100 \\
\bottomrule
\end{tabular}
\end{table*}

\begin{table*}[]
    \centering
        \caption{Photometry available for our targets.}
        \label{tab:phot}
    \begin{tabular}{lccccc}
    \hline
         Band    & MVlup            & NOlup        & MTlup        & THA15-36     & 2MASSJ16090850 \\
         \hline
            U       & 11.98 $^{(2)}$     & 15.13 $^{(2)}$ & ...          & ...          & ...            \\
            B       & 12.59 $^{(3)}$     & 14.65$^{(2)}$  & ...          & 15.30 $^{(6)}$ & 18.00 $^{(8)}$   \\
            V       & 11.28$^{(3)}$      & 12.86 $^{(5)}$ & 12.32 $^{(6)}$ & 13.89 $^{(6)}$ & 16.60 $^{(8)}$   \\
            R       & 10.02 $^{(2)}$     & 12.14 $^{(2)}$ & 12.10 $^{(6)}$ & 13.57 $^{(6)}$ & 16.43 $^{(8)}$   \\
            G       & 11.04 $^{(1)}$     & 12.21 $^{(1)}$ & 11.99$^{(1)}$  & ...          & 15.66 $^{(1)}$   \\
            I       & 9.5210.02 $^{(4)}$ & 11.15 $^{(5)}$ & 11.01 $^{(5)}$ & 11.68 $^{(7)}$ & 14.16 $^{(8)}$   \\
            J       & 9.56 $^{(4)}$      & 9.90 $^{(4)}$  & 10.11 $^{(4)}$ & 10.47 $^{(4)}$ & 12.20 $^{(4)}$   \\
            H       & 9.02 $^{(4)}$      & 9.29 $^{(4)}$  & 9.57 $^{(4)}$  & 9.77$^{(4)}$   & 11.64 $^{(4)}$   \\
            Ks      & 8.88 $^{(4)}$      & 9.12 $^{(4)}$  & 9.38 $^{(4)}$  & 9.53 $^{(4)}$  & 11.36 $^{(4)}$   \\
            SDSS\_g & ...              & ...          & 12.90 $^{(6)}$ & 14.61 $^{(6)}$ & ...            \\
            SDSS\_r & ...              & ...          & 11.86 $^{(6)}$ & 13.29 $^{(6)}$ & ...            \\
            SDSS\_i & ...              & ...          & 11.42 $^{(6)}$ & 12.26 $^{(6)}$ & ...  \\
        \hline
    \end{tabular}
    \tablefoot{(1): \citet{gaiaEDR3}, (2): \citet{Makarov2007}, (3): \citet{Hog2000A}, (4): \citet{Cutri2003} (5): \citet{Kiraga2012} (6): \citet{Zacharias2012} (7): \citet{Cieza2007} (8): \citet{merin2008}}
    
\end{table*}

\begin{table*}
\caption{Signal to noise ratios at different wavelengths.}
    \begin{tabular}{lrrrrrrrrr}
\toprule
                    name &  416 nm &448 nm &  490 nm &701 nm & 750 nm & 801nm &  1217.5 nm &  1600 nm & 2214 nm \\
\midrule
          RXJ0445.8+1556 &    15.93 &    12.99 &    14.77 &    56.26 &    53.54 &   115.46 &           48.87 &     14.27 &    273.74 \\
          RXJ1508.6-4423 &    16.24 &    14.61 &    16.63 &    69.86 &    50.61 &   136.19 &           50.14 &     30.65 &     88.64 \\
          RXJ1526.0-4501 &     6.60 &     8.07 &     7.86 &    24.60 &    21.29 &   116.23 &          106.00 &     16.40 &    131.51 \\
                  HBC407 &     6.06 &     7.47 &     7.57 &    19.43 &    16.57 &   104.87 &           84.71 &     14.47 &    160.83 \\
    PZ99J160843.4-260216 &     7.63 &     8.95 &     8.65 &    28.02 &    24.43 &   135.80 &          109.30 &     19.15 &    150.05 \\
            CD-31\_12522 &     7.67 &     9.09 &     8.26 &    24.71 &    21.82 &    74.21 &          123.74 &     16.71 &    367.12 \\
          RXJ1515.8-3331 &     6.24 &     7.26 &     7.15 &    22.46 &    18.48 &   109.48 &           72.32 &     20.20 &    239.62 \\
    PZ99J160550.5-253313 &     5.73 &     6.83 &     6.73 &    19.68 &    16.02 &    91.52 &           94.63 &     11.15 &    124.80 \\
          RXJ0457.5+2014 &     6.75 &     7.53 &     7.41 &    25.51 &    22.08 &   105.95 &          101.53 &     13.02 &    172.28 \\
RXJ0438.6+1546 &     6.15 &     7.26 &     7.07 &    20.07 &    19.44 &    61.56 &           95.07 &     16.04 &    252.33 \\ 
          RXJ1608.9-3905 &     7.97 &     8.52 &     7.83 &    25.72 &    24.26 &    96.94 &          147.80 &     15.28 &    409.61 \\
                   MV Lup &     5.28 &     6.05 &     6.33 &    17.62 &    15.12 &    72.69 &          113.82 &     13.89 &    237.25 \\
          RXJ1547.7-4018 &     5.50 &     6.03 &     6.20 &    18.46 &    15.93 &    84.74 &           96.32 &     11.79 &    189.26 \\
          RXJ1538.6-3916 &     5.45 &     5.69 &     6.04 &    18.39 &    14.74 &    76.15 &          103.16 &     15.17 &    209.29 \\
          MTlup &     5.41 &     5.55 &     6.06 &    19.81 &    16.27 &    86.89 &          101.78 &     12.81 &    150.80 \\
  2MASSJ15552621-3338232 &     6.73 &     6.37 &     6.72 &    18.14 &    15.27 &    71.02 &          141.01 &     15.20 &    229.93 \\
          RXJ1540.7-3756 &     5.65 &     5.54 &     6.24 &    20.54 &    16.85 &    87.05 &          200.39 &     16.13 &    271.96 \\
          RXJ1543.1-3920 &     5.61 &     5.57 &     6.17 &    19.65 &    16.16 &    85.94 &          166.84 &     17.40 &    204.82 \\
                   MX Lup &     6.41 &     6.38 &     6.69 &    18.25 &    15.91 &    71.51 &          152.44 &     14.45 &    287.35 \\
                   SO879 &     6.77 &     6.87 &     7.57 &    23.79 &    19.75 &    33.73 &           67.35 &     18.16 &    115.75 \\
                    TWA6 &     8.59 &     7.95 &     8.20 &    40.78 &    30.25 &    56.04 &           80.51 &     18.99 &    161.06 \\
            CD -36 7429A &     5.77 &     5.56 &     6.36 &    18.44 &    15.67 &    68.95 &           49.40 &     13.86 &     94.58 \\
          RXJ1607.2-3839 &     7.56 &     7.23 &     7.70 &    24.94 &    22.70 &    62.65 &          158.47 &     19.88 &    230.31 \\
                   MW Lup &     7.15 &     7.02 &     7.75 &    18.97 &    18.36 &    69.86 &          198.35 &     17.78 &    228.41 \\
                    NOlup &     6.21 &     5.67 &     6.77 &    18.57 &    18.27 &    66.65 &          118.13 &     16.40 &    156.53 \\
            Tyc7760283\_1 &     6.97 &     6.55 &     7.76 &    16.85 &    17.27 &    61.08 &           98.96 &     16.05 &    100.23 \\
                   TWA14 &     8.51 &     7.49 &     8.99 &    28.03 &    26.01 &    60.53 &           79.91 &     22.04 &    143.18 \\
    THA15-36A &     8.39 &     7.51 &     8.59 &    15.02 &    19.72 &    52.67 &          102.13 &     24.23 &    167.92 \\
    RXJ1121.3-3447\_app2 &     7.39 &     6.92 &     8.24 &    15.38 &    18.34 &    59.87 &           61.83 &     19.41 &     62.27 \\
     RXJ1121.3-3447\_app1 &     7.57 &     6.88 &     8.13 &    14.52 &    18.62 &    50.31 &           88.64 &     14.27 &     52.72 \\
    THA15-36B &     9.08 &     8.21 &     9.49 &    14.36 &    20.78 &    42.33 &           79.87 &     34.78 &    108.86 \\
             CD -29 8887A &     9.39 &     8.53 &    10.00 &    15.82 &    20.39 &    57.08 &           63.04 &     20.19 &    122.24 \\
              CD -36 7429B &    14.08 &    10.62 &    10.78 &     9.69 &    19.74 &    34.59 &           59.83 &     31.37 &     91.81 \\
              TWA15\_app2 &     9.73 &     7.95 &     9.99 &    12.17 &    20.60 &    42.61 &           80.82 &     23.36 &     82.30 \\
                    TWA7 &    11.32 &     9.45 &    10.25 &     9.82 &    20.29 &    33.24 &           85.80 &     22.48 &    103.95 \\
                    Sz67 &    15.69 &    13.26 &    14.02 &    26.27 &    34.31 &    56.15 &          154.49 &     49.16 &    185.14 \\
                    RECX-6 &    14.06 &    11.61 &    11.95 &    12.88 &    24.79 &    39.64 &          132.75 &     34.53 &    161.71 \\
              TWA15\_app1 &    10.47 &     7.19 &    10.73 &    16.40 &    23.91 &    39.71 &           66.87 &     24.27 &    106.87 \\
                    Sz94 &    14.37 &    12.51 &    12.55 &    16.18 &    27.21 &    53.03 &           82.15 &     38.01 &    101.69 \\
                   SO797 &     7.47 &     8.30 &    11.13 &    17.07 &    29.16 &    33.55 &           40.57 &     15.84 &     45.29 \\
                    Par\_Lup3\_2 &    12.00 &    12.77 &     9.96 &    13.89 &    23.39 &    26.01 &           85.31 &     35.86 &     58.28 \\
  2MASSJ16090850-3903430 &     2.75 &     4.80 &     4.02 &     9.46 &    21.75 &    23.83 &           87.33 &     60.66 &     75.65 \\
  SO925 &     2.48 &     4.84 &     6.99 &    11.99 &    21.53 &    23.61 &           55.29 &     14.36 &      7.40 \\
                   SO999 &     4.13 &     6.82 &     8.69 &    12.86 &    22.97 &    24.16 &           43.29 &     36.90 &     10.63 \\
                   SO641 &     7.66 &     7.24 &     9.12 &    11.82 &    22.64 &    26.69 &           52.31 &     41.93 &     43.13 \\
                V1191Sco &     6.73 &     8.55 &     8.03 &    12.52 &    22.06 &    20.13 &          102.46 &     49.53 &     86.91 \\
  2MASSJ16091713-3927096 &     0.73 &     1.87 &     4.68 &    13.13 &    21.43 &    23.64 &           68.44 &     54.13 &     49.19 \\
                   Sz107 &     8.35 &    10.41 &    10.47 &    21.05 &    30.13 &    34.14 &           87.44 &     39.34 &     71.01 \\
              Par\_Lup3\_1 &     0.41 &     0.97 &     1.59 &     4.83 &    14.26 &    12.26 &           54.47 &     31.02 &     56.42 \\
                   LM717 &     0.55 &     1.06 &     2.94 &     7.90 &    16.36 &    11.87 &           63.38 &     38.59 &     58.48 \\
       J11195652-7504529 &     0.17 &     0.13 &     0.45 &     4.13 &    13.42 &     6.24 &           52.85 &     39.90 &     38.90 \\
                   LM601 &     0.00 &     0.19 &     0.54 &     5.10 &    11.83 &     5.44 &           48.58 &     42.03 &     46.00 \\
               CHSM17173 &     0.22 &     0.10 &     0.97 &     8.49 &    12.82 &     6.17 &           42.83 &     33.14 &     48.77 \\
                   TWA26 &     0.23 &     0.60 &     2.50 &    10.46 &    11.32 &     9.03 &           41.68 &     41.30 &     57.69 \\
               DENIS1245 &     0.20 &     0.12 &     0.41 &     6.25 &     8.96 &     8.13 &           35.15 &     46.19 &     40.09 \\
\bottomrule
\end{tabular}

\end{table*}

\newpage
\newpage

The sources MV Lup, NO Lup, and MT Lup  were not observed using slits of $5.0"$. The narrow slit spectra were therefore flux calibrated by using the available photometry (see Table \ref{tab:phot}). This was done by converting the photometry in a given band to a corresponding flux density. The flux density in the same bands is also computed for the narrow-slit spectra by integrating them after being multiplied to the filter bandpass. The ratio of these fluxes allows us to rescale the spectra in absolute units correcting for slit losses.  
For MV Lup  and NO Lup  the spectrum in the UVB arm is matched to the photometric flux using a factor constant with wavelength. whereas VIS and NIR arm spectra are rescaled using a factor with a linear dependency on wavelength. This was done because variation of the seeing disk with wavelengths in the wide ranges of VIS and NIR arms and atmospheric refraction can produce a wavelength dependent light loss. For MT Lup a constant factor was used for the entire spectrum to preserve continuity between the UVB and VIS arms.

2MASSJ16090850-3903430 and THA15-36 were both observed consecutively during the night of 23/24-Jun 2022. Both are about $\sim 0.7$ mag less bright than the archival photometry. The spectrum of 2MASSJ16090850-3903430 whas therefore re-scaled with a constant factor to match the photometry (see Table \ref{tab:phot}). 

THA15-36 is a binary system with a separation of $\sim 2 ''$ that was aligned-with and resolved-in the X-Shooter slits. The traces were fitted with two Gaussians at a number of wavelengths. This customized spectra extraction allowed us to estimate the flux ratio between both components as a function of wavelength. 
The latter enabled us to calculate the total flux in the broad-slit spectra for both components. In the narrow slit observations components were extracted from the rectified, flux calibrated 2d spectra using an extraction window with a width of $0.8 ''$ in all arms in order to avoid contamination from its companion.
The extracted narrow slit spectra were then rescaled to their respective flux contributions to correct for wavelength dependent flux loss in the same way as mentioned before. The spectra were then multiplied by the same constant factor such that the sum of both components matches the available photometry (see Table \ref{tab:phot}).

For RXJ1607.2-3839 we noticed noticed that the continuum in the NIR arm bent upwards at shorter wavelengths and was less bright than the archival photometry. The bend in the continuum is likely caused by an issue with the used flux standard. The flux standard taken later that night was therefore used instead. This mitigated the upward bend of the VIS arm. After this correction, the overall flux density in the arm was still lower compared to the VIS arm and the NIR photometry. This was rectified by re-scaling the NIR to the VIS arm such that their overlapping regions have the same median flux density. This also improved the match between the NIR arm and the available 2MASS photometry.

The NIR arm of RECX-6 displayed a similar upward bend as RXJ1607.2-3839. To mitigate this issue the flux standard of the night before was used for the reduction of the NIR arm spectrum of RECX-6.

CD-31\_12522 was observed in poor seeing conditions. This caused the slit losses in the UVB narrow slit observation to become wavelength depended. We, therefore, used a linear relation rather than the mean flux ratio to rescale the UVB narrow slit to the broad slit observation.

\section{2MASSJ16075888-3924347}\label{2mass888}
The spectrum of 2MASSJ16075888-3924347 appears to have spectral features typical of an accreting T Tauri star rather than a non-accreting Class III object. For example, the $\rm H\alpha$ line luminosity appears much larger than in the other Class III templates of similar spectral type (M5.5). We measure an equivalent width of $\rm EW  = -3.914 \pm 0.093 nm$. This would classify the target as a CTTS rather than a WTTS according to the classification scheme of \citet{Barrado2003,white2003}. Similarly, the \ion{Ca}{ii} IR triplet is detected for this target but not in comparable Class III templates. This result is puzzling considering that the SED of this target presented by \cite{Lovell2021}, does not show any excess emission at longer wavelengths and is consistent with that of other Class III objects. Moreover, \citet{Lovell2021} did not detect any flux at $856 \,\rm\mu m$, giving an upper limit of $<0.11 \:\rm mJy$. No previous $\rm H\alpha$ equivalent widths are available in the literature for this target. A detailed analysis of this star is beyond the scope of this work and will be presented by Stelzer et al. in prep..

\section{Uncertainties on extinction.}\label{app:uncertainties}

\begin{figure*}[th!]
     \centering
     \includegraphics[width =0.49\textwidth]{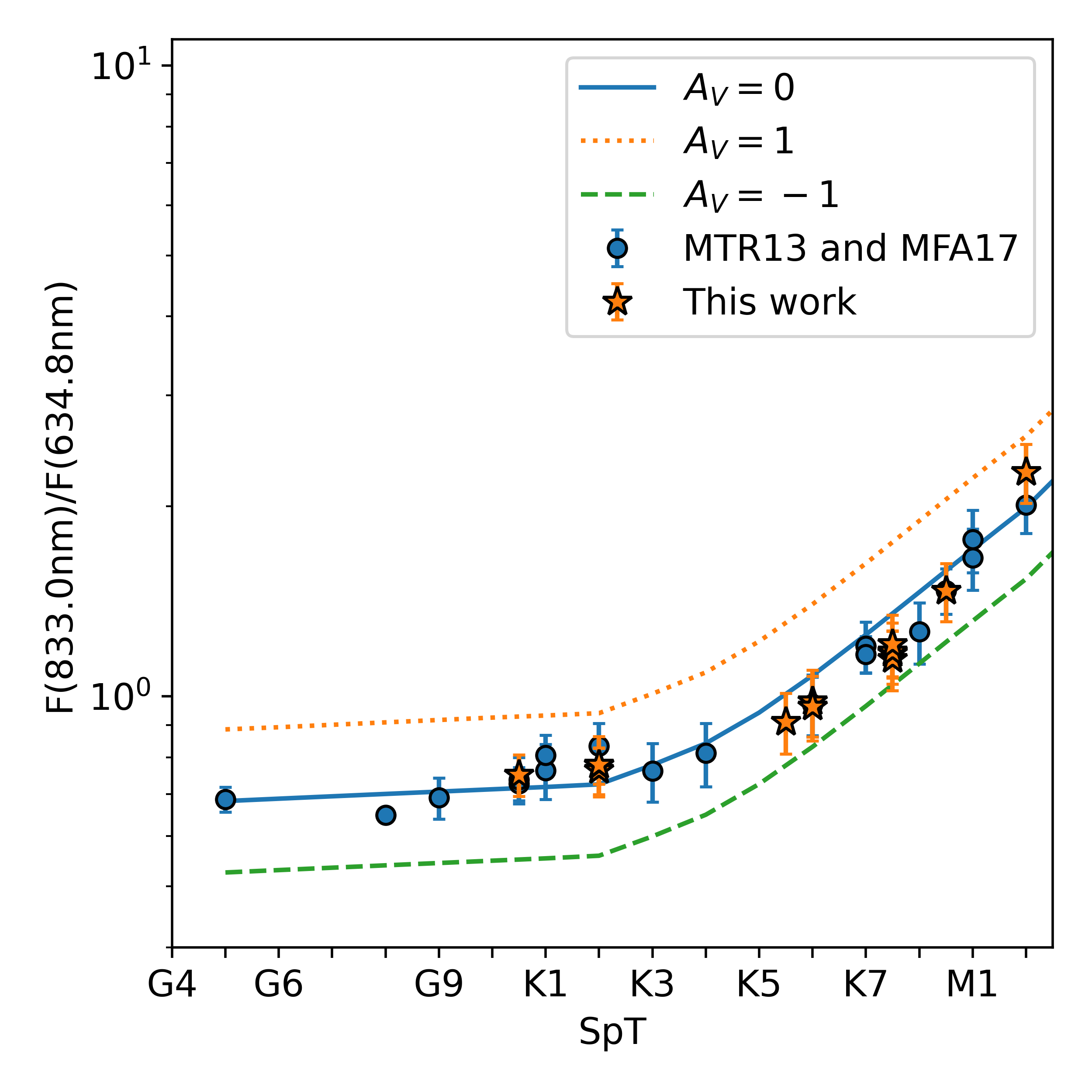}
     \includegraphics[width =0.49\textwidth]{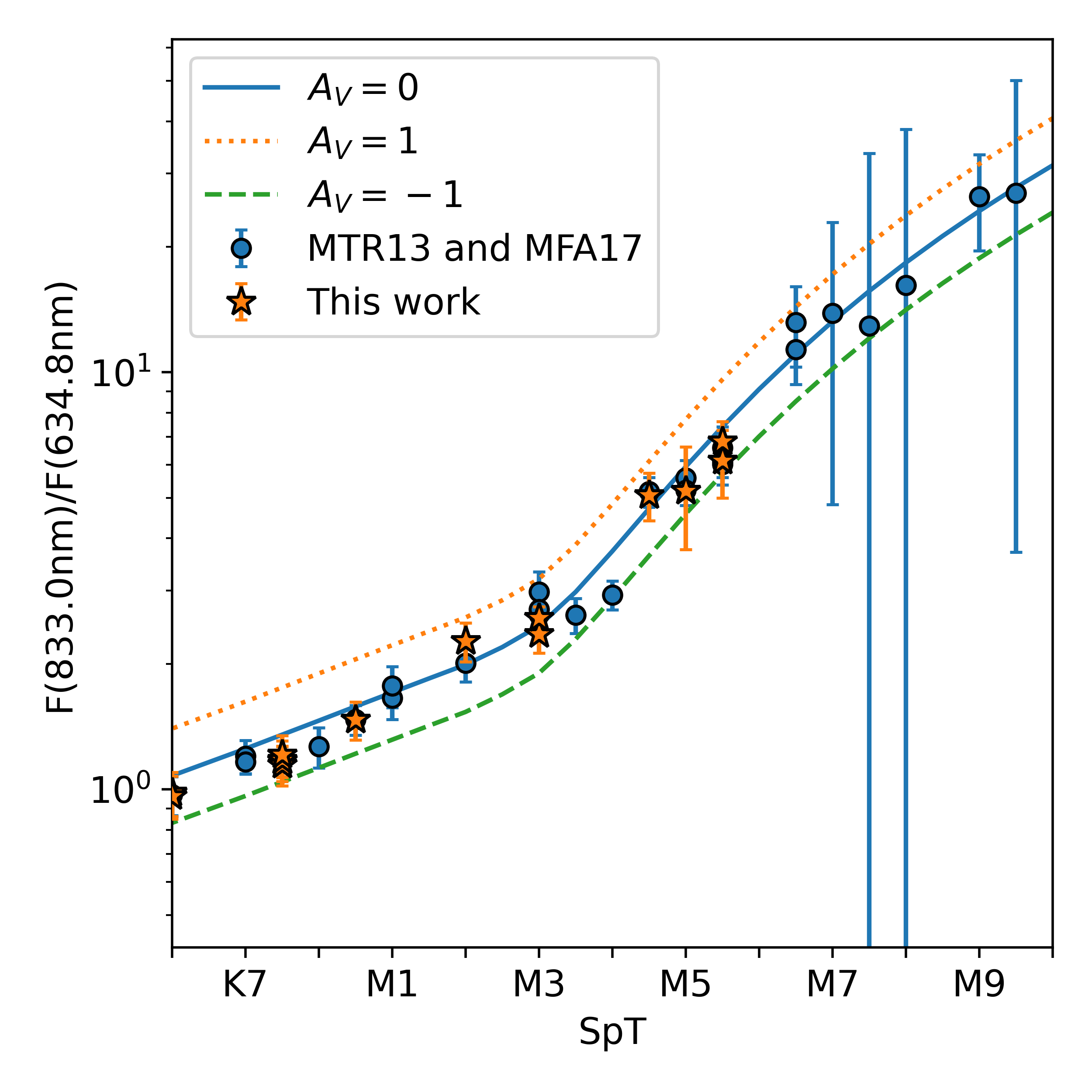}
     \caption{$F_{red} = F(833nm)/F(634.8nm)$ ratio calculated for the Class III templates analysed here (orange stars) as well as for the samples presented by \citetalias{manara13a} and \citetalias{manara17b} (blue circles). The error bars represent the uncertainties due to the noise in the spectra and do not include the uncertainty on the extinction correction.  The blue line shows the value of $F_{red}$ for  $A_V = 0$  as a function of spectral type as derived by \citetalias{manara17b}. The green and orange lines are obtained by deredening and reddening the relationship of \citetalias{manara17b} by 1 mag respectively.}
     \label{fig:AvRat1}
 \end{figure*}

An independent confirmation of the extinction can be found through the ratio $F_{\rm red} = F(833.0nm)/F(634.8nm)$, as first suggested by \citetalias{HH14}. This ratio is sensitive to both extinction and spectral type. Here we adopt the relationship of \citetalias{manara17b}, who re-calibrated the relations of \citetalias{HH14} using observations of Class III YSO's. \citetalias{manara17b} adjusted the relationships of \citetalias{HH14} to account for differences in observed M-type spectra and the corresponding synthetic and dwarf spectra on which \citepalias{HH14} based their relations. Fig. \ref{fig:AvRat1}  shows these ratios for our observations after dereddening. Here it can be seen that our sample has a similar agreement with $A_V = 0$ as the samples of \citetalias{manara13a} and \citetalias{manara17b}.

\section{Luminosity calculation}\label{app:BolCor}

\begin{figure}[t]
    \centering
    \includegraphics[width =0.5\textwidth]{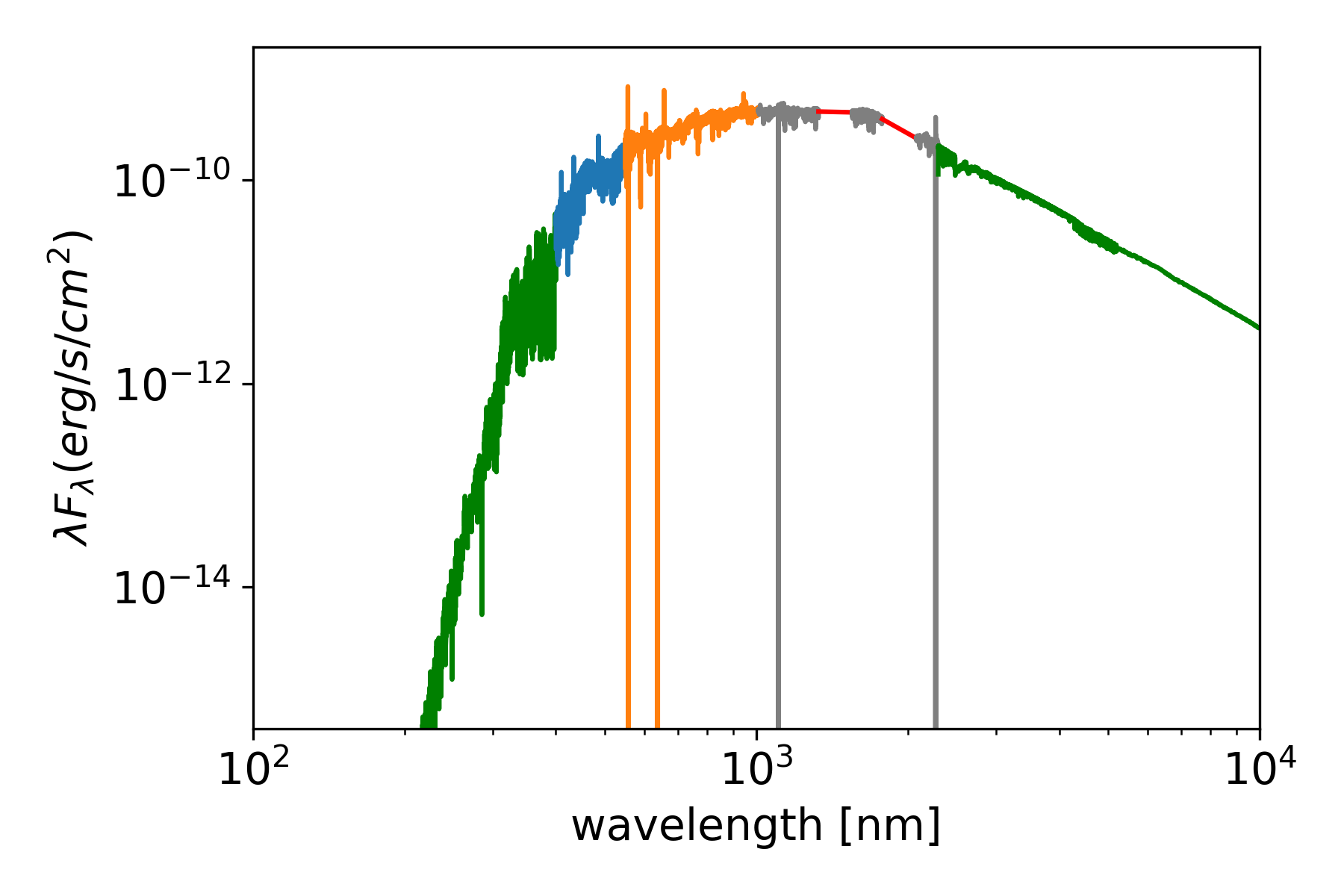}
    \caption{Example of a flux calibrated X-Shooter spectrum extrapolated with a BT-Settl model spectrum. The model segments (dark green) are scaled at the blue and red edges of the X-Shooter spectrum. The UVB, VIS, and NIR arms are plotted in blue orange and gray, respectively. The strong telluric bands in the NIR arm are replaced by a linear interpolation (red lines).  In this figure, we show the example of RXJ1607.2-3839 which has been extended with a BT-Settl model of $T_{\rm eff}= 4000$ K. The BT-Settl model spectra have been convolved with a Gaussian kernel for clarity.}
    \label{fig:LumCalc}
\end{figure}

Fig. \ref{fig:LumCalc} shows an example of a extrapolated X-Shooter spectrum used in our luminosity calculation.

\begin{figure}[th!]
    \centering
    \includegraphics[width =0.5\textwidth]{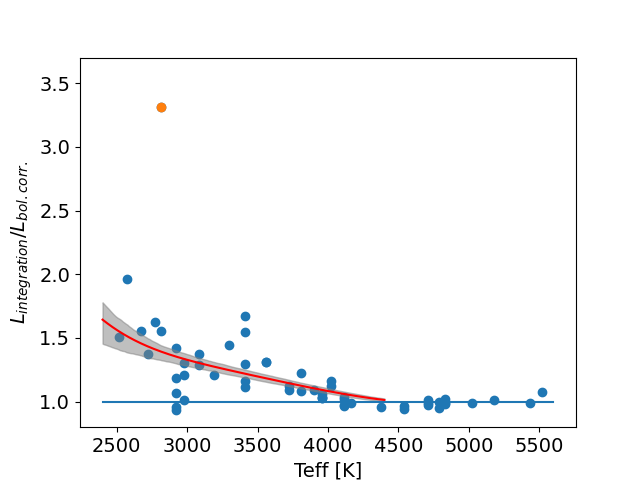}
    \caption{Ratio of the luminosities obtained from our spectral integration method and the value computed using the bolometric correction of \citepalias{HH14}. The blue dots indicate the values obtained for the spectra presented here aswell as in \citepalias{manara13a} and \citepalias{manara17b}. The red line indicates the non-parametric fit, and the shaded area indicates the corresponding 1$\sigma$ uncertainty interval obtained from bootstrapping the non-parametric fitting procedure. The orange point is Par-Lup3-1, which was excluded from this procedure.} 
    \label{fig:LintOverLbol}
\end{figure}

We notice a small but systematic discrepancy between the luminosities obtained by integrating our extrapolated X-Shooter spectra and the values found by using the bolometric correction of \citetalias{HH14}. Fig. \ref{fig:LintOverLbol} shows that the ratio between the two luminosity estimates increases for later spectral types.
The correction of \citetalias{HH14} was based on the same BT-Settl models that we used to extrapolate our X-Shooter spectra. \citetalias{HH14} mention that for targets with $T_{\rm eff} < 3500$ K they interpolate over a VO feature that is stronger in the BT-Settl models compared to their observations. We apply the same interpolation when using the bolometric correction.

\begin{figure}[th!]
    \centering
    \includegraphics[width =0.5\textwidth]{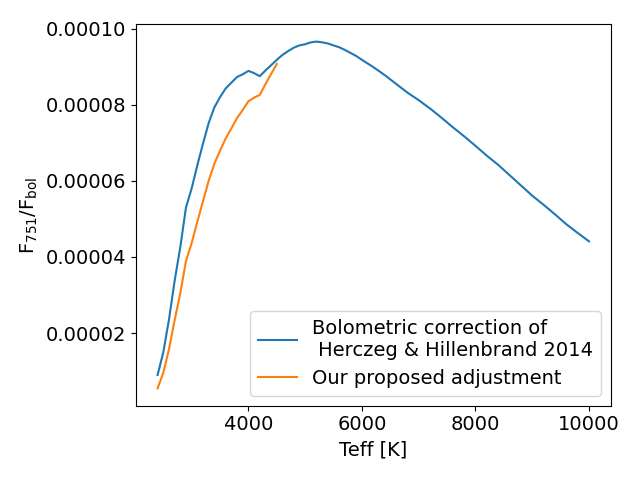}
    \caption{The bolometric correction ($F_{751}/F_{Bol}$) of \citetalias{HH14} and our suggested adjustment for the additional flux present in observations of Class III YSOs.}
    \label{fig:BolCorr}
\end{figure}

The reason for the discrepancy appears to be due to additional flux in the NIR part of the observed spectra compared to the BT-Settl models. A likely cause for this could be the presence of cold spots on the stellar surface, which was discussed in Sect. \ref{spectralTyping}.
We therefore propose a correction to the relationship provided by \citetalias{HH14}, which we expect to account for the typical contribution of stellar spots to the total luminosity. We compute this temperature dependent correction factor by using the non-parametric fit displayed on Fig. \ref{fig:LintOverLbol}. During this fitting procedure, we excluded Par Lup3 1, since it is an extreme outlier, with an extremely bright NIR flux for its SpT, which affects the resulting best fit. The adjusted bolometric correction can be found is illustrated in Fig. \ref{fig:BolCorr} and its values are listed in Table \ref{tab:corrtobollcorr}.

\begin{table}[]

    \centering
    \caption{The bolometric correction of \citetalias{HH14} and our adjustment to it.}
    \label{tab:corrtobollcorr}
    \begin{tabular}{c|cc}
\toprule
$\rm T_{\rm eff}$ & $F_{751}/F_{bol} [\AA ^-1]$ HH14 & $F_{751}/F_{bol} [\AA ^-1]$ Adjusted \\
\midrule
    2400 & $8.98\cdot 10^{-6}$ & $5.60\cdot 10^{-6}$ \\
 2500 & $1.49\cdot 10^{-5}$ & $9.76\cdot 10^{-6} $\\
 2600 & $2.35\cdot 10^{-5}$ & $1.60\cdot 10^{-5}$ \\
 2700 &$ 3.37\cdot 10^{-5}$ & $2.38\cdot 10^{-5}$ \\
 2800 & $4.26\cdot 10^{-5}$ & $3.10\cdot 10^{-5}$ \\
 2900 & $5.30\cdot 10^{-5}$ & $3.95\cdot 10^{-5}$ \\
 3000 & $5.80\cdot 10^{-5}$ & $4.41\cdot 10^{-5}$ \\
 3100 & $6.41\cdot 10^{-5}$ & $4.97\cdot 10^{-5}$ \\
 3200 & $6.98\cdot 10^{-5}$ & $5.51\cdot 10^{-5}$ \\
 3300 & $7.52\cdot 10^{-5}$ & $6.04\cdot 10^{-5}$ \\
 3400 &$ 7.93\cdot 10^{-5}$ & $6.48\cdot 10^{-5}$ \\
 3500 & $8.20\cdot 10^{-5}$ & $6.82\cdot 10^{-5}$ \\
 3600 & $8.43\cdot 10^{-5}$ & $7.14\cdot 10^{-5}$ \\
 3700 & $8.58\cdot 10^{-5}$ & $7.39\cdot 10^{-5}$ \\
 3800 & $8.73\cdot 10^{-5}$ & $7.66\cdot 10^{-5}$ \\
 3900 & $8.80\cdot 10^{-5}$ & $7.86\cdot 10^{-5}$ \\
 4000 & $8.89\cdot 10^{-5}$ & $8.09\cdot 10^{-5} $\\
 4100 & $8.83\cdot 10^{-5}$ & $8.18\cdot 10^{-5}$ \\
 4200 & $8.75\cdot 10^{-5} $& $8.25\cdot 10^{-5}$ \\
 4300 & $8.90\cdot 10^{-5}$ & $8.53\cdot 10^{-5}$ \\
 4400 & $9.04\cdot 10^{-5}$ & $8.80\cdot 10^{-5}$ \\
 4500 & $9.18\cdot 10^{-5}$ &        = \\
 4600 & $9.31\cdot 10^{-5}$ &        = \\
 4700 & $9.41\cdot 10^{-5}$ &        = \\
 4800 & $9.50\cdot 10^{-5}$ &        = \\
 4900 & $9.56\cdot 10^{-5}$ &        = \\
 5000 & $9.59\cdot 10^{-5}$ &        = \\
 5100 & $9.64\cdot 10^{-5}$ &        = \\
 5200 & $9.66\cdot 10^{-5}$ &        = \\
 5300 & $9.64\cdot 10^{-5}$ &        = \\
 5400 & $9.61\cdot 10^{-5}$ &        = \\
 5500 & $9.56\cdot 10^{-5} $&        = \\
 5600 & $9.51\cdot 10^{-5}$ &        = \\
 5700 & $9.44\cdot 10^{-5}$ &        = \\
 5800 & $9.36\cdot 10^{-5}$ &        = \\
 5900 & $9.28\cdot 10^{-5}$ &        = \\
 6000 & $9.18\cdot 10^{-5}$ &        = \\
 6200 & $8.99\cdot 10^{-5}$ &        = \\
 6400 & $8.78\cdot 10^{-5}$ &        = \\
 6600 & $8.55\cdot 10^{-5}$ &        = \\
 6800 & $8.32\cdot 10^{-5}$&        = \\
 7000 & $8.12\cdot 10^{-5}$ &        = \\
 7200 & $7.90\cdot 10^{-5}$ &        = \\
 7400 & $7.66\cdot 10^{-5}$ &        = \\
 7600 & $7.41\cdot 10^{-5}$ &        = \\
 7800 & $7.17\cdot 10^{-5}$ &        = \\
 8000 & $6.92\cdot 10^{-5}$ &        = \\
 8200 & $6.66\cdot 10^{-5}$ &        = \\
 8400 & $6.42\cdot 10^{-5} $&        = \\
 8600 & $6.15\cdot 10^{-5}$ &        = \\
 8800 & $5.88\cdot 10^{-5}$ &        = \\
 9000 & $5.61\cdot 10^{-5}$ &        = \\
 9200 & $5.37\cdot 10^{-5}$ &        = \\
 9400 & $5.12\cdot 10^{-5}$ &        = \\
 9600 & $4.86\cdot 10^{-5}$ &        = \\
 9800 & $4.63\cdot 10^{-5}$ &        = \\
10000 & $4.41\cdot 10^{-5}$ &        = \\
\bottomrule
\end{tabular}
\end{table}

\section{Notable results for the Chameleon I sample}\label{FitChaI}
\paragraph{WZ Cha:} For WZ Cha our method obtains a SpT of M5 whereas the fit to the individual Class III templates is degenerate with local solutions at M3 and M5. The local solution at SpT M3 appeared due to the CD -36 7429B template and since the solution appears to better reproduce the veiling in the Ca II line at $\sim 420 {\rm nm}$. CD 36 7429B does however appear as an outlier among the M3 templates in that its TiO features around $710 {\rm nm}$ appear to be deeper than templates of the same SpT. When removing this template the degeneracy disappears. We therefore prefer the solution presented here.

\paragraph{CV Cha:} The best fit of CV Cha displays a strong degeneracy in terms of SpT, stretching from $\sim$K4 to $\sim$G9. The solution found here results in a SpT of K3. This degeneracy is a likely consequence of the poor constraint on early spectral types in our model spectra. 

\paragraph{VZ Cha:} For VZ Cha we find an extinction that is $\Delta A_V =0.5$ mag lower than obtained using the method of \citet{manara13b}. This differing result appears to be a consequence of the stricter constraint imposed on the Paschen continuum by the best fit metric used in our method, resulting in a better fit of this region. Despite the large difference in obtained $A_V$, the accretion and stellar properties still agree within errors.

\begin{figure}
    \centering
    \includegraphics[width = 0.49\textwidth]{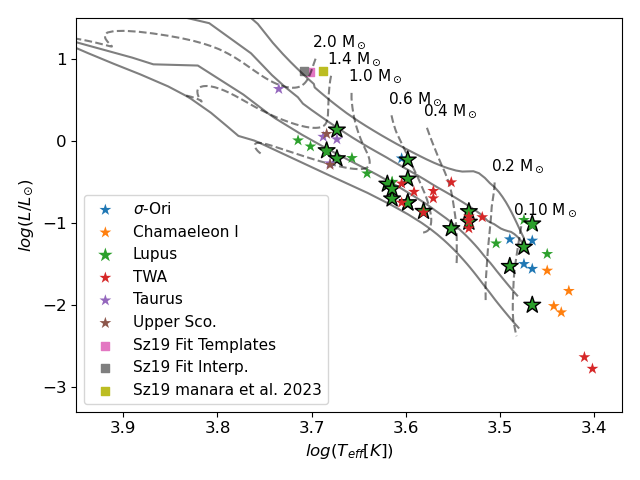}
    \caption{HRD containing the different results obtained for Sz19 for comparison the objects analysed here (highlighted with larger black outlined markers) and those analysed by MTR13 and MFA17 are displayed. The model isochrones and evolutionary tracks by \citet{Feiden2016} are also shown. The isochrones are the 1.2,3,10 and 30 Myrs ones}
    \label{fig:HRD_Cl3}
\end{figure}

\paragraph{Sz 19:} The luminosity of Sz19 appears large for its spectral type, resulting in a high mass estimate. Fig. \ref{fig:HRD_Cl3} illustrates this by showing Sz19 and our Class III templates on an HRD.
The spectra of this target agree with the available photometry within $<2 \,$mag. The results presented here agree with those found by \citet{manara14} (although the slightly later SpT results in a very different stellar mass). An analysis of the emission lines results in accretion luminosities that are in agreement with those found from our analysis of the UV excess. A detailed investigation into why this source appears as an outlier is beyond the scope of this paper.

\newpage
\newpage
\section{Plots of the best fitting models}\label{FitChaIBalmerPaschen}
\begin{figure*}
    \centering
    \includegraphics[width =0.47\textwidth]{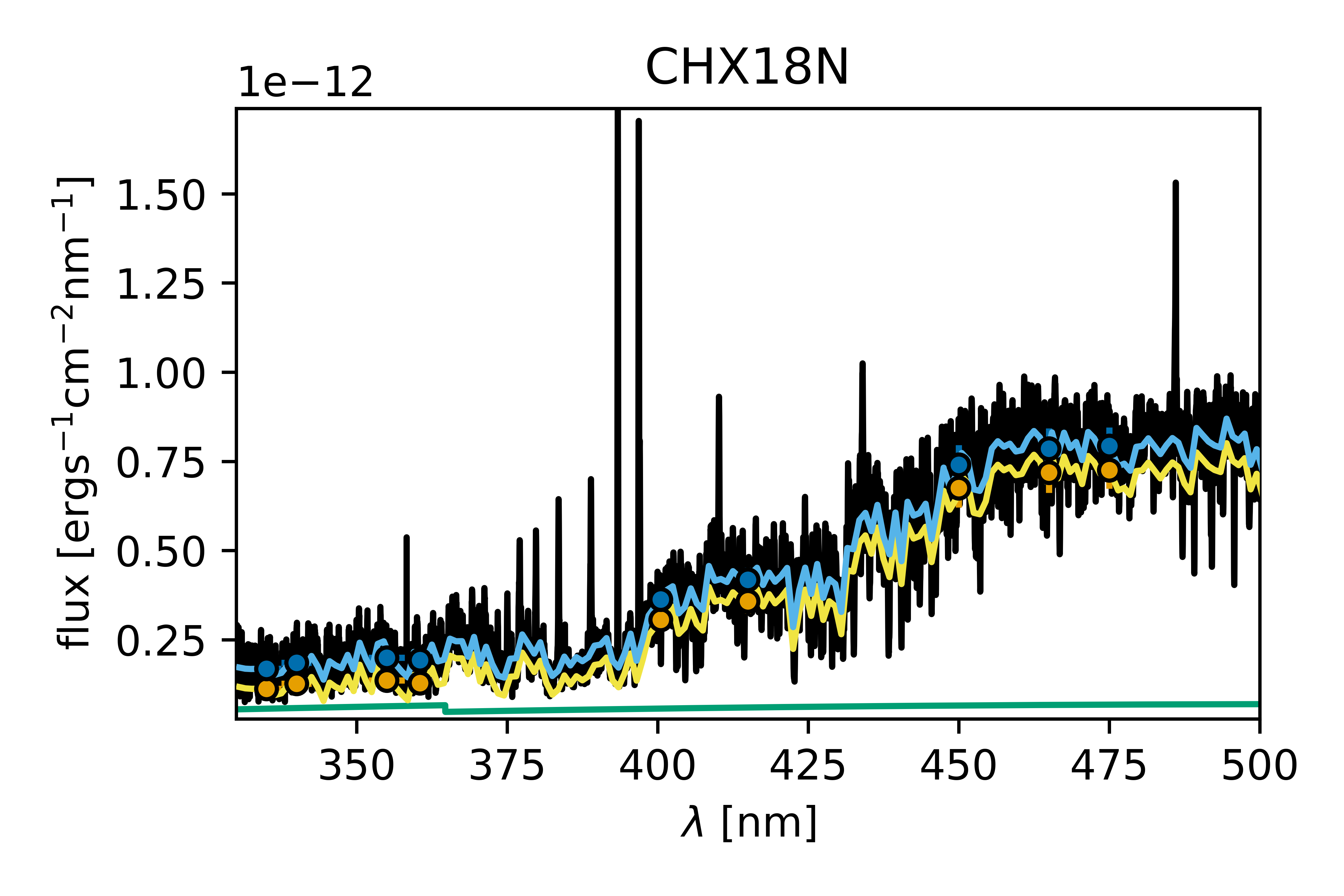}
    \includegraphics[width =0.47\textwidth]{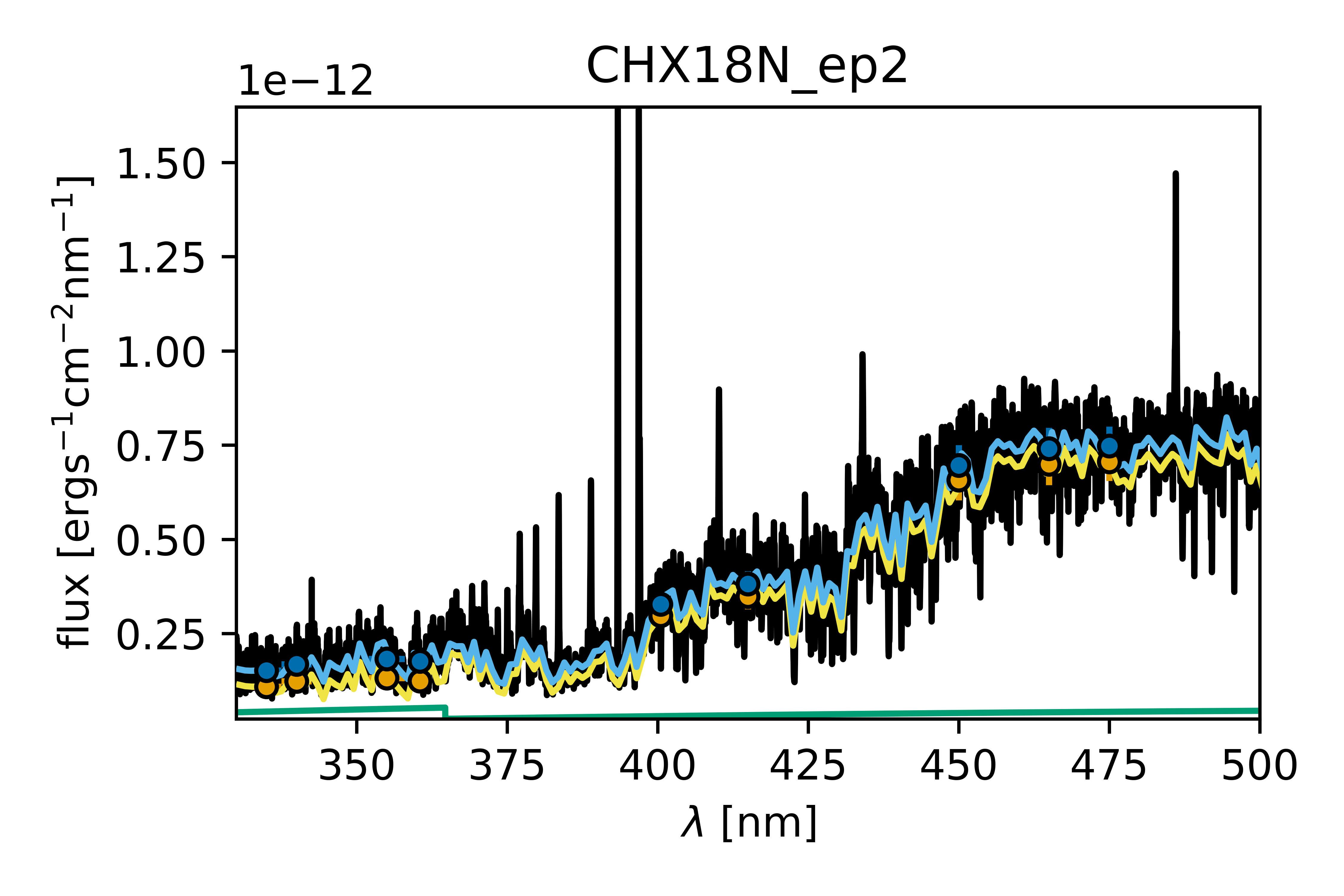}
    \includegraphics[width =0.47\textwidth]{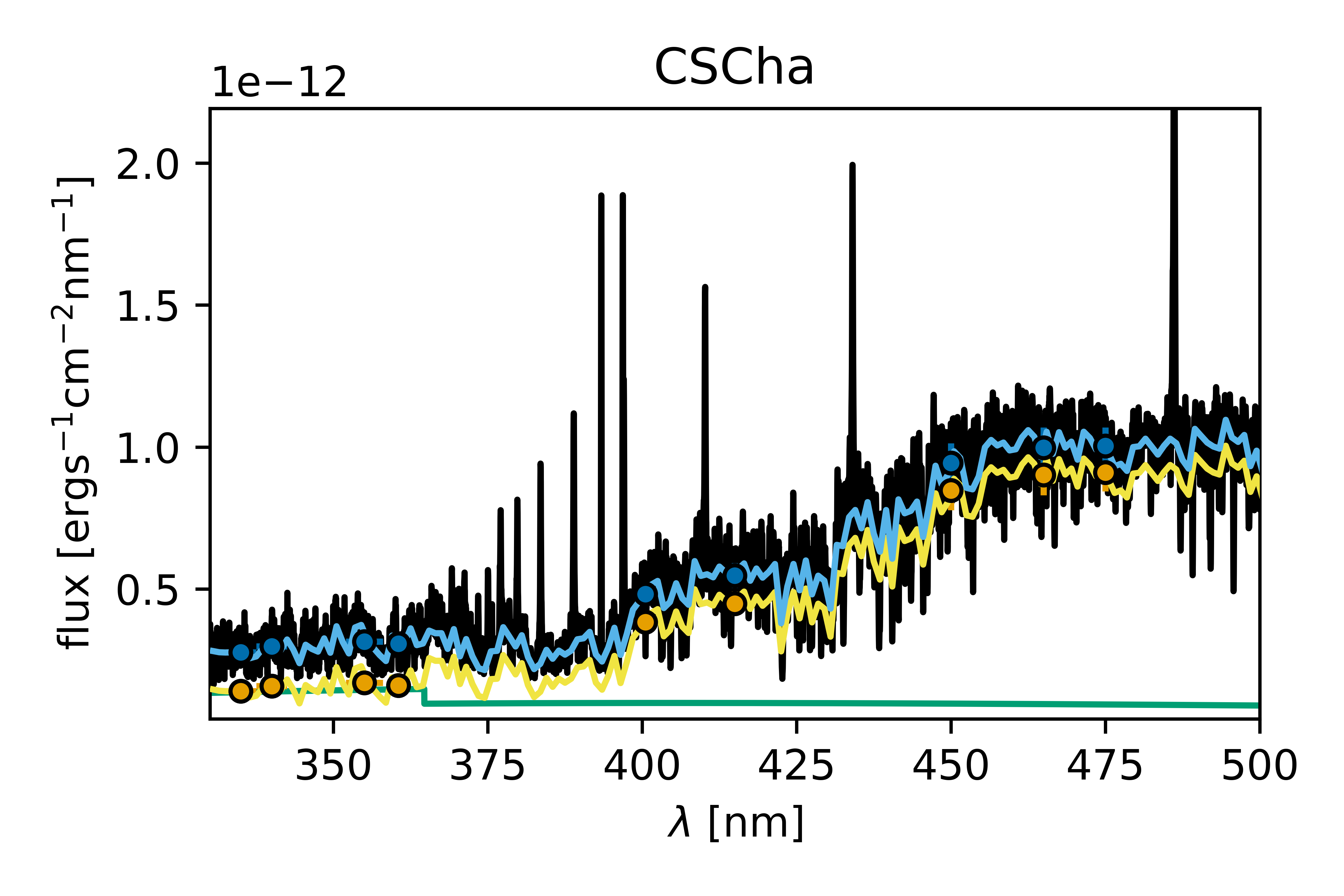}
    \includegraphics[width =0.47\textwidth]{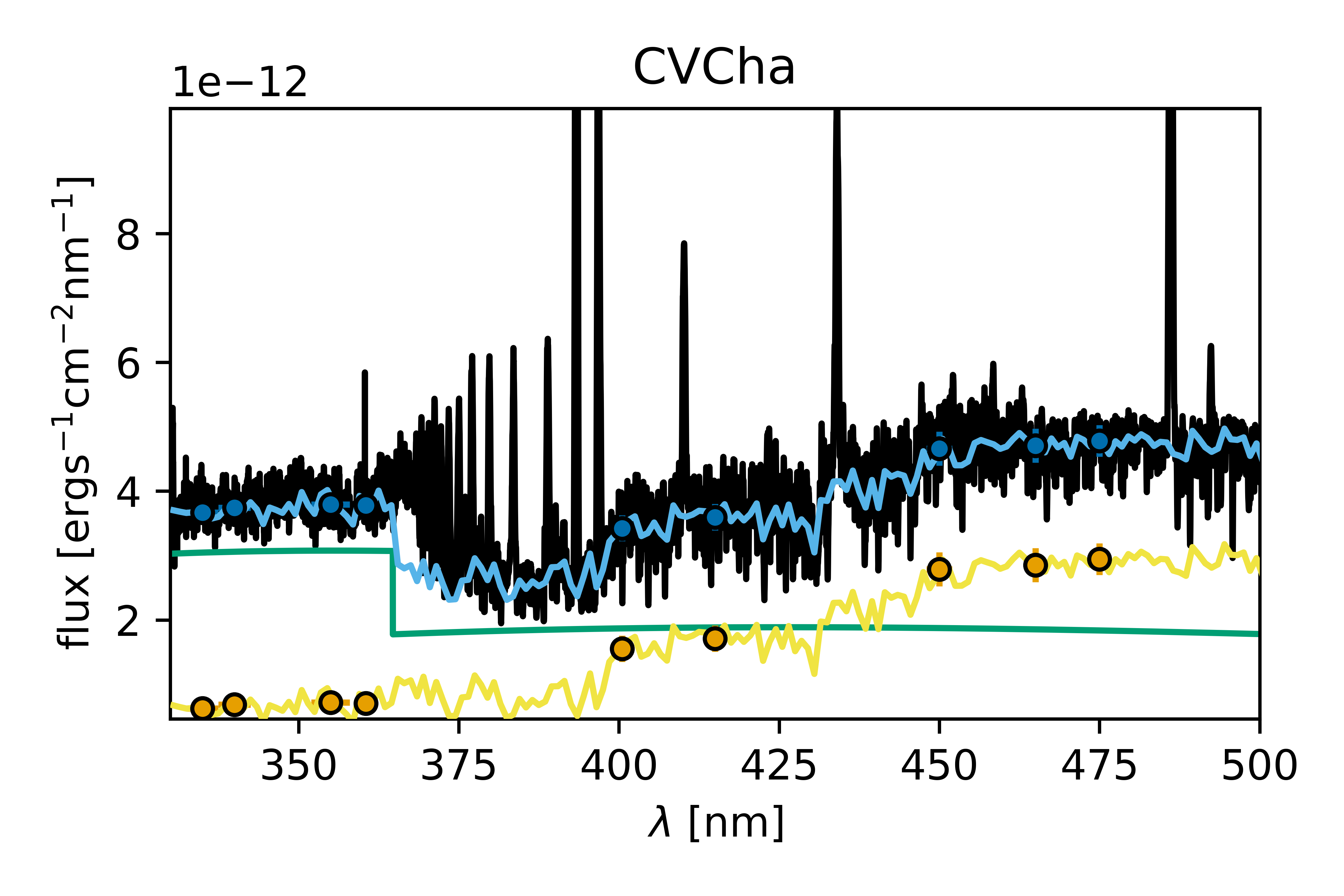}
    \includegraphics[width =0.47\textwidth]{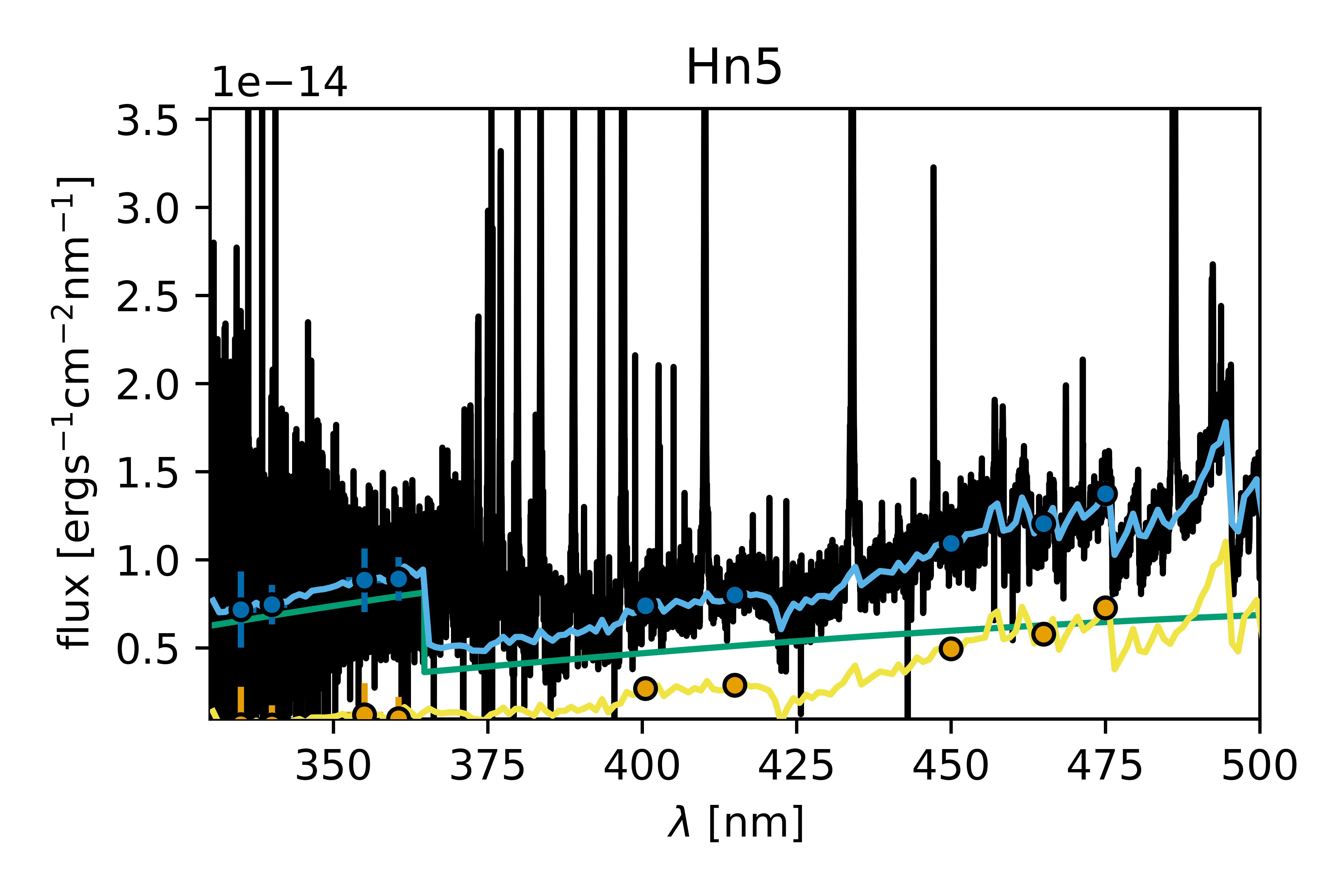}
     \includegraphics[width =0.47\textwidth]{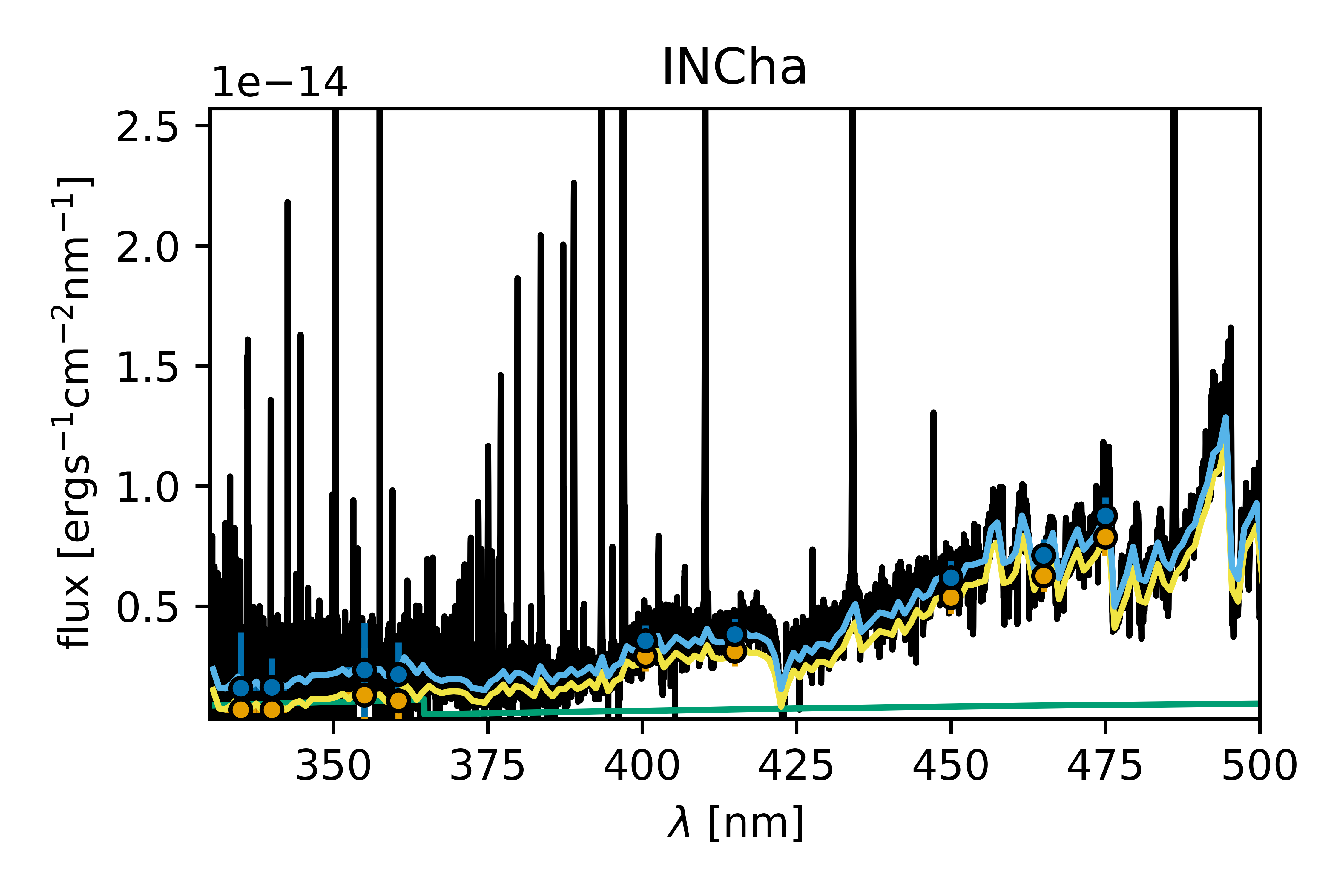}
     \includegraphics[width =0.47\textwidth]{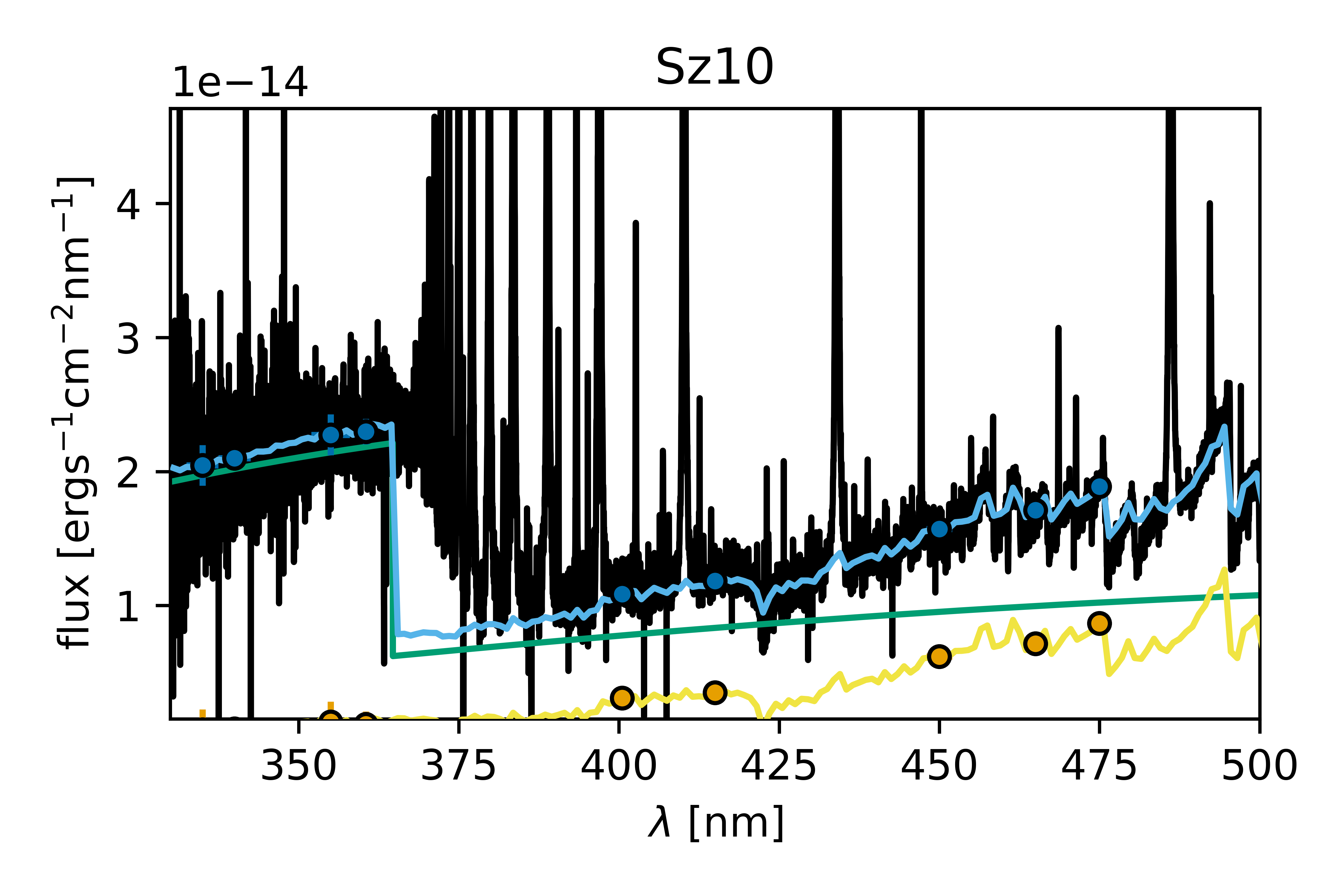}
     \includegraphics[width =0.47\textwidth]{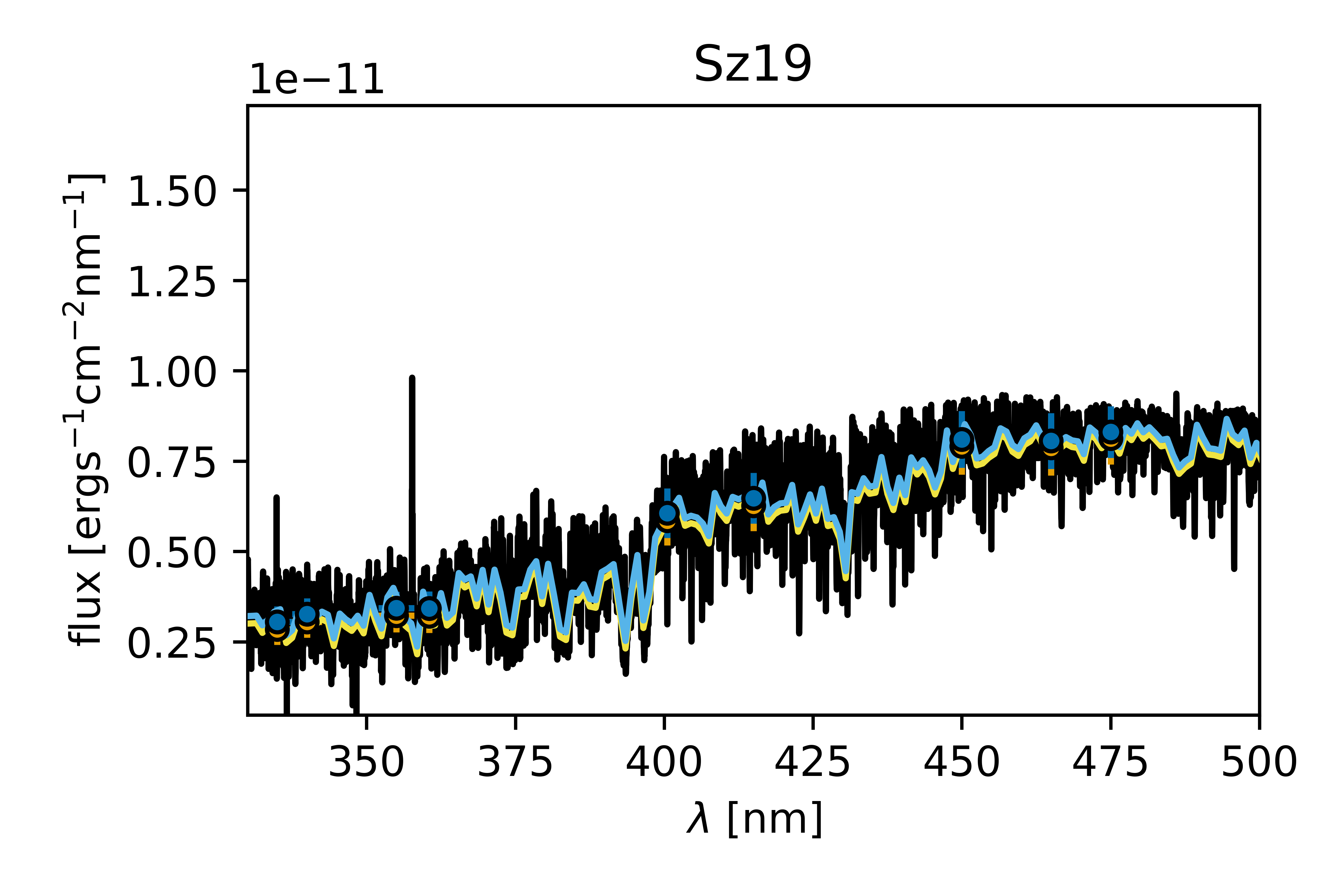}
    \caption{Best fits obtained with FRAPPE for the Balmer and Paschen continuum region for the targets in the Chamaeleon I association. The black line is the dereddened observed spectrum. The green line is the accretion slab model. The yellow line is the interpolated spectrum at the best fitting SpT. The blue line is the best fit. The points of the same but darker color highlight the wavelengths used in the best fit determination.}
    \label{fig:ChaIBestFitsContFitter}
\end{figure*}

\begin{figure*}
    \centering
    \includegraphics[width =0.49\textwidth]{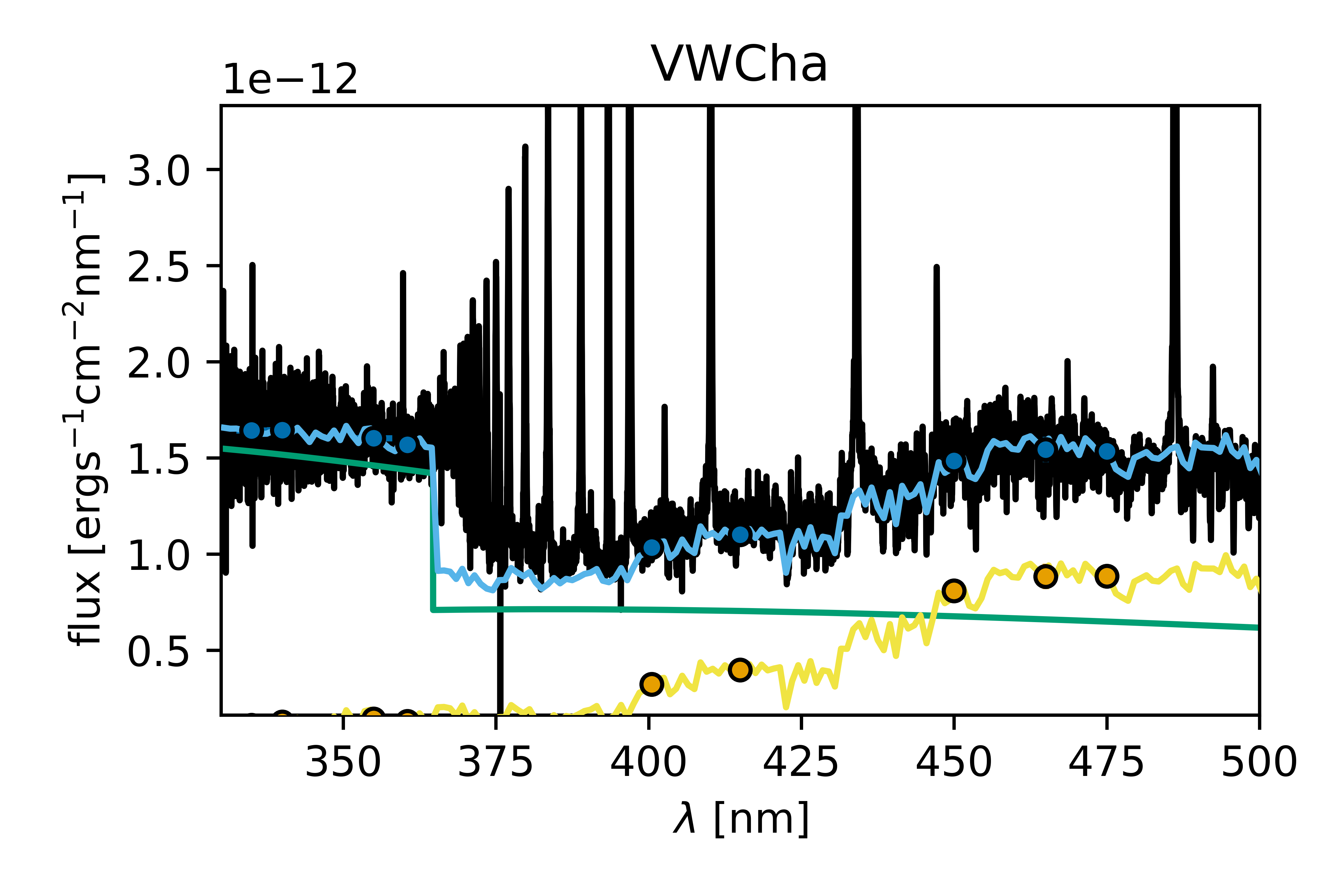}
    \includegraphics[width =0.49\textwidth]{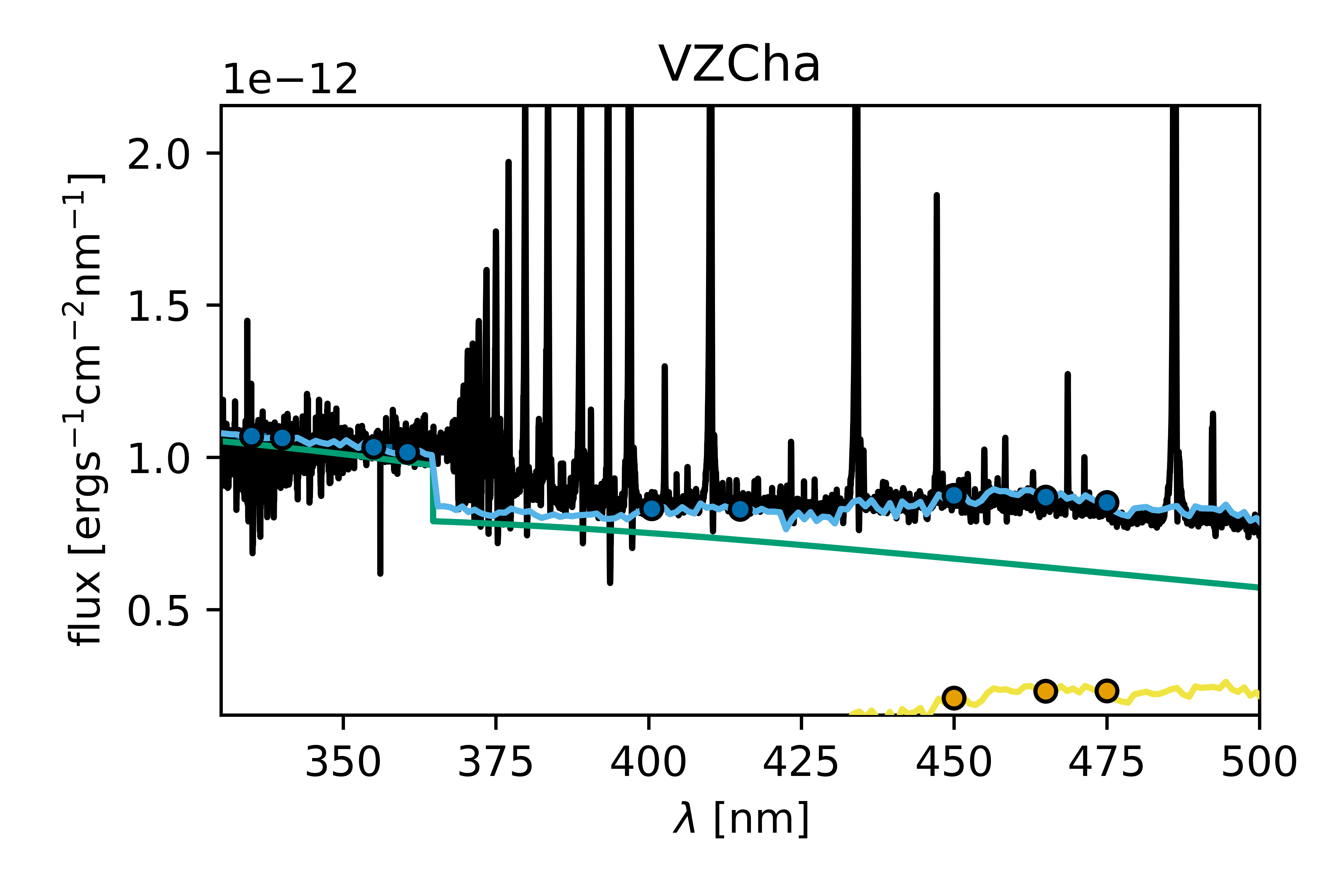}
    \includegraphics[width =0.49\textwidth]{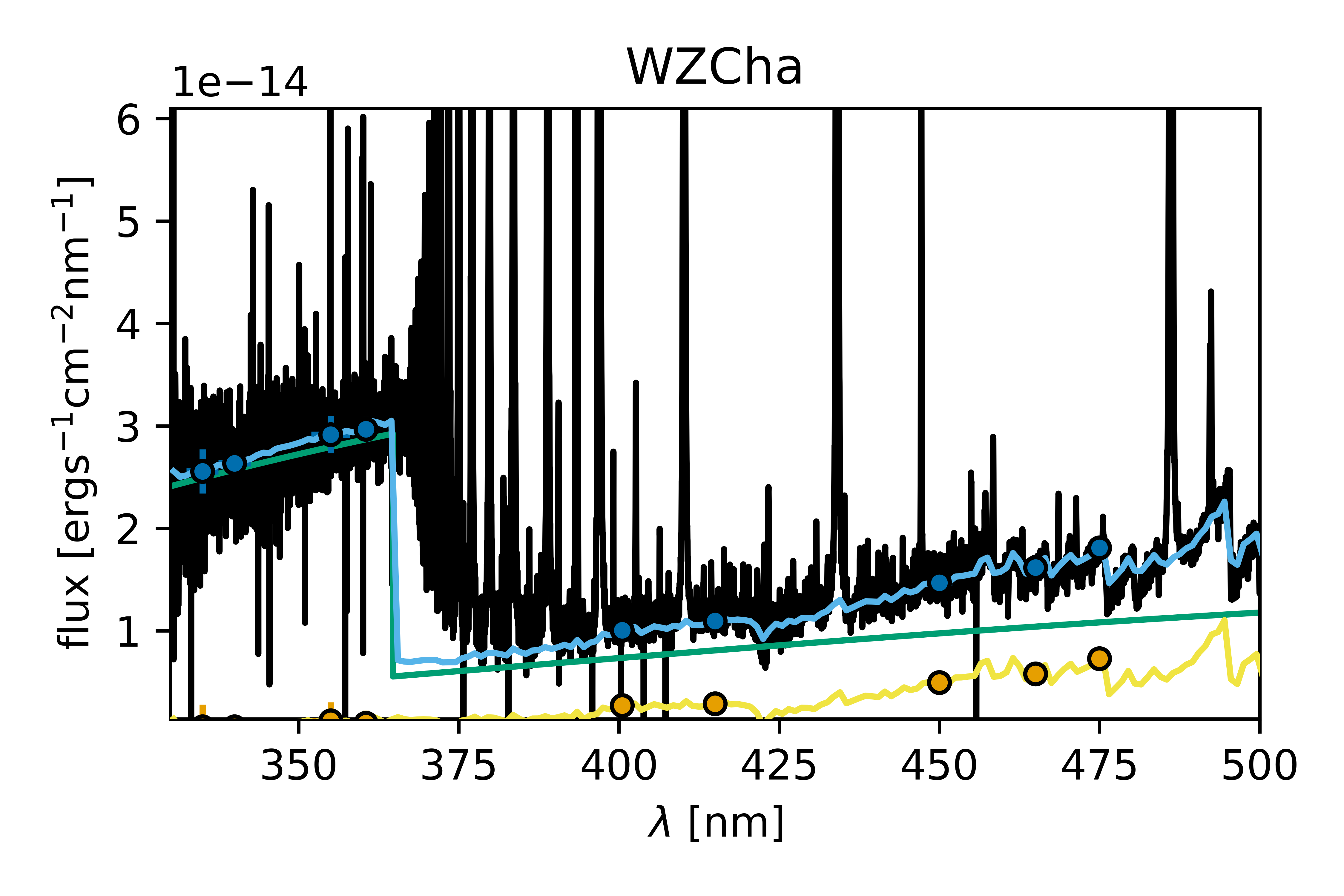}
    \includegraphics[width =0.49\textwidth]{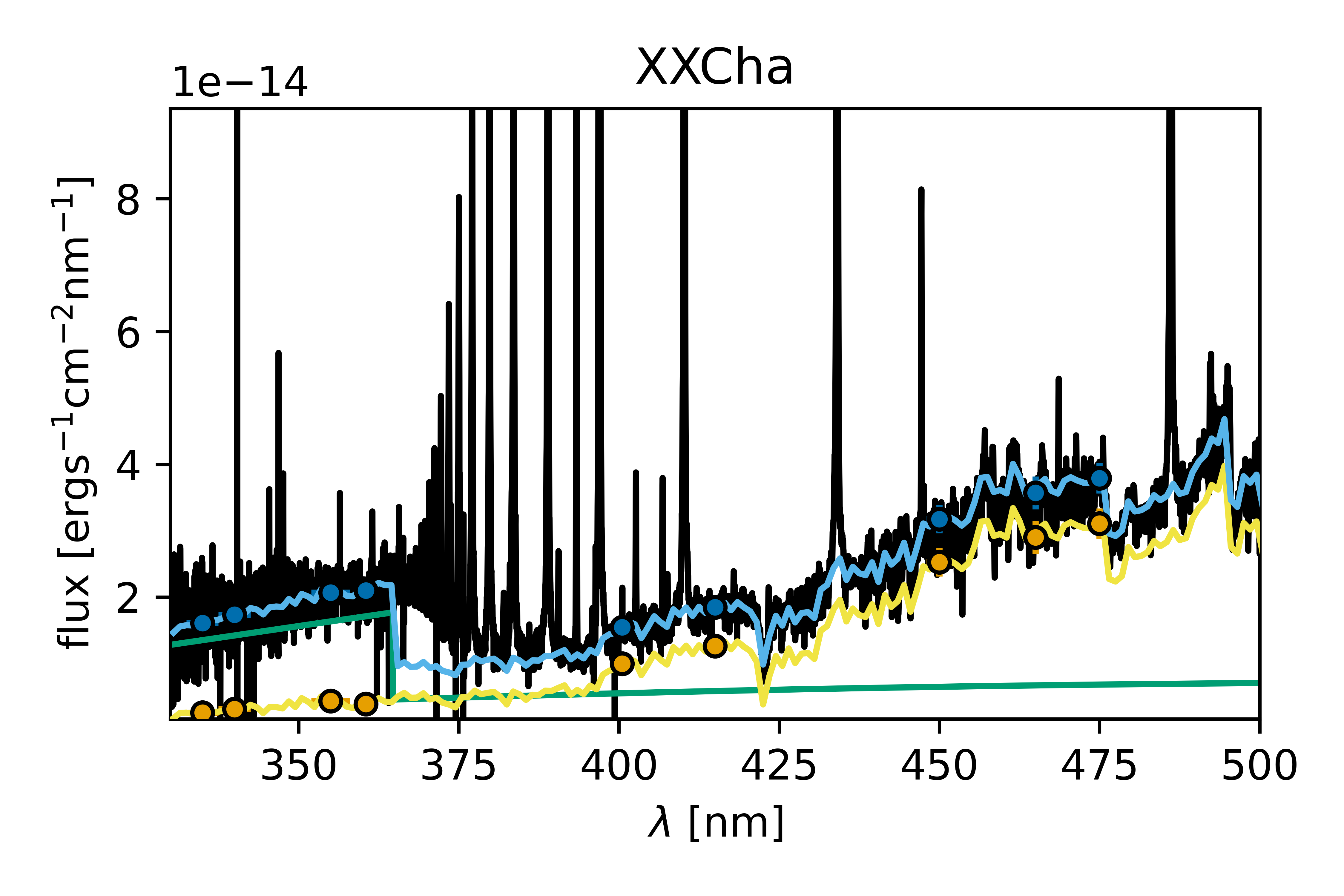}
    \includegraphics[width =0.49\textwidth]{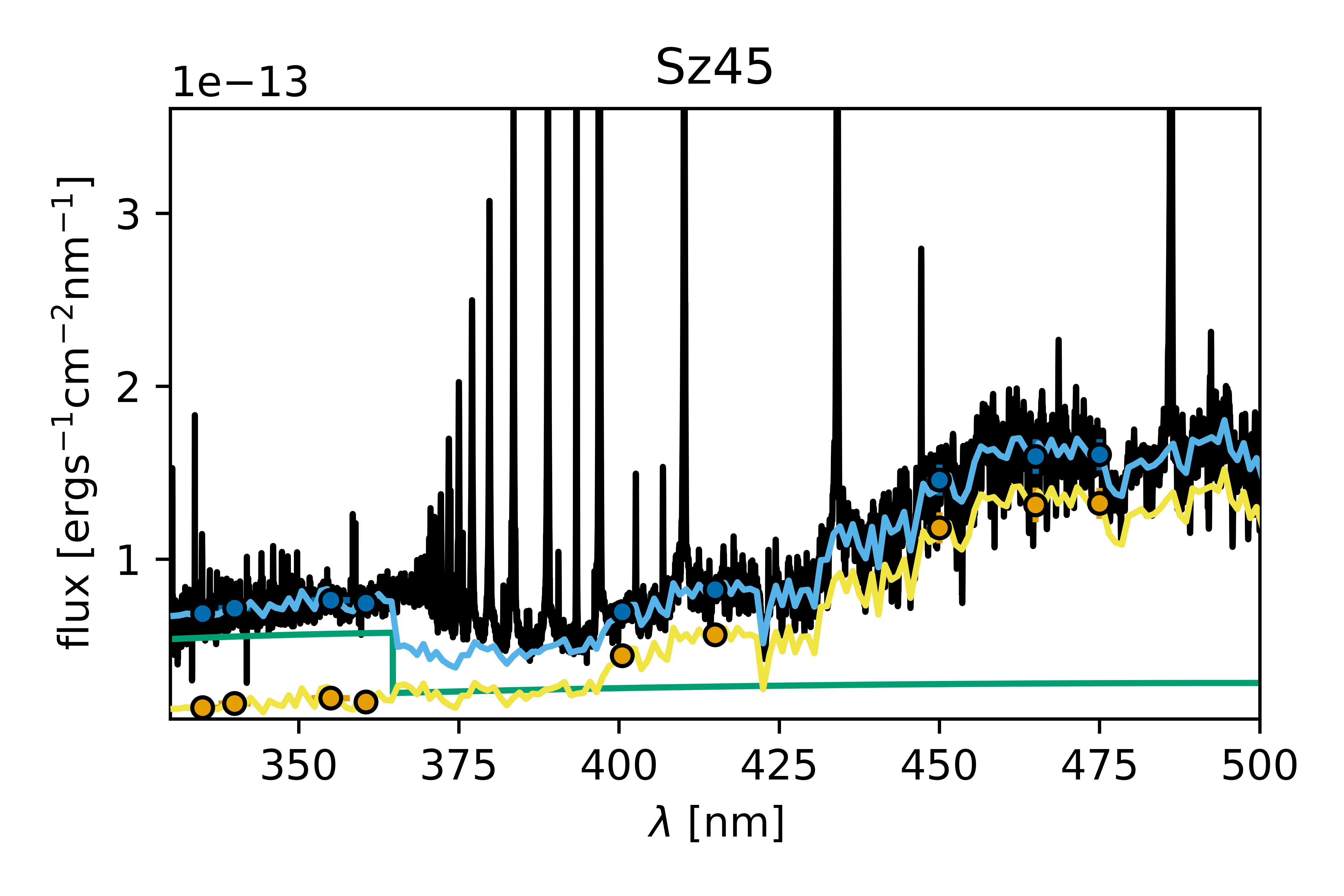}
     \includegraphics[width =0.49\textwidth]{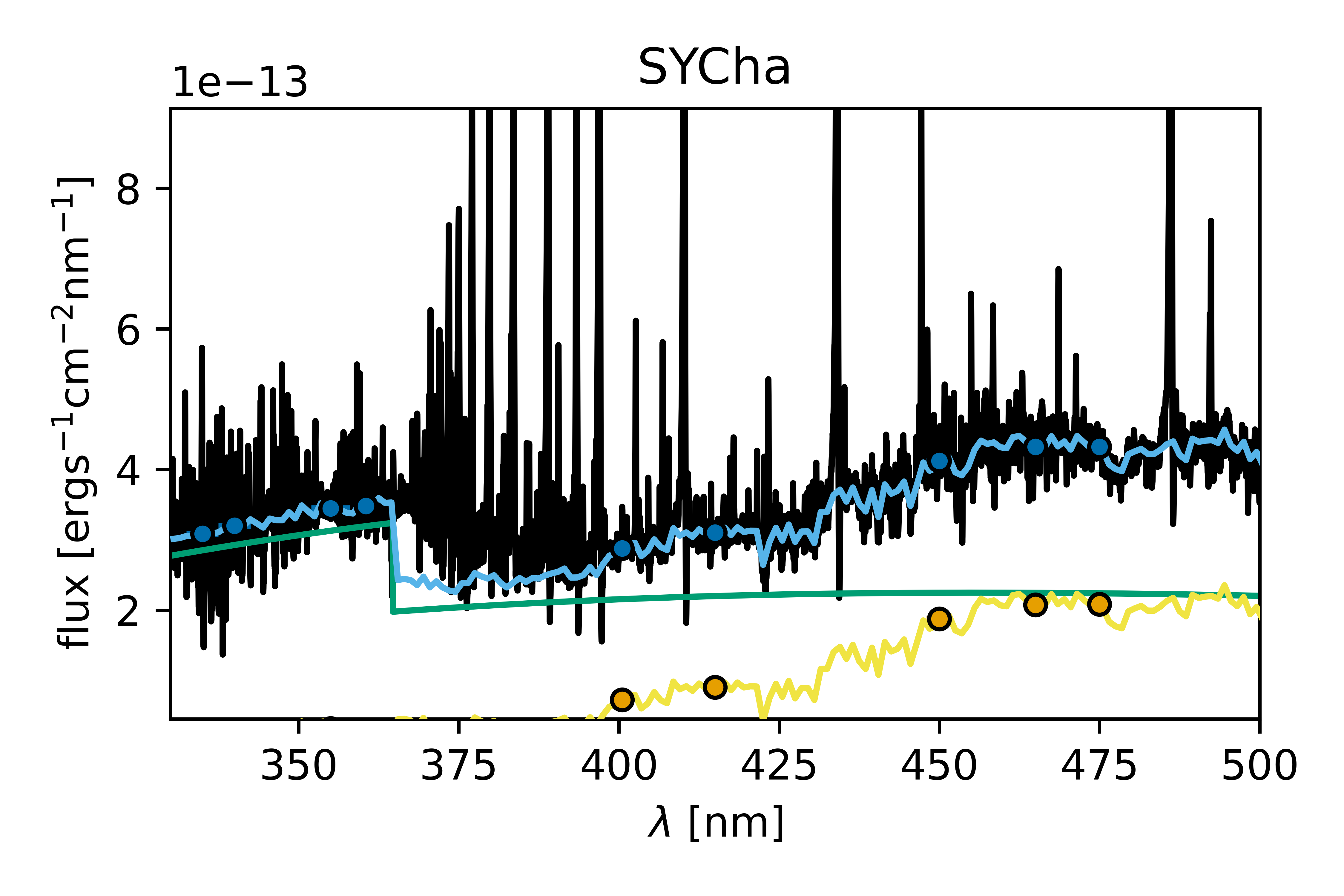}
    \caption{Best fits obtained with FRAPPE for the Balmer and Paschen continuum region for the targets in the Chamaeleon I association. Colors as in Fig \ref{fig:ChaIBestFitsContFitter}}
\end{figure*}

\begin{figure*}
    \centering
    \includegraphics[width =0.49\textwidth]{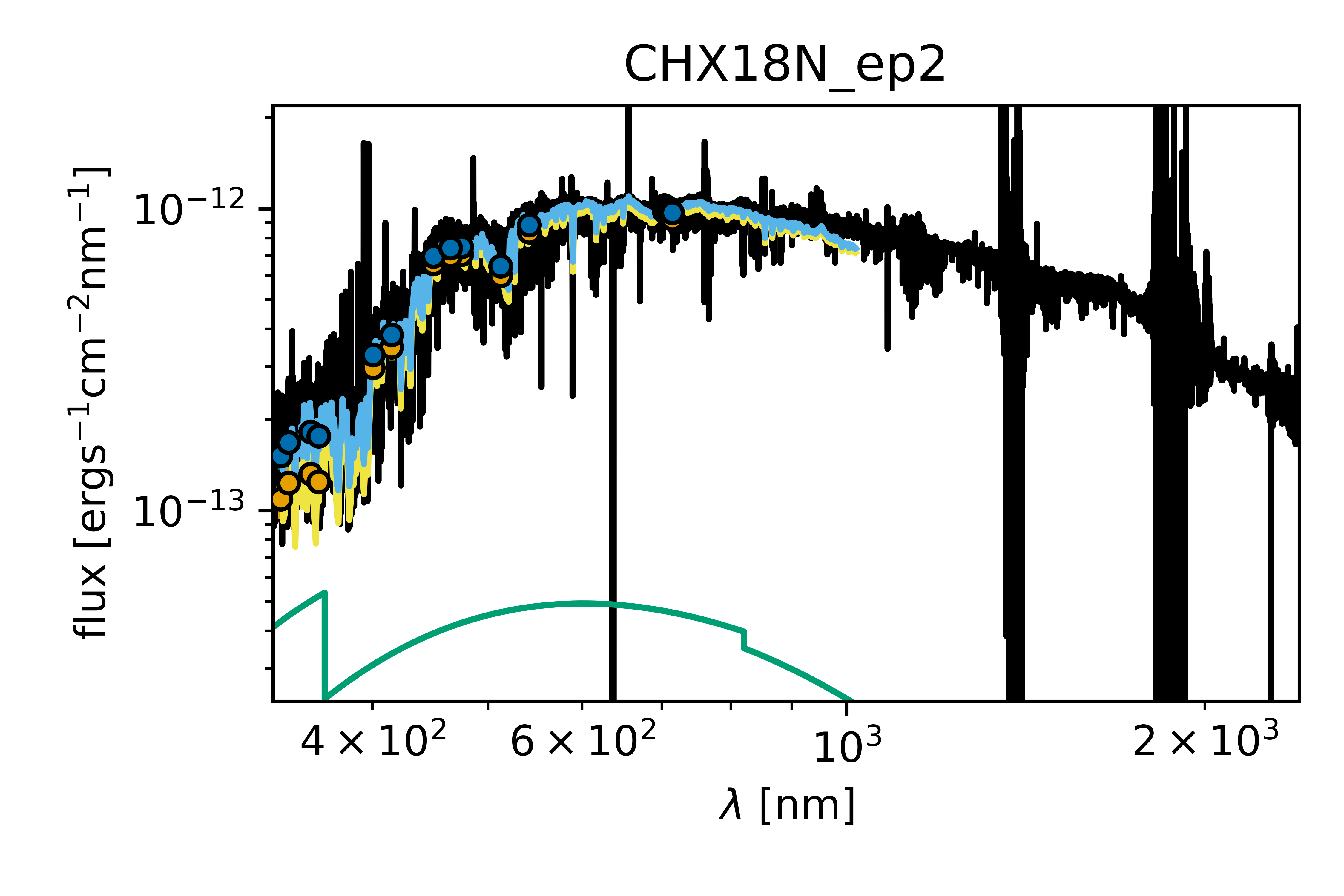}
    \includegraphics[width =0.49\textwidth]{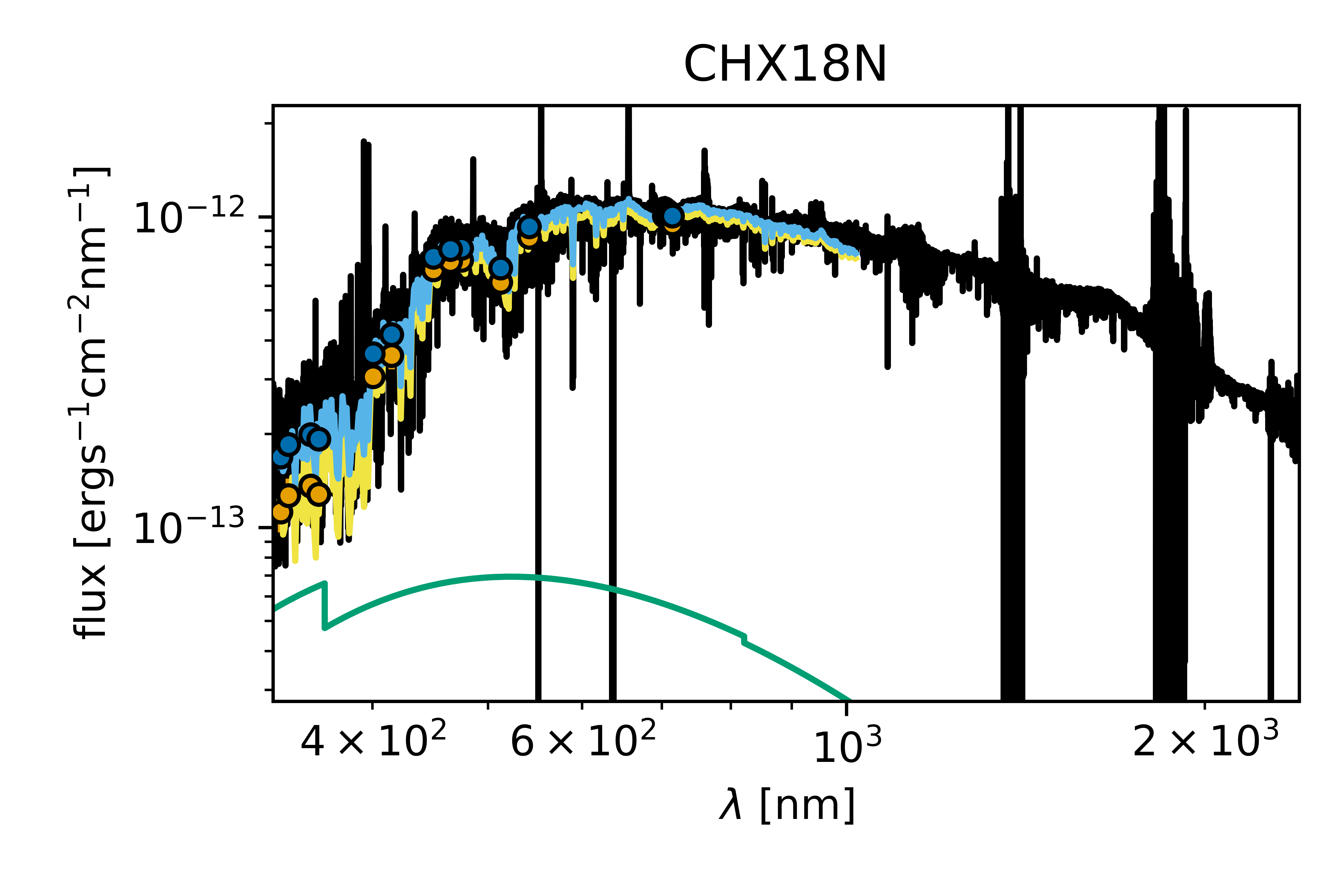}
    \includegraphics[width =0.49\textwidth]{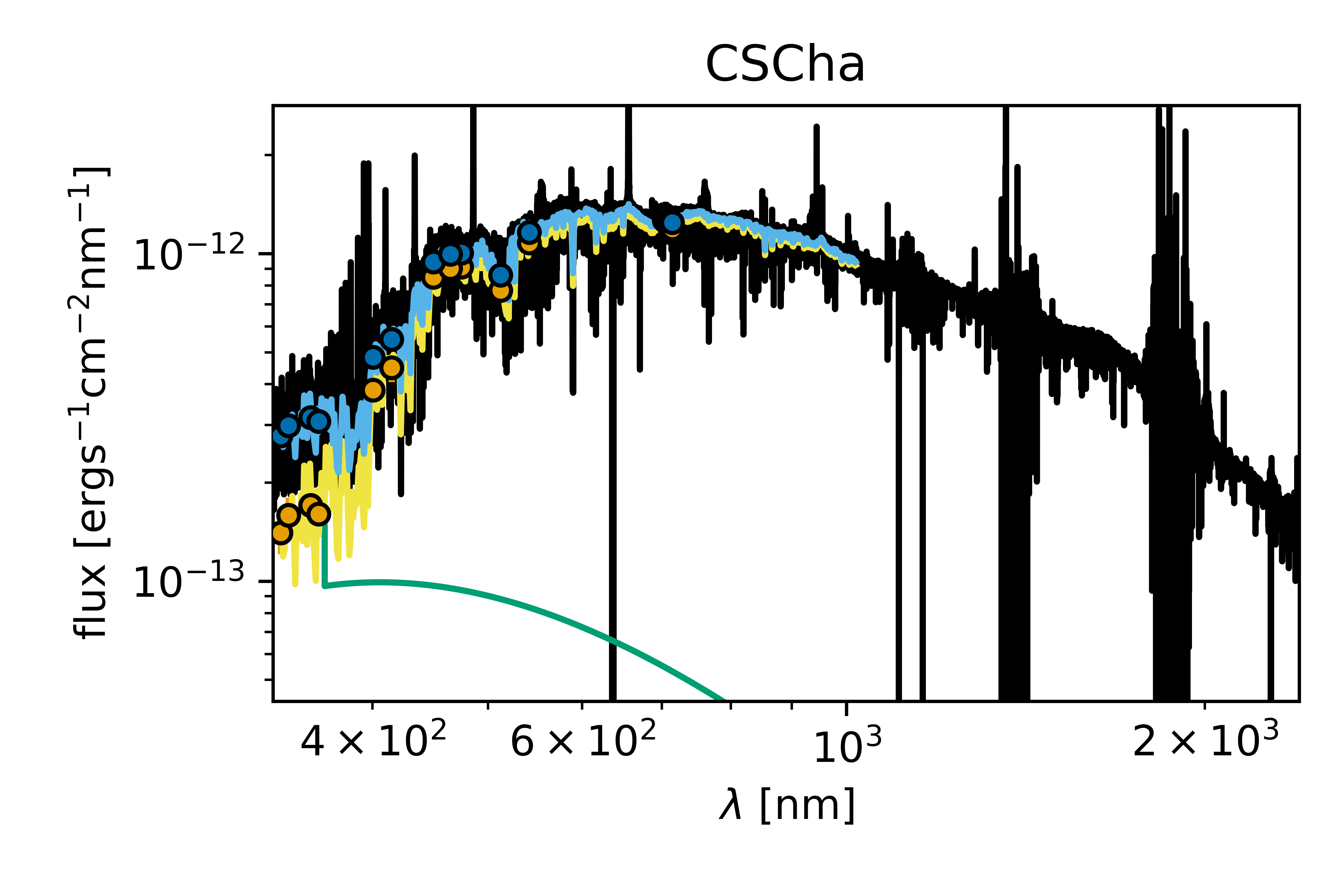}
    \includegraphics[width =0.49\textwidth]{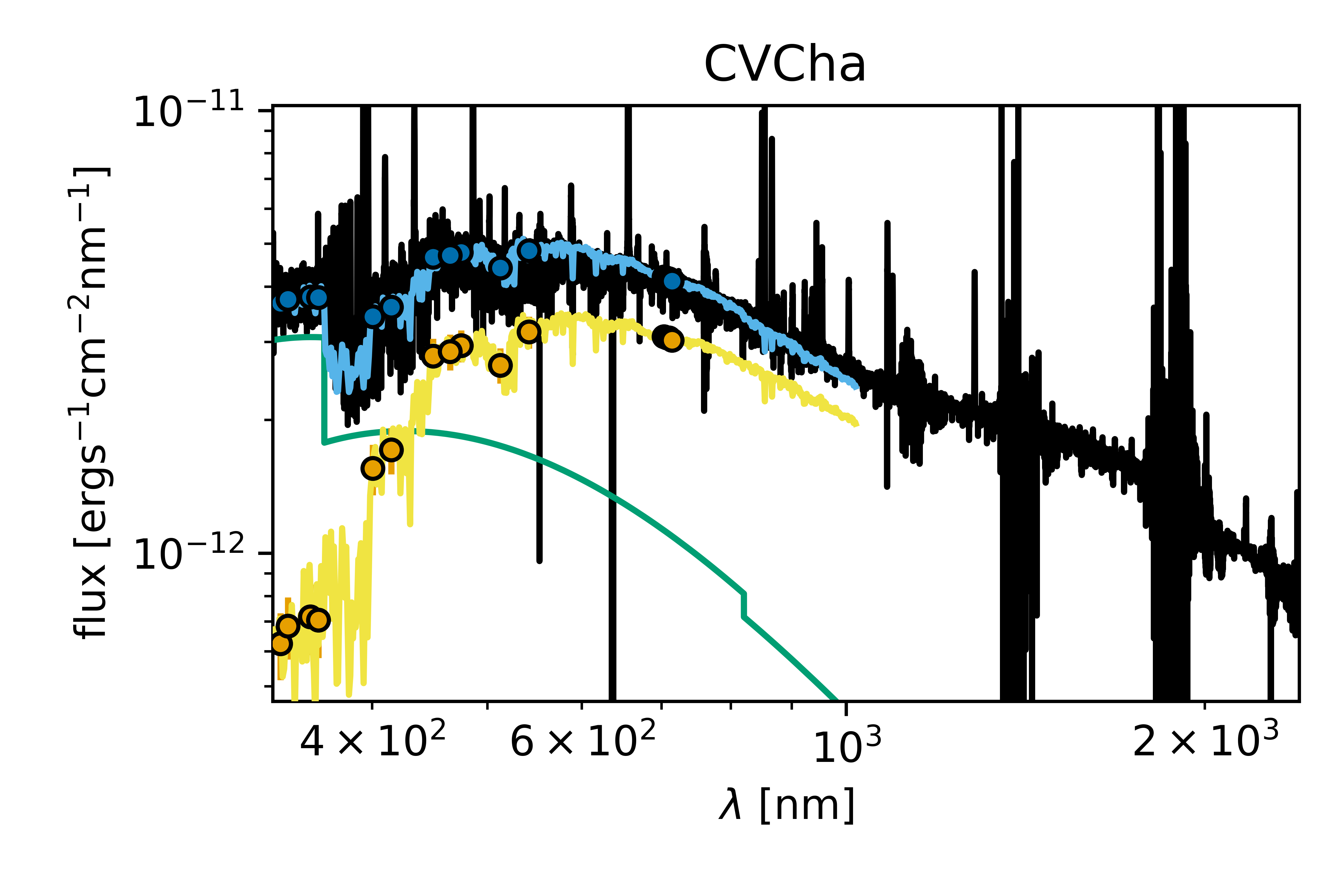}
    \includegraphics[width =0.49\textwidth]{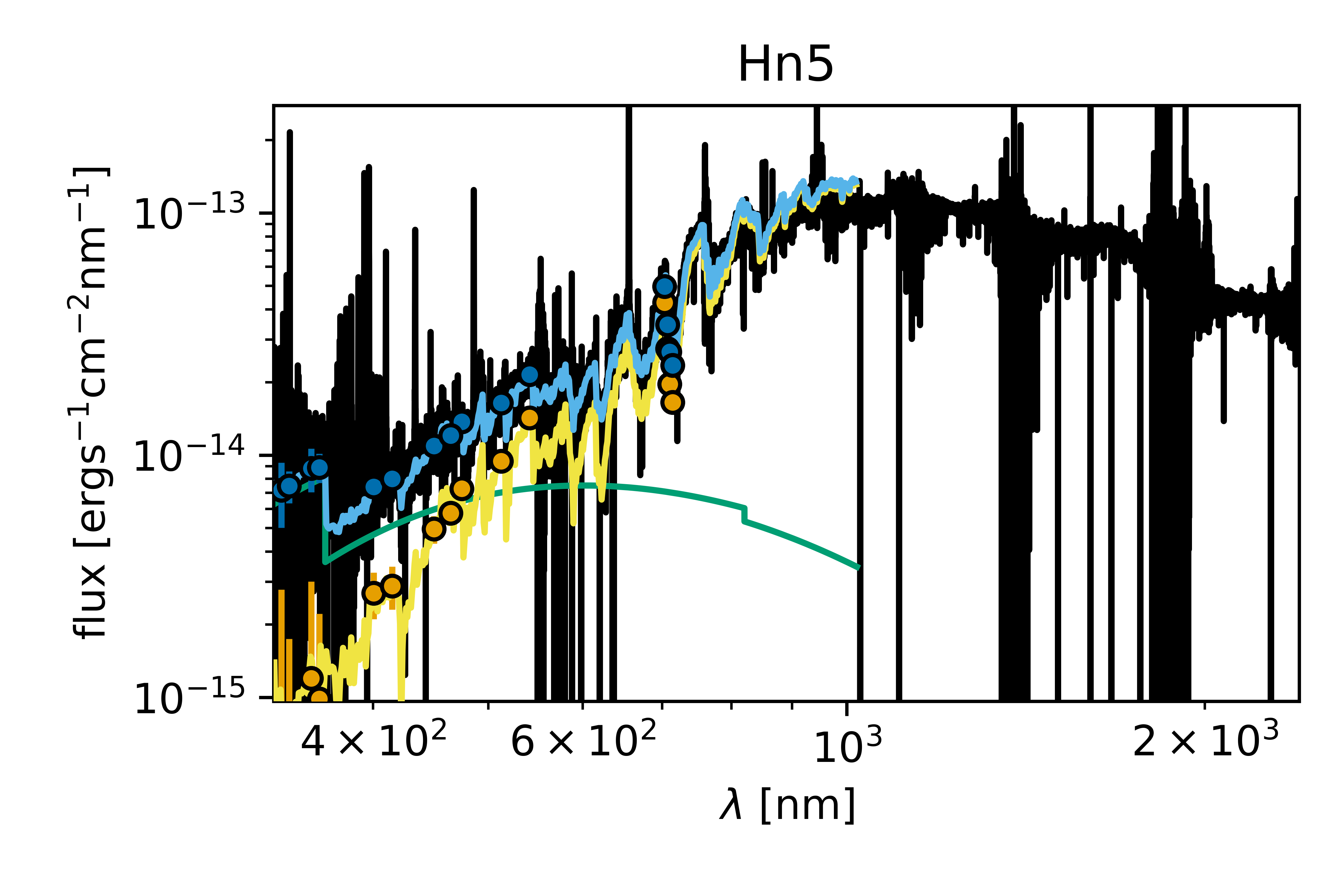}
     \includegraphics[width =0.49\textwidth]{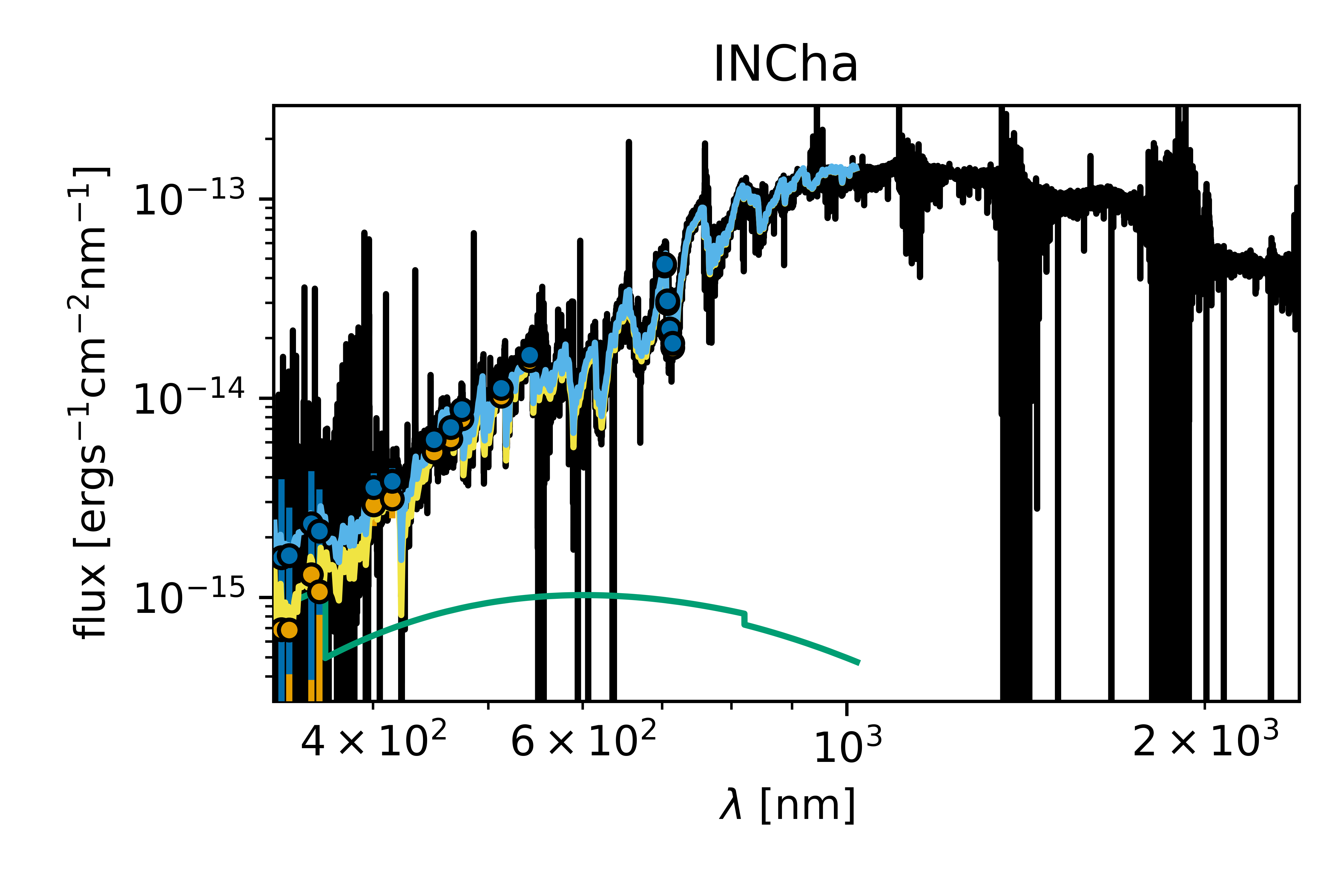}
     \includegraphics[width =0.49\textwidth]{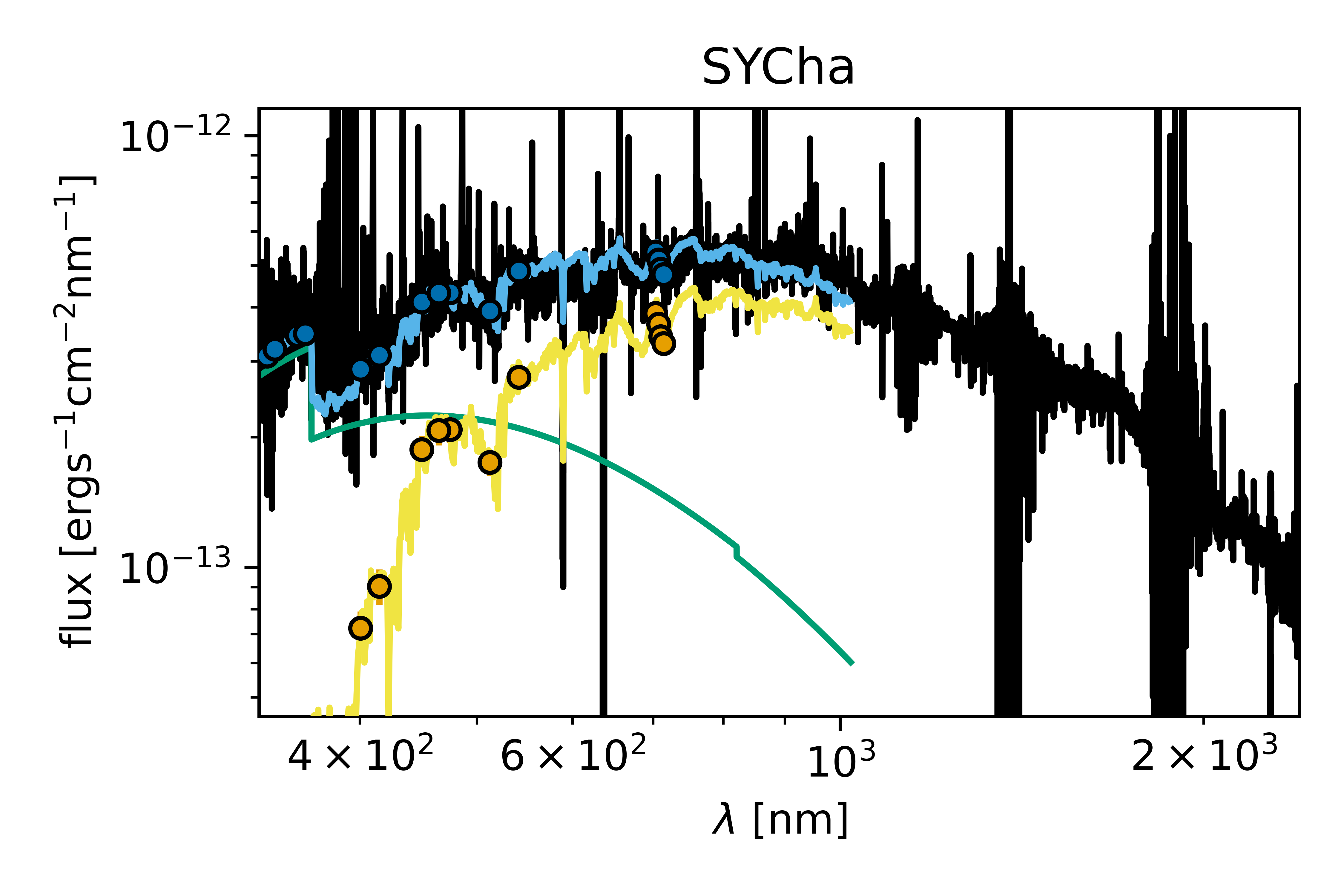}
     \includegraphics[width =0.49\textwidth]{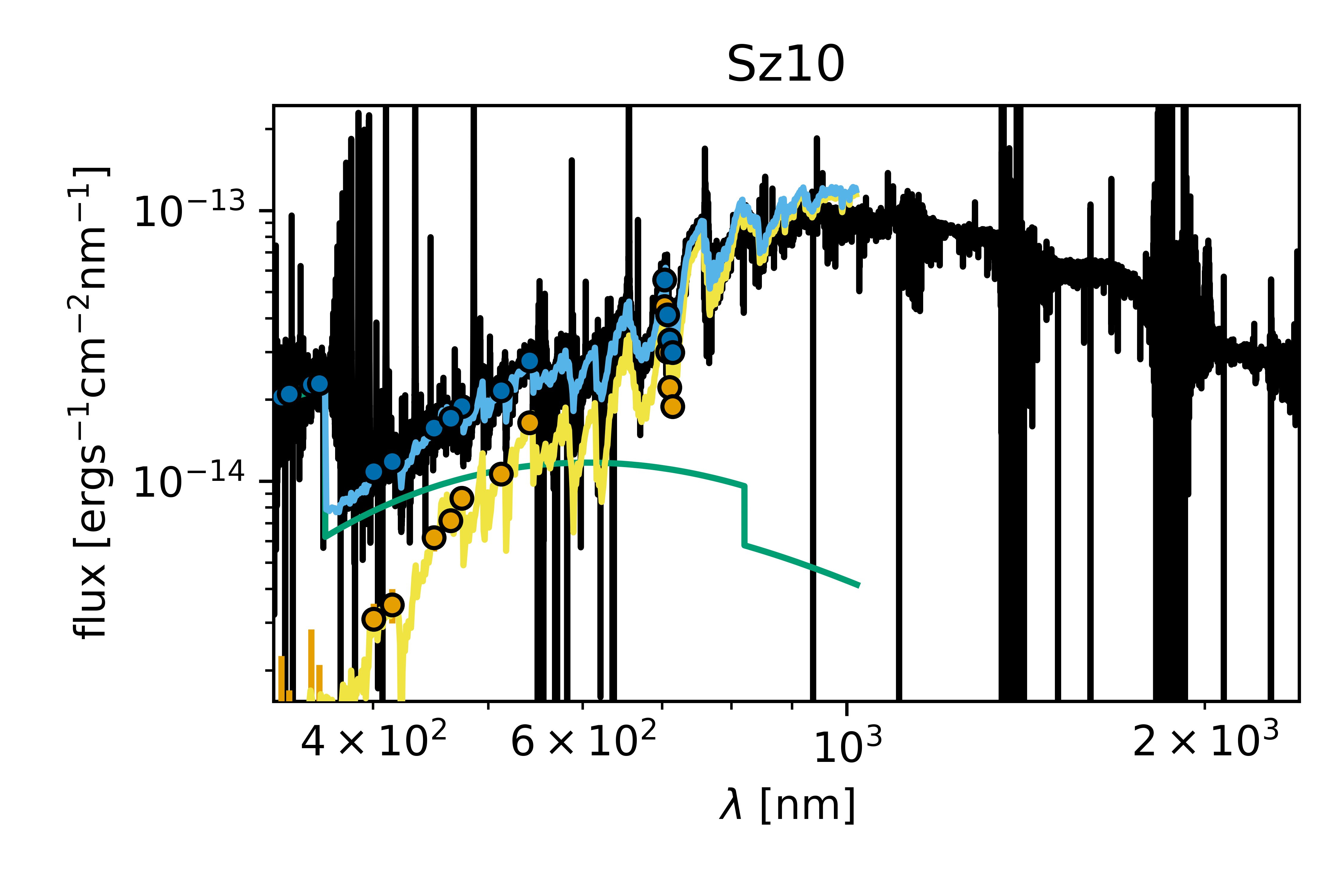}
    \caption{Best fits obtained with FRAPPE for the Balmer and Paschen continuum region for the targets in the Chamaeleon I association. Colors as in Fig \ref{fig:ChaIBestFitsContFitter}}
\end{figure*}

\begin{figure*}
    \centering
    \includegraphics[width =0.49\textwidth]{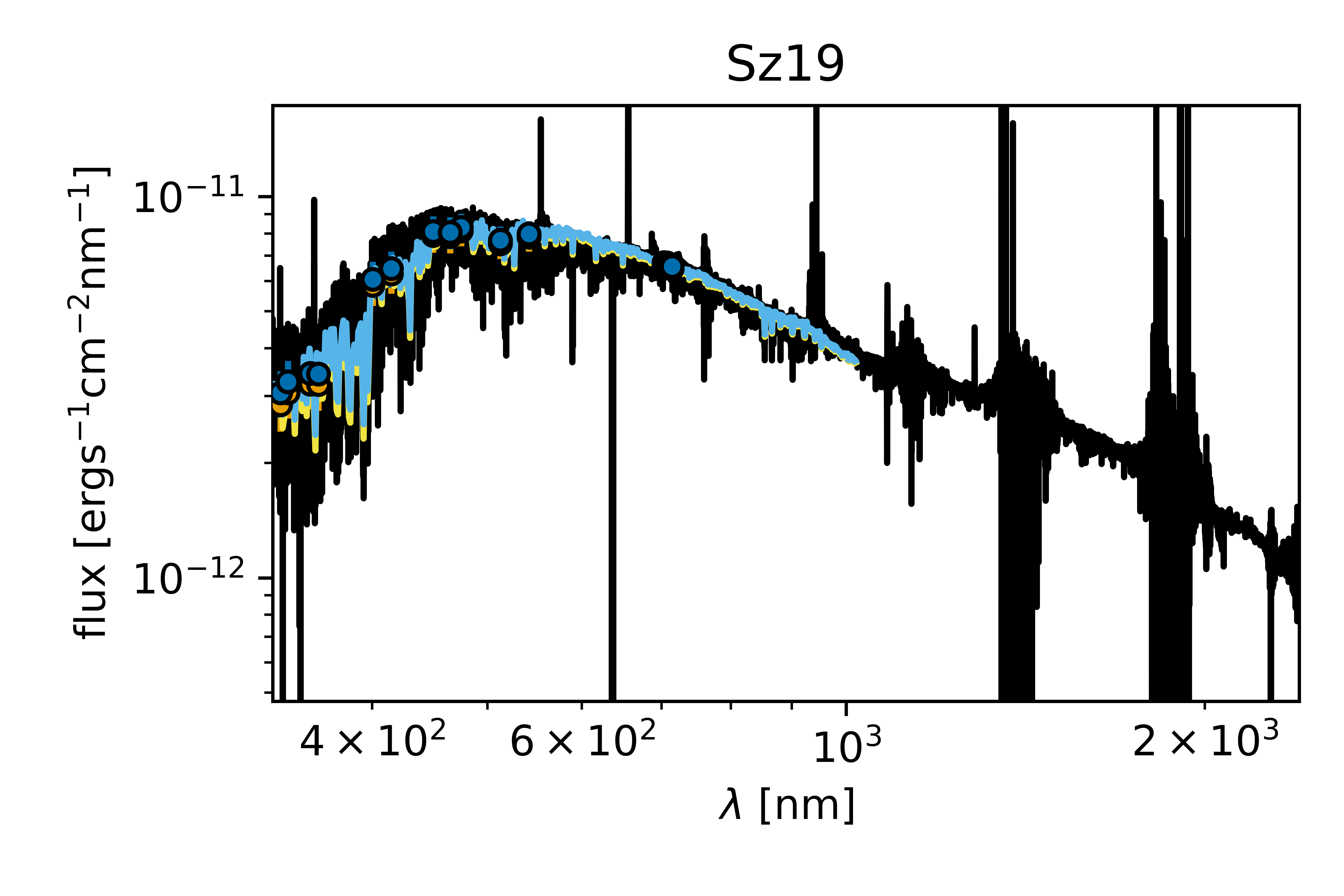}
    \includegraphics[width =0.49\textwidth]{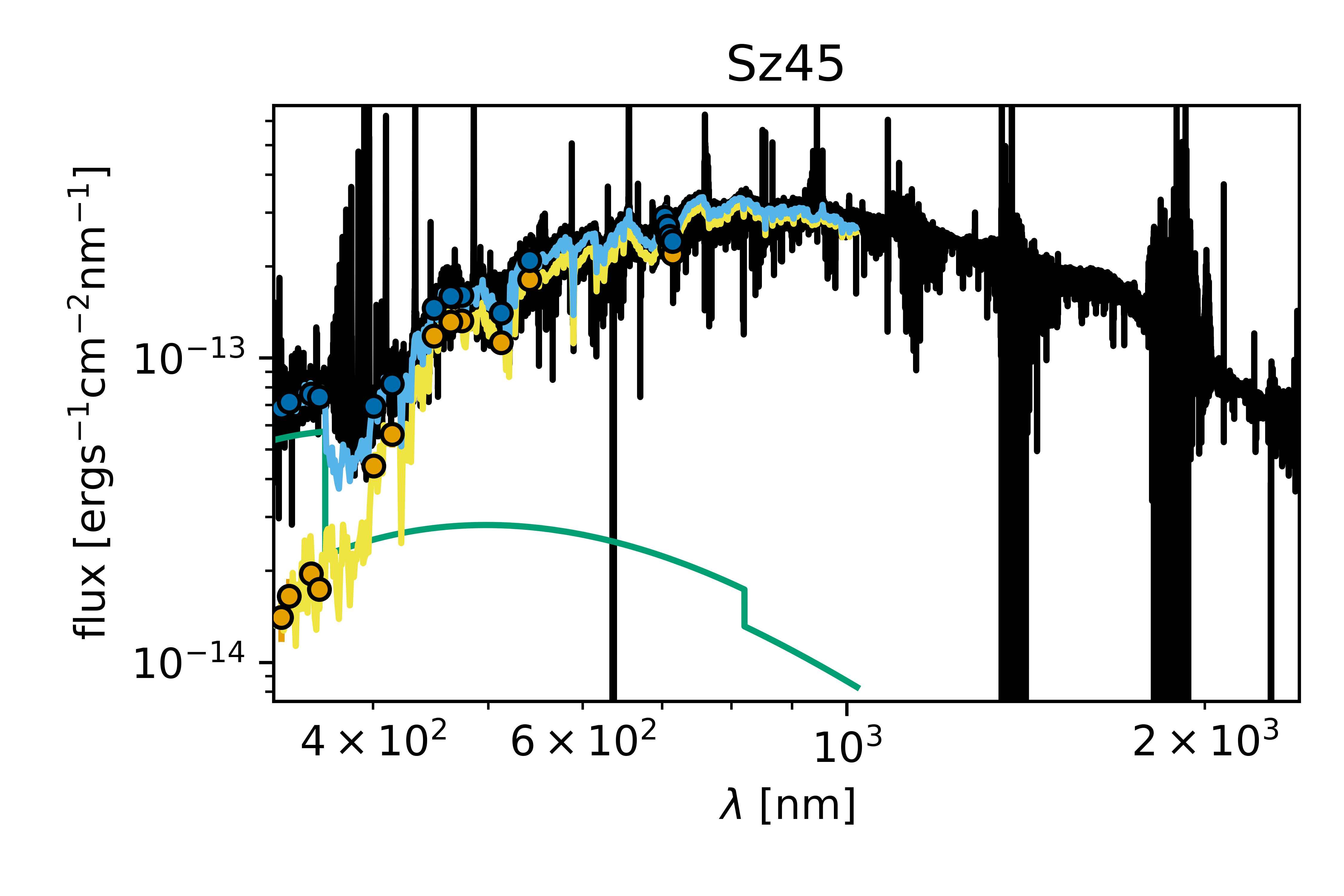}
    \includegraphics[width =0.49\textwidth]{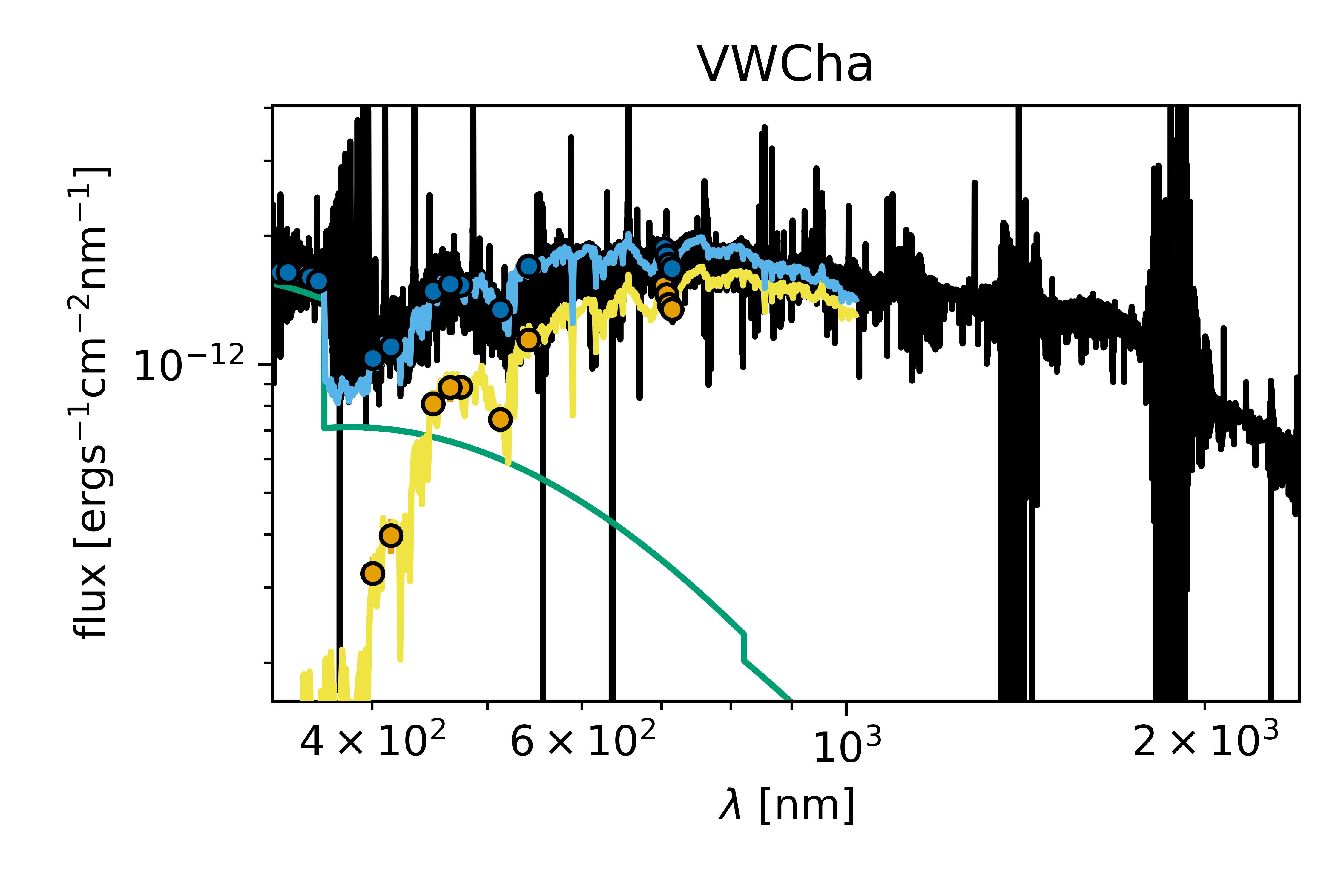}
    \includegraphics[width =0.49\textwidth]{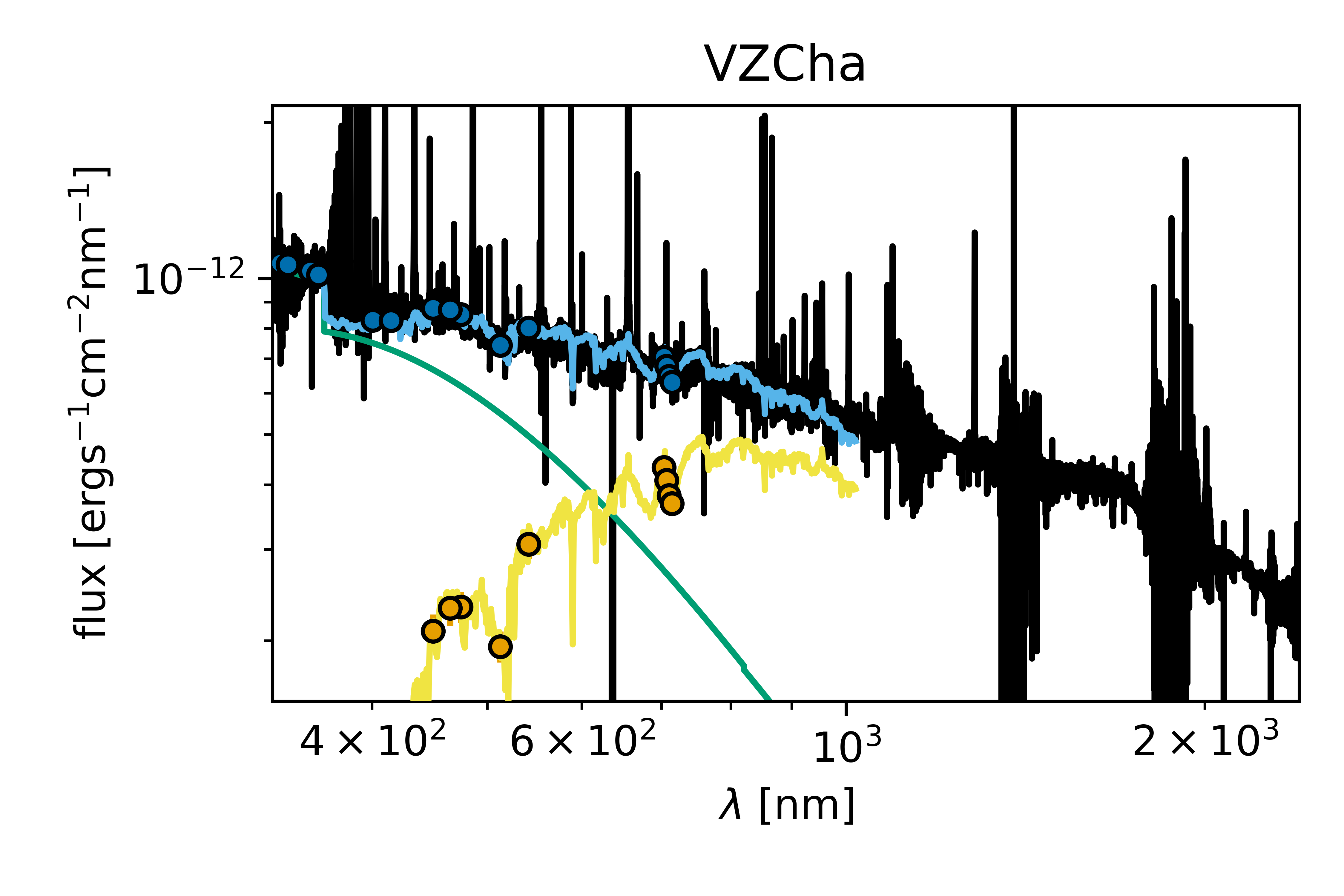}
    \includegraphics[width =0.49\textwidth]{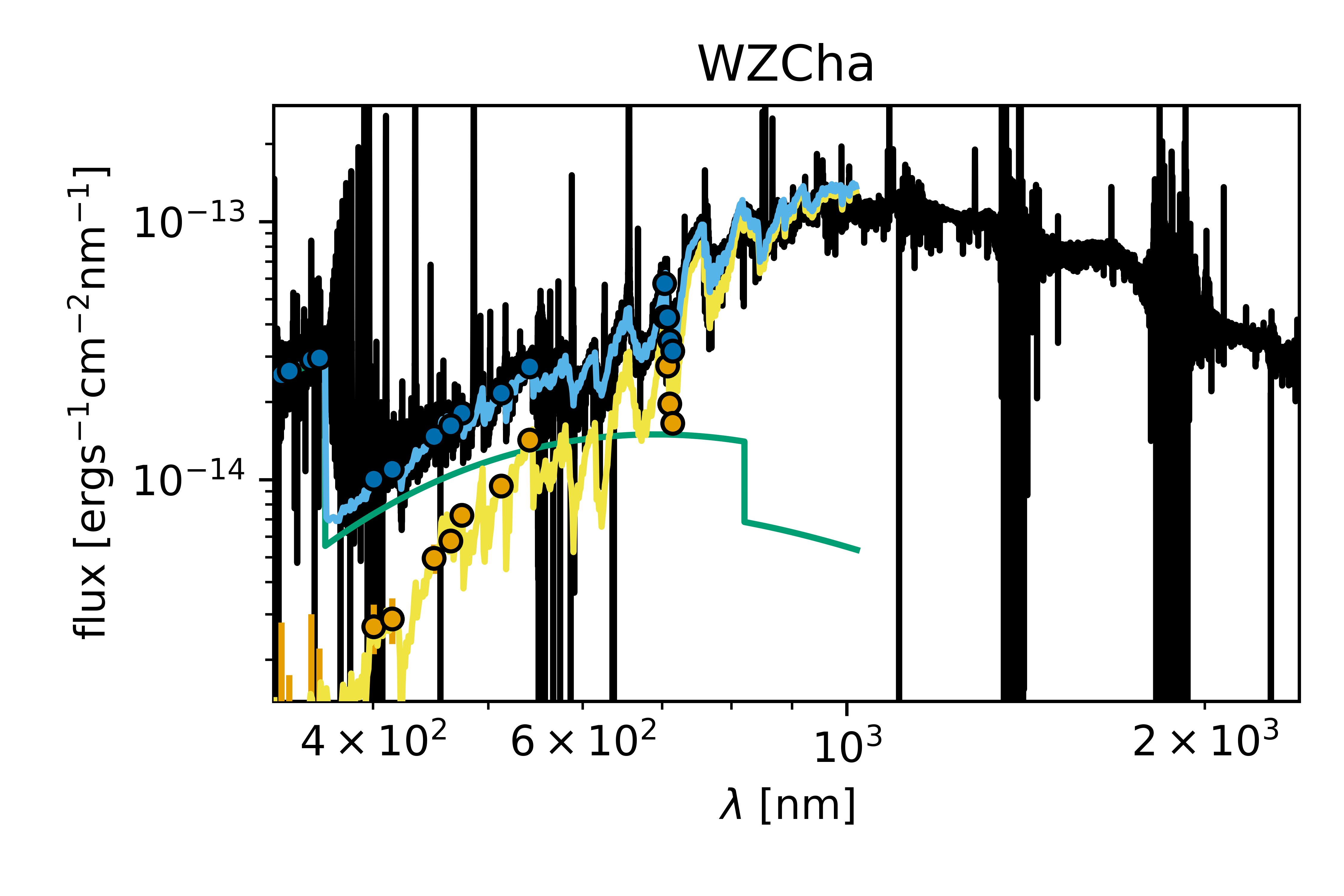}
     \includegraphics[width =0.49\textwidth]{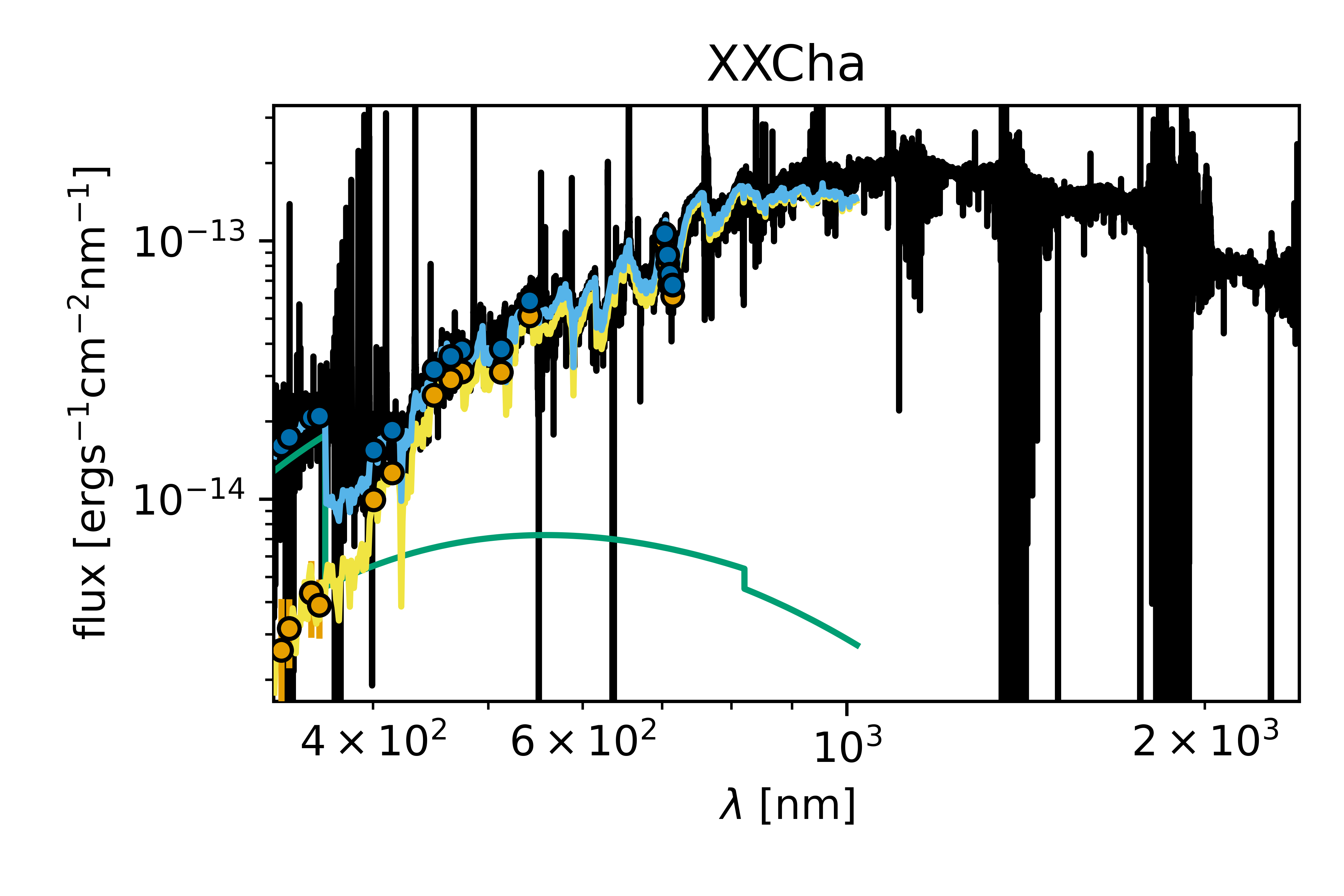}
    \caption{Best fits obtained with FRAPPE for the Balmer and Paschen continuum region for the targets in the Chamaeleon I association. Colors as in Fig \ref{fig:ChaIBestFitsContFitter}}
\end{figure*}
\newpage
\section{Plots of the Class III spectra}\label{app:luhman}


\begin{figure*}[th!]
    \centering
    \includegraphics[width =\textwidth]{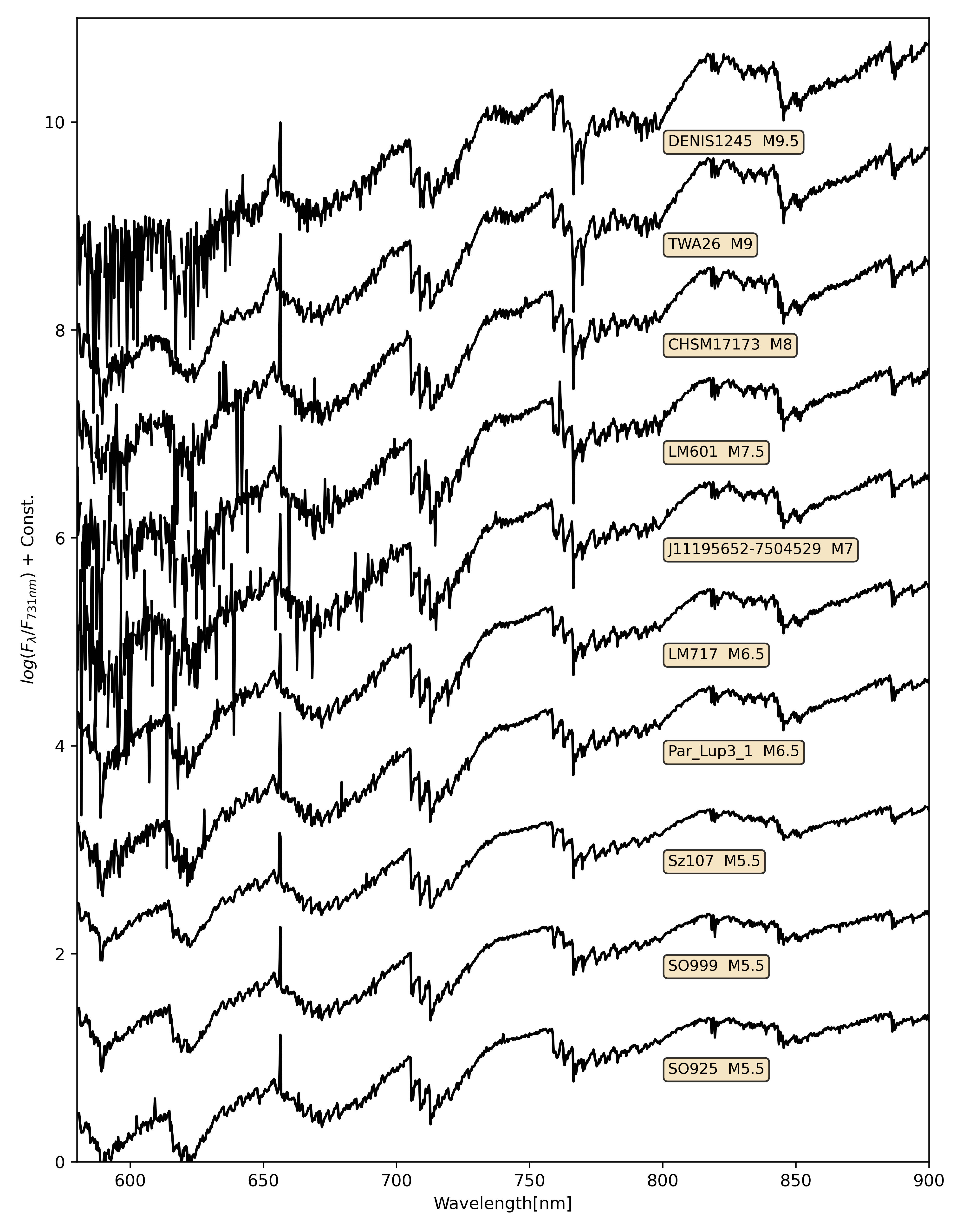}
    \caption{X Shooter spectra of Class III YSO with spectral types ranging from M9.5 to M5.5 SpT. All the spectra are normalized at 733 nm and offset in the vertical direction by 0.5 for clarity. The spectra are also smoothed to the resolution of 2500 at 750 nm to make the identification of the molecular features easier.}
    \label{fig:Luh750M9.5toM5.5}
\end{figure*}

\begin{figure*}[th!]
    \centering
    \includegraphics[width =\textwidth]{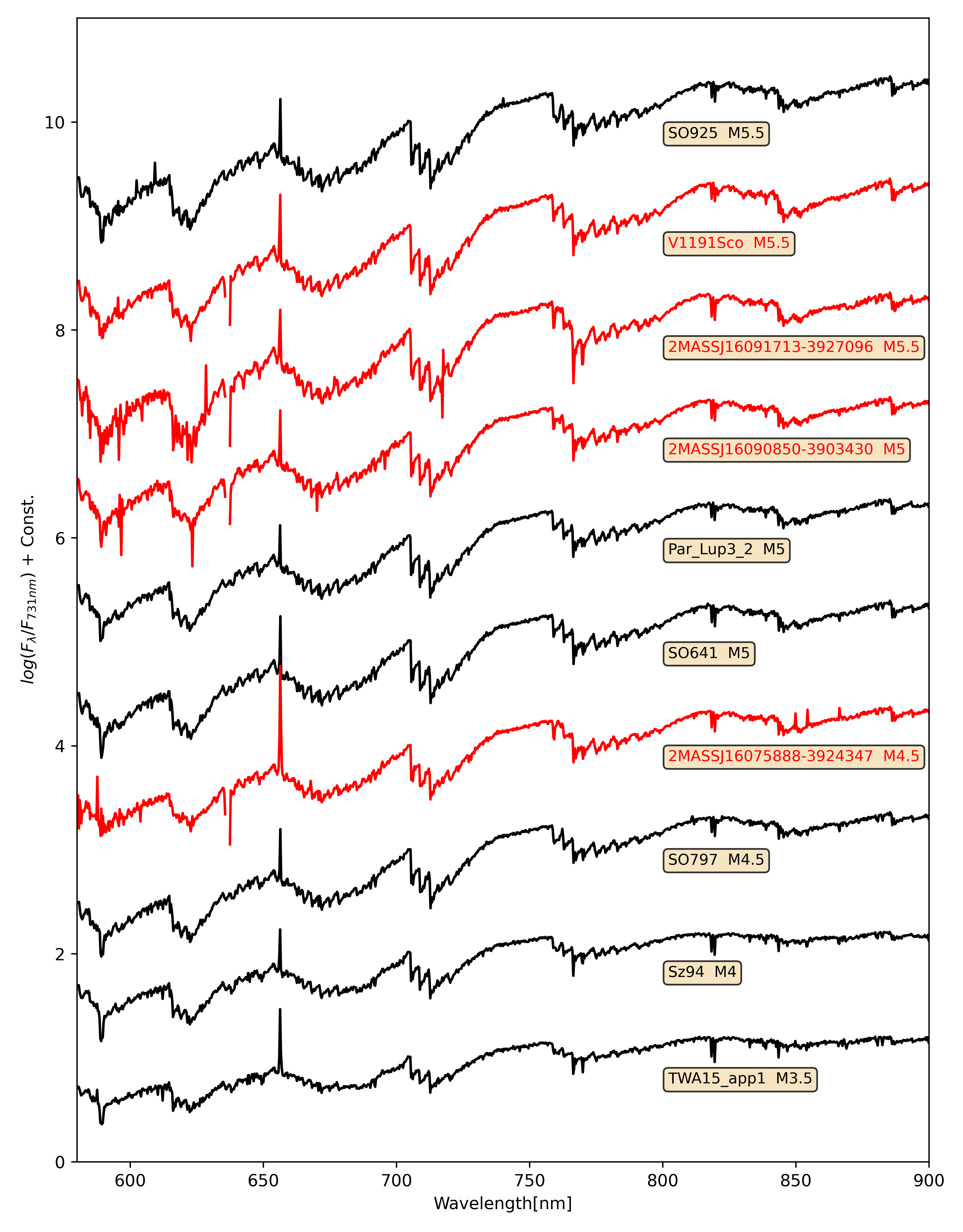}
    \caption{X Shooter spectra of Class III YSO with spectral types ranging from M5.5 to M3.5 SpT All the spectra are normalized at 733 nm and offset in the vertical direction by 0.5 for clarity. TThe spectra are also smoothed to the resolution of 2500 at 750 nm to make the identification of the molecular features easier.}
\end{figure*}

\begin{figure*}[th!]
    \centering
    \includegraphics[width =\textwidth]{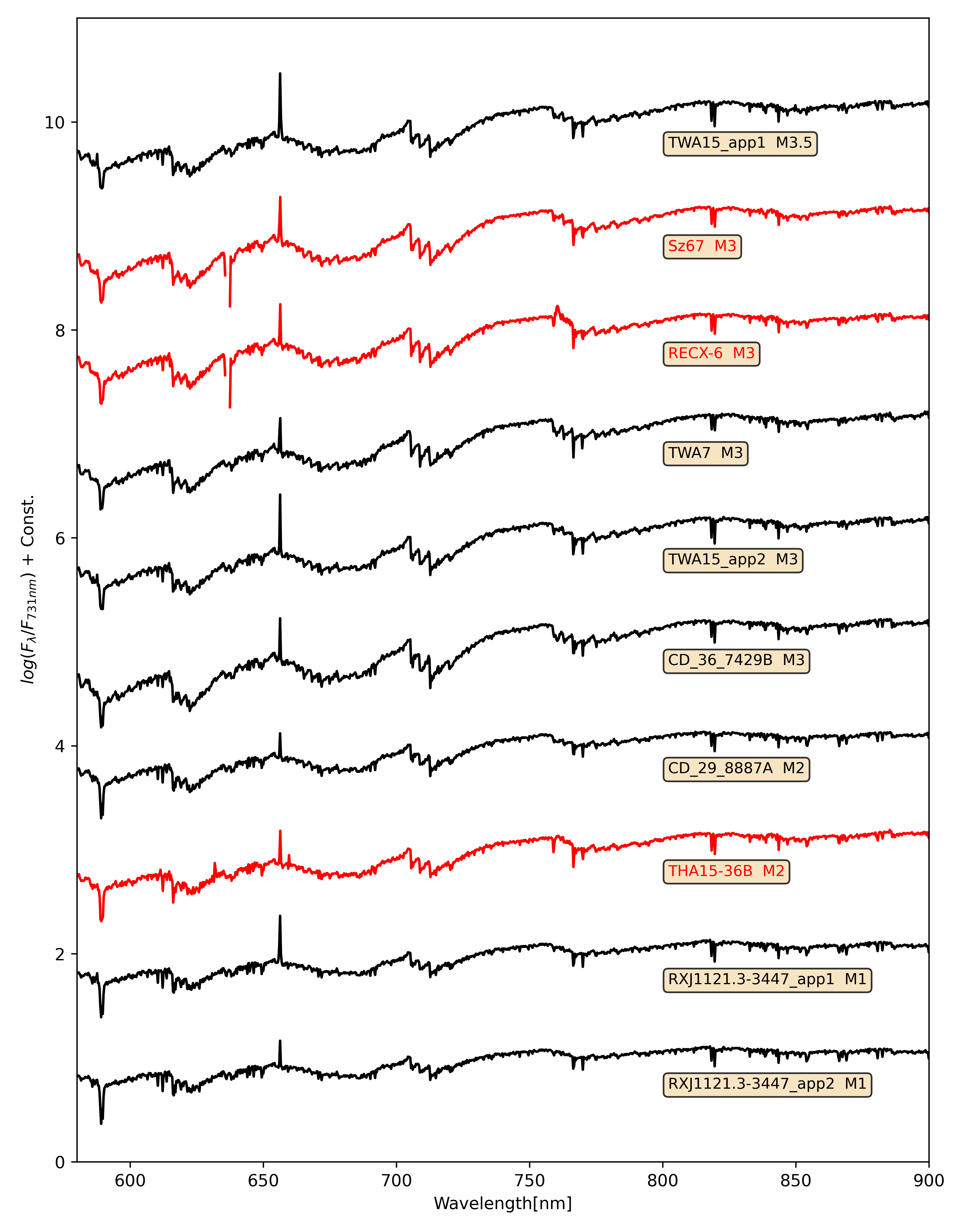}
    \caption{X Shooter spectra of Class III YSO with spectral types ranging from M3.5 to M1 SpT. All the spectra are normalized at 733 nm and offset in the vertical direction by 0.5 for clarity. The spectra are also smoothed to the resolution of 2500 at 750 nm to make the identification of the molecular features easier.}
\end{figure*}

\begin{figure*}[th!]
    \centering
    \includegraphics[width =\textwidth]{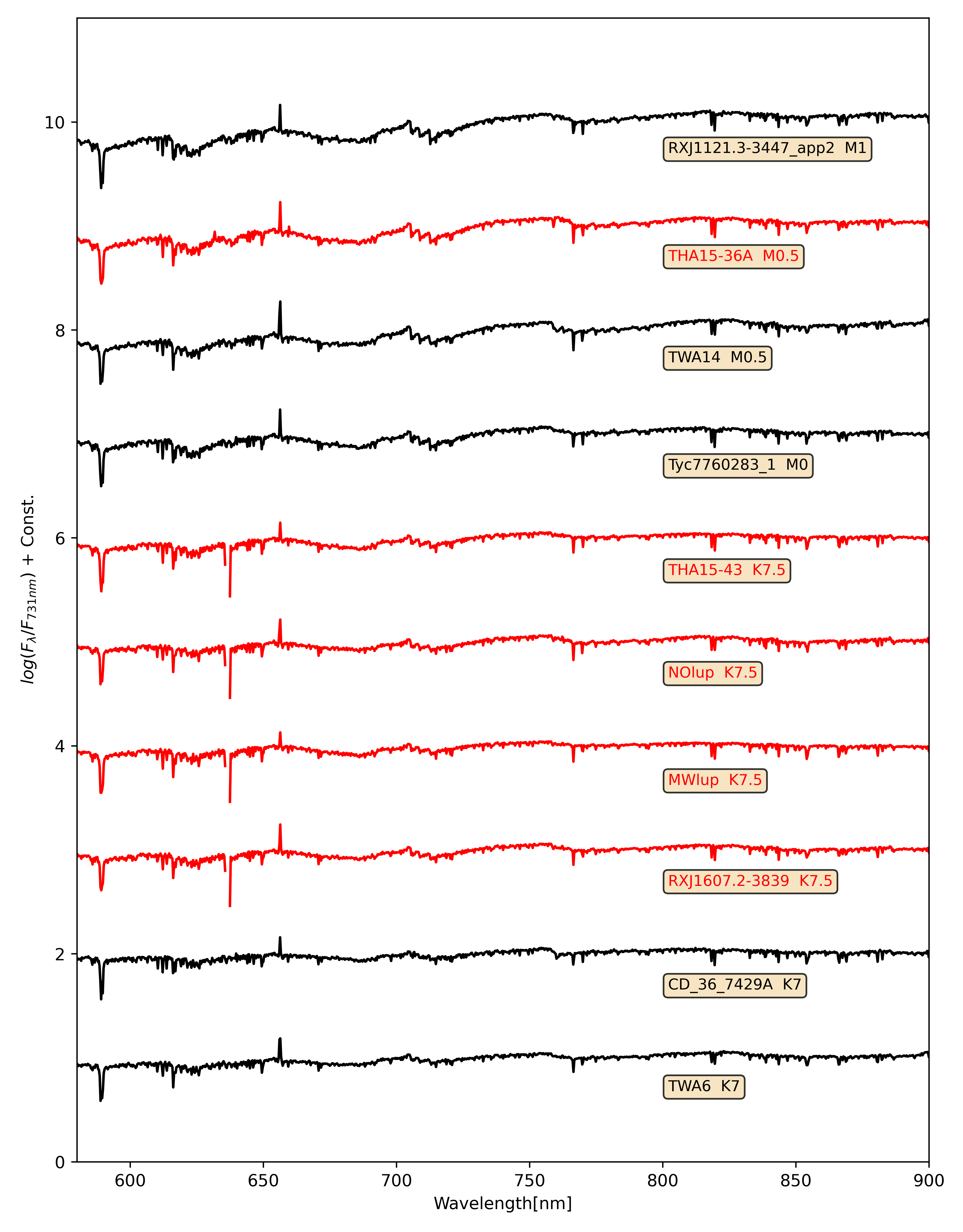}
    \caption{X Shooter spectra of Class III YSO with spectral types ranging from M1 to K7. All the spectra are normalized at 733 nm and offset in the vertical direction by 0.5 for clarity. The spectra are also smoothed to the resolution of 2500 at 750 nm to make the identification of the molecular features easier.}
\end{figure*}

\begin{figure*}[th!]
    \centering
    \includegraphics[width =\textwidth]{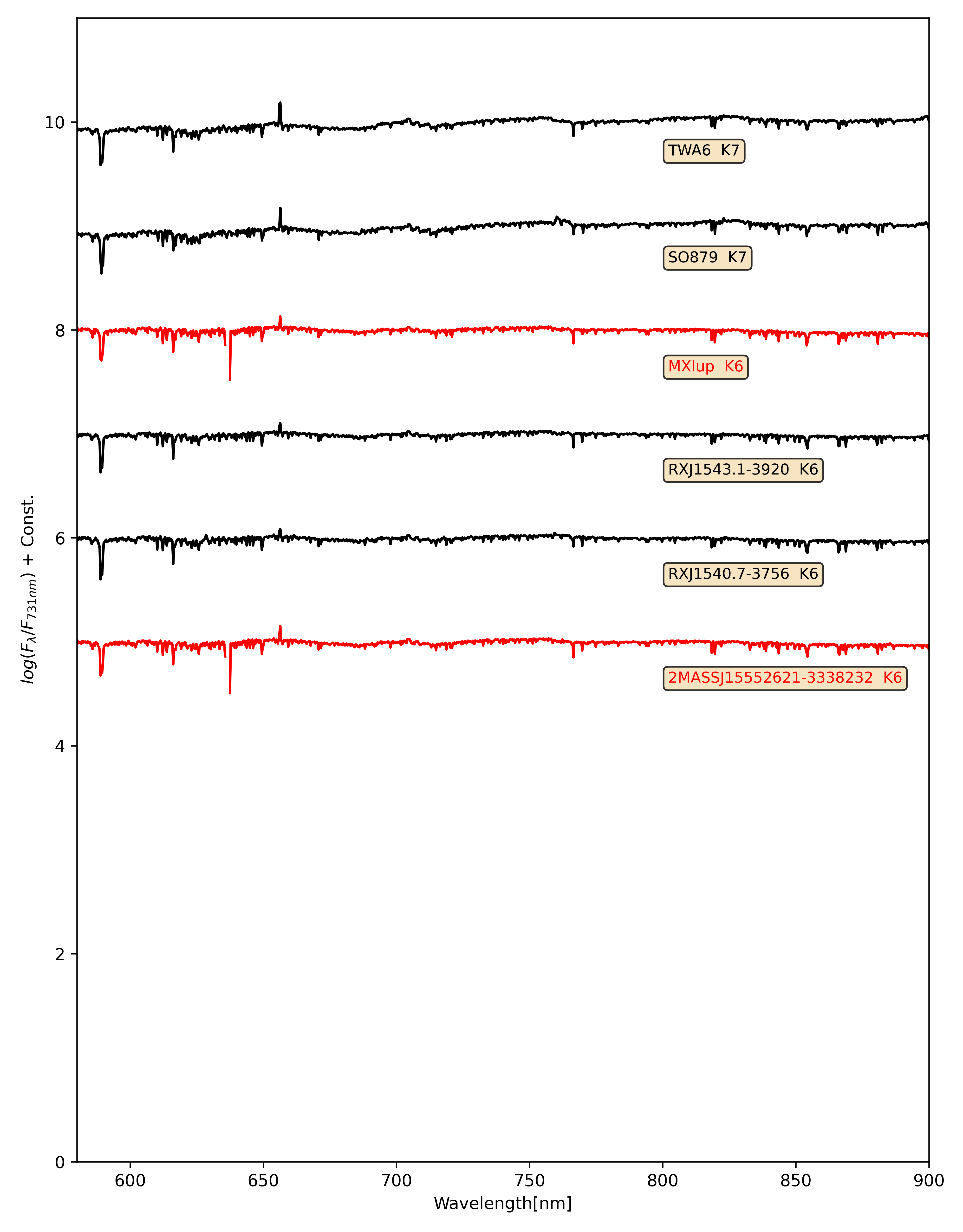}
    \caption{X Shooter spectra of Class III YSO with spectral types ranging from K7 to K6 ordered according to SpT and normalized at 733 nm}
\end{figure*}


\begin{figure*}[th!]
    \centering
    \includegraphics[width =\textwidth]{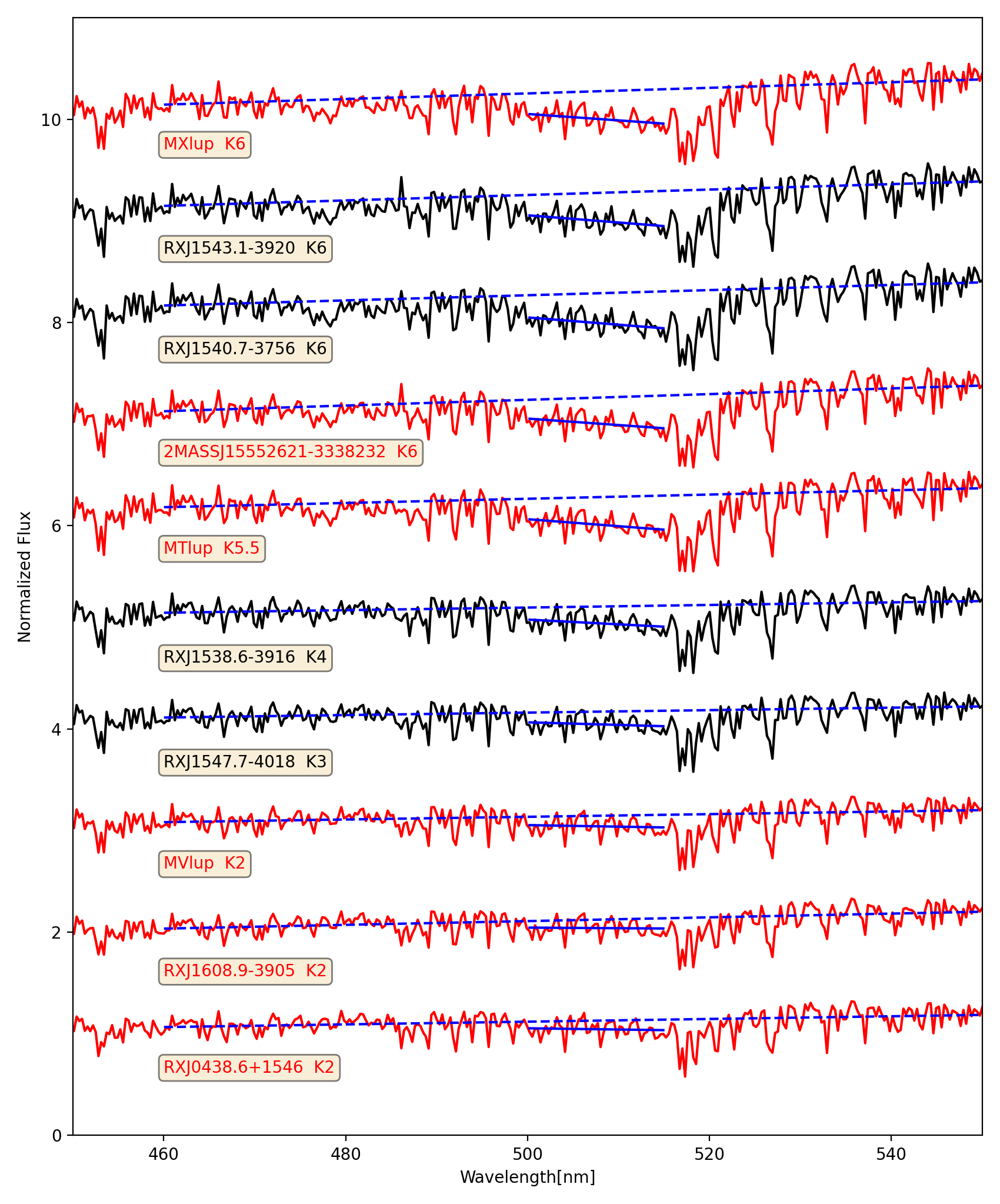}
    \caption{X Shooter spectra of Class III YSO with spectral types ranging from K6 to K2 ordered according to SpT and normalized at 520 nm. (continued on Fig. \ref{fig:LuhmanUVB5101}). Spectra presented in this work are indicated in red and spectra from the samples of \citetalias{manara13a} and \citetalias{manara17b} are indicated in black.}
    \label{fig:LuhmanUVB5100}
\end{figure*}

\begin{figure*}[th!]
    \centering
    \includegraphics[width =\textwidth]{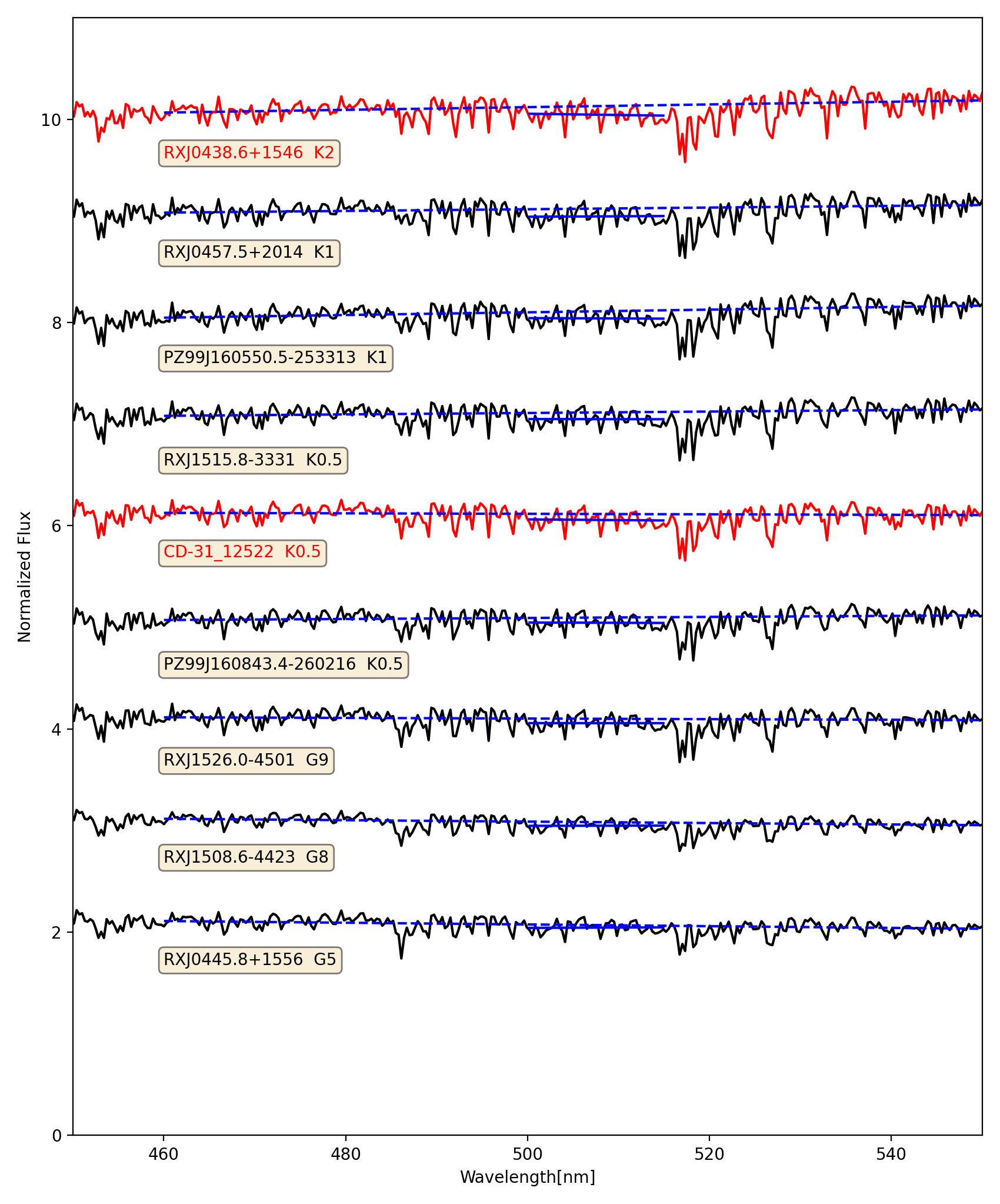}
    \caption{X Shooter spectra of Class III YSO with spectral types ranging from K2 to G5 ordered according to SpT and normalized at 520 nm. (continued on Fig. \ref{fig:LuhmanUVB5101}). Spectra presented in this work are indicated in red and spectra from the samples of \citetalias{manara13a} and \citetalias{manara17b} are indicated in black.}
    \label{fig:LuhmanUVB5101}
\end{figure*}

\end{document}